\documentclass[longauth]{aa}

\usepackage{graphicx}
\usepackage{natbib}
\usepackage{scalerel}
\usepackage{xspace}
\usepackage{caption}
\usepackage{comment}

\newcommand{\Eridanus}{\mbox{Eridanus~V}}
\newcommand{\Fornax}{\mbox{Fornax~7}}
\newcommand{\UMi}{\mbox{Ursa Minor~II}}
\newcommand{\EuclidI}{\mbox{Euclid~LVDw~DR1~2}}
\newcommand{\EuclidII}{\mbox{Euclid~LVDw~DR1~1}}
\newcommand{\EuclidIII}{\mbox{Euclid~LVDw~DR1~12}}
\newcommand{\EuclidIV}{\mbox{Euclid~LVDw~DR1~3}}
\newcommand{\EuclidV}{\mbox{Euclid~LVDw~DR1~9}}
\newcommand{\EuclidVI}{\mbox{Euclid~LVDw~DR1~6}}
\newcommand{\EuclidVII}{\mbox{Euclid~LVDw~DR1~5}}
\newcommand{\EuclidVIII}{\mbox{Euclid~LVDw~DR1~7}}
\newcommand{\EuclidIX}{\mbox{Euclid~LVDw~DR1~11}}
\newcommand{\EuclidX}{\mbox{Euclid~LVDw~DR1~8}}
\newcommand{\EuclidXI}{\mbox{Euclid~LVDw~DR1~10}}
\newcommand{\EuclidXII}{\mbox{Euclid~LVDw~DR1~4}}
\newcommand{\feh}{\mbox{$\rm [Fe/H]$}}
\newcommand{\magarc}{mag\,arcsec$^{-2}$}

\usepackage[table]{xcolor}
\usepackage{rotating}
\usepackage{txfonts}
\usepackage[pdfencoding=auto,psdextra]{hyperref}
\hypersetup{
    colorlinks=true,
    linkcolor=blue,
    filecolor=magenta,      
    urlcolor=blue,
    citecolor=blue
}
\def\massi#1{#1}

\makeatletter
\renewcommand*\aa@pageof{, page \thepage{} of \pageref*{LastPage}}
\makeatother

\usepackage[utf8]{inputenc}

\renewcommand{\linenumbers}[0]{}

\usepackage{euclid}

\begin{document}
%
% Put the title and authors of your (Standard Project) paper here
%

\title{Euclid Data Release 1 (DR1)}
\subtitle{Pushing back the borders of the invisible with \Euclid. Discovery of fifteen resolved low-surface-brightness stellar systems from the Milky Way to the Local Volume}
%\thanks{This paper is published on
%       behalf of the Euclid Consortium}}

%% please do not edit the author list once you copy it from the
%% Publication Portal -- contact ECEB Bureau for changes

%% please do not edit the author list -- contact ECEB Bureau for changes
% \newcommand{\orcid}[1]{} %% define as link to https://orcid.org/#1 if needed
\newcommand{\orcid}[1]{} %% if already defined in aa.cls: comment, or use renewcommand			   

\author{Euclid Collaboration: M.~Gatto\orcid{0000-0003-4636-6457}\thanks{\email{massimiliano.gatto@inaf.it}}\inst{\ref{aff1}}
\and S.~Rusterucci\orcid{0009-0006-4379-9934}\inst{\ref{aff2},\ref{aff3}}
\and M.~Bellazzini\orcid{0000-0001-8200-810X}\inst{\ref{aff4}}
\and N.~F.~Martin\orcid{0000-0002-1349-202X}\inst{\ref{aff2},\ref{aff5}}
\and A.~M.~N.~Ferguson\orcid{0000-0001-7934-1278}\inst{\ref{aff6}}
\and M.~Libralato\orcid{0000-0001-9673-7397}\inst{\ref{aff7}}
\and R.~Pascale\orcid{0000-0002-6389-6268}\inst{\ref{aff4}}
\and L.~R.~Bedin\orcid{0000-0003-4080-6466}\inst{\ref{aff7}}
\and T.~Saifollahi\orcid{0000-0002-9554-7660}\inst{\ref{aff2},\ref{aff8}}
\and C.~Tortora\orcid{0000-0001-7958-6531}\inst{\ref{aff1}}
\and E.~Balbinot\orcid{0000-0002-1322-3153}\inst{\ref{aff3},\ref{aff9}}
\and M.~Griggio\orcid{0000-0002-5060-1379}\inst{\ref{aff10}}
\and F.~Annibali\inst{\ref{aff4}}
\and G.~Battaglia\orcid{0000-0002-6551-4294}\inst{\ref{aff11},\ref{aff12}}
\and M.~L.~M.~Collins\orcid{0000-0002-1693-3265}\inst{\ref{aff13}}
\and M.~Correnti\orcid{0000-0001-6464-3257}\inst{\ref{aff14},\ref{aff15}}
\and E.~Dalessandro\orcid{0000-0003-4237-4601}\inst{\ref{aff4}}
\and A.~Doliva-Dolinsky\orcid{0000-0001-9775-9029}\inst{\ref{aff13}}
\and P.-A.~Duc\orcid{0000-0003-3343-6284}\inst{\ref{aff2}}
\and D.~Erkal\orcid{0000-0002-8448-5505}\inst{\ref{aff13}}
\and P.~Jablonka\orcid{0000-0002-9655-1063}\inst{\ref{aff16}}
\and J.~Junais\orcid{0000-0002-7016-4532}\inst{\ref{aff17}}
\and M.~Marconi\orcid{0000-0002-1330-2927}\inst{\ref{aff1}}
\and F.~R.~Marleau\orcid{0000-0002-1442-2947}\inst{\ref{aff18}}
\and D.~Mart\'inez-Delgado\inst{\ref{aff19},\ref{aff20}}
\and D.~Massari\orcid{0000-0001-8892-4301}\inst{\ref{aff4}}
\and I.~Musella\orcid{0000-0001-5909-6615}\inst{\ref{aff1}}
\and N.~R.~Napolitano\orcid{0000-0003-0911-8884}\inst{\ref{aff21},\ref{aff1},\ref{aff22}}
\and F.~Niederhofer\orcid{0000-0002-4341-9819}\inst{\ref{aff17}}
\and Y.~Revaz\orcid{0000-0002-6227-0108}\inst{\ref{aff16}}
\and V.~Ripepi\orcid{0000-0003-1801-426X}\inst{\ref{aff1}}
\and J.~D.~Sakowska\orcid{0000-0002-1594-1466}\inst{\ref{aff23}}
\and S.~E.~T.~Smith\orcid{0000-0002-6946-8280}\inst{\ref{aff24}}
\and E.~Starkenburg\orcid{0000-0003-4501-103X}\inst{\ref{aff3}}
\and S.~Taibi\orcid{0000-0001-6469-8805}\inst{\ref{aff16}}
\and G.~Thomas\orcid{0000-0002-2468-5521}\inst{\ref{aff12},\ref{aff11}}
\and K.~Voggel\orcid{0000-0001-6215-0950}\inst{\ref{aff2}}
\and B.~Altieri\orcid{0000-0003-3936-0284}\inst{\ref{aff25}}
\and S.~Andreon\orcid{0000-0002-2041-8784}\inst{\ref{aff26}}
\and N.~Auricchio\orcid{0000-0003-4444-8651}\inst{\ref{aff4}}
\and S.~Awan\inst{\ref{aff27}}
\and C.~Baccigalupi\orcid{0000-0002-8211-1630}\inst{\ref{aff28},\ref{aff29},\ref{aff30},\ref{aff31}}
\and M.~Baldi\orcid{0000-0003-4145-1943}\inst{\ref{aff32},\ref{aff4},\ref{aff33}}
\and S.~Bardelli\orcid{0000-0002-8900-0298}\inst{\ref{aff4}}
\and P.~Battaglia\orcid{0000-0002-7337-5909}\inst{\ref{aff4}}
\and R.~Bender\orcid{0000-0001-7179-0626}\inst{\ref{aff34},\ref{aff35}}
\and A.~Biviano\orcid{0000-0002-0857-0732}\inst{\ref{aff29},\ref{aff28}}
\and E.~Branchini\orcid{0000-0002-0808-6908}\inst{\ref{aff36},\ref{aff37},\ref{aff26}}
\and M.~Brescia\orcid{0000-0001-9506-5680}\inst{\ref{aff38},\ref{aff1}}
\and J.~Brinchmann\orcid{0000-0003-4359-8797}\inst{\ref{aff39},\ref{aff40},\ref{aff41}}
\and S.~Camera\orcid{0000-0003-3399-3574}\inst{\ref{aff42},\ref{aff43},\ref{aff44}}
\and G.~Ca\~nas-Herrera\orcid{0000-0003-2796-2149}\inst{\ref{aff9}}
\and V.~Capobianco\orcid{0000-0002-3309-7692}\inst{\ref{aff44}}
\and C.~Carbone\orcid{0000-0003-0125-3563}\inst{\ref{aff45}}
\and J.~Carretero\orcid{0000-0002-3130-0204}\inst{\ref{aff46},\ref{aff47}}
\and M.~Castellano\orcid{0000-0001-9875-8263}\inst{\ref{aff14}}
\and G.~Castignani\orcid{0000-0001-6831-0687}\inst{\ref{aff4}}
\and S.~Cavuoti\orcid{0000-0002-3787-4196}\inst{\ref{aff1},\ref{aff22}}
\and K.~C.~Chambers\orcid{0000-0001-6965-7789}\inst{\ref{aff48}}
\and A.~Cimatti\inst{\ref{aff49}}
\and C.~Colodro-Conde\inst{\ref{aff11}}
\and G.~Congedo\orcid{0000-0003-2508-0046}\inst{\ref{aff6}}
\and C.~J.~Conselice\orcid{0000-0003-1949-7638}\inst{\ref{aff50}}
\and L.~Conversi\orcid{0000-0002-6710-8476}\inst{\ref{aff51},\ref{aff25}}
\and Y.~Copin\orcid{0000-0002-5317-7518}\inst{\ref{aff52}}
\and A.~Costille\inst{\ref{aff53}}
\and F.~Courbin\orcid{0000-0003-0758-6510}\inst{\ref{aff54},\ref{aff55},\ref{aff56}}
\and H.~M.~Courtois\orcid{0000-0003-0509-1776}\inst{\ref{aff57}}
\and M.~Cropper\orcid{0000-0003-4571-9468}\inst{\ref{aff27}}
\and H.~Degaudenzi\orcid{0000-0002-5887-6799}\inst{\ref{aff58}}
\and G.~De~Lucia\orcid{0000-0002-6220-9104}\inst{\ref{aff29}}
\and C.~Dolding\orcid{0009-0003-7199-6108}\inst{\ref{aff27}}
\and H.~Dole\orcid{0000-0002-9767-3839}\inst{\ref{aff59}}
\and F.~Dubath\orcid{0000-0002-6533-2810}\inst{\ref{aff58}}
\and X.~Dupac\inst{\ref{aff25}}
\and S.~Dusini\orcid{0000-0002-1128-0664}\inst{\ref{aff60}}
\and S.~Escoffier\orcid{0000-0002-2847-7498}\inst{\ref{aff61}}
\and M.~Farina\orcid{0000-0002-3089-7846}\inst{\ref{aff62}}
\and R.~Farinelli\inst{\ref{aff4}}
\and F.~Faustini\orcid{0000-0001-6274-5145}\inst{\ref{aff14}}
\and S.~Ferriol\inst{\ref{aff52}}
\and M.~Frailis\orcid{0000-0002-7400-2135}\inst{\ref{aff29}}
\and E.~Franceschi\orcid{0000-0002-0585-6591}\inst{\ref{aff4}}
\and M.~Fumana\orcid{0000-0001-6787-5950}\inst{\ref{aff45}}
\and L.~Gabarra\orcid{0000-0002-8486-8856}\inst{\ref{aff63}}
\and S.~Galeotta\orcid{0000-0002-3748-5115}\inst{\ref{aff29}}
\and K.~George\orcid{0000-0002-1734-8455}\inst{\ref{aff64}}
\and B.~Gillis\orcid{0000-0002-4478-1270}\inst{\ref{aff6}}
\and C.~Giocoli\orcid{0000-0002-9590-7961}\inst{\ref{aff4},\ref{aff33}}
\and J.~Gracia-Carpio\orcid{0000-0003-4689-3134}\inst{\ref{aff34}}
\and A.~Grazian\orcid{0000-0002-5688-0663}\inst{\ref{aff7}}
\and F.~Grupp\inst{\ref{aff34},\ref{aff35}}
\and L.~Guzzo\orcid{0000-0001-8264-5192}\inst{\ref{aff65},\ref{aff26},\ref{aff66}}
\and W.~G.~Hartley\orcid{0000-0002-4563-5101}\inst{\ref{aff58}}
\and S.~V.~H.~Haugan\orcid{0000-0001-9648-7260}\inst{\ref{aff67}}
\and S.~Hemmati\orcid{0000-0003-2226-5395}\inst{\ref{aff68}}
\and H.~Hoekstra\orcid{0000-0002-0641-3231}\inst{\ref{aff9}}
\and W.~Holmes\orcid{0009-0007-8554-4646}\inst{\ref{aff69}}
\and I.~M.~Hook\orcid{0000-0002-2960-978X}\inst{\ref{aff70}}
\and F.~Hormuth\inst{\ref{aff71}}
\and A.~Hornstrup\orcid{0000-0002-3363-0936}\inst{\ref{aff72},\ref{aff73}}
\and M.~Huertas-Company\orcid{0000-0002-1416-8483}\inst{\ref{aff11},\ref{aff74},\ref{aff75}}
\and K.~Jahnke\orcid{0000-0003-3804-2137}\inst{\ref{aff5}}
\and M.~Jhabvala\inst{\ref{aff76}}
\and S.~Kermiche\orcid{0000-0002-0302-5735}\inst{\ref{aff61}}
\and A.~Kiessling\orcid{0000-0002-2590-1273}\inst{\ref{aff69}}
\and B.~Kubik\orcid{0009-0006-5823-4880}\inst{\ref{aff52}}
\and K.~Kuijken\orcid{0000-0002-3827-0175}\inst{\ref{aff9}}
\and M.~K\"ummel\orcid{0000-0003-2791-2117}\inst{\ref{aff35}}
\and M.~Kunz\orcid{0000-0002-3052-7394}\inst{\ref{aff77}}
\and H.~Kurki-Suonio\orcid{0000-0002-4618-3063}\inst{\ref{aff78},\ref{aff79}}
\and A.~M.~C.~Le~Brun\orcid{0000-0002-0936-4594}\inst{\ref{aff80}}
\and D.~Le~Mignant\orcid{0000-0002-5339-5515}\inst{\ref{aff53}}
\and P.~Liebing\orcid{0009-0005-8043-7150}\inst{\ref{aff27}}
\and S.~Ligori\orcid{0000-0003-4172-4606}\inst{\ref{aff44}}
\and P.~B.~Lilje\orcid{0000-0003-4324-7794}\inst{\ref{aff67}}
\and V.~Lindholm\orcid{0000-0003-2317-5471}\inst{\ref{aff78},\ref{aff79}}
\and I.~Lloro\orcid{0000-0001-5966-1434}\inst{\ref{aff81}}
\and M.~Magliocchetti\orcid{0000-0001-9158-4838}\inst{\ref{aff62}}
\and G.~Mainetti\orcid{0000-0003-2384-2377}\inst{\ref{aff82}}
\and O.~Mansutti\orcid{0000-0001-5758-4658}\inst{\ref{aff29}}
\and O.~Marggraf\orcid{0000-0001-7242-3852}\inst{\ref{aff83}}
\and M.~Martinelli\orcid{0000-0002-6943-7732}\inst{\ref{aff14},\ref{aff84}}
\and N.~Martinet\orcid{0000-0003-2786-7790}\inst{\ref{aff53}}
\and F.~Marulli\orcid{0000-0002-8850-0303}\inst{\ref{aff85},\ref{aff4},\ref{aff33}}
\and R.~J.~Massey\orcid{0000-0002-6085-3780}\inst{\ref{aff86}}
\and N.~Mauri\orcid{0000-0001-8196-1548}\inst{\ref{aff49},\ref{aff33}}
\and E.~Medinaceli\orcid{0000-0002-4040-7783}\inst{\ref{aff4}}
\and S.~Mei\orcid{0000-0002-2849-559X}\inst{\ref{aff87},\ref{aff88}}
\and M.~Melchior\inst{\ref{aff89}}
\and M.~Meneghetti\orcid{0000-0003-1225-7084}\inst{\ref{aff4},\ref{aff33}}
\and E.~Merlin\orcid{0000-0001-6870-8900}\inst{\ref{aff7}}
\and G.~Meylan\orcid{0000-0001-6503-0209}\inst{\ref{aff16}}
\and P.~Monaco\orcid{0000-0003-2083-7564}\inst{\ref{aff90},\ref{aff29},\ref{aff30},\ref{aff28}}
\and A.~Mora\orcid{0000-0002-1922-8529}\inst{\ref{aff91}}
\and M.~Moresco\orcid{0000-0002-7616-7136}\inst{\ref{aff85},\ref{aff4}}
\and C.~Moretti\orcid{0000-0003-3314-8936}\inst{\ref{aff29},\ref{aff28},\ref{aff30}}
\and L.~Moscardini\orcid{0000-0002-3473-6716}\inst{\ref{aff85},\ref{aff4},\ref{aff33}}
\and R.~Nakajima\orcid{0009-0009-1213-7040}\inst{\ref{aff83}}
\and C.~Neissner\orcid{0000-0001-8524-4968}\inst{\ref{aff92},\ref{aff47}}
\and S.-M.~Niemi\orcid{0009-0005-0247-0086}\inst{\ref{aff93}}
\and C.~Padilla\orcid{0000-0001-7951-0166}\inst{\ref{aff92}}
\and S.~Paltani\orcid{0000-0002-8108-9179}\inst{\ref{aff58}}
\and F.~Pasian\orcid{0000-0002-4869-3227}\inst{\ref{aff29}}
\and W.~J.~Percival\orcid{0000-0002-0644-5727}\inst{\ref{aff94},\ref{aff95},\ref{aff96}}
\and V.~Pettorino\orcid{0000-0002-4203-9320}\inst{\ref{aff93}}
\and A.~Pezzotta\orcid{0000-0003-0726-2268}\inst{\ref{aff26}}
\and S.~Pires\orcid{0000-0002-0249-2104}\inst{\ref{aff97}}
\and G.~Polenta\orcid{0000-0003-4067-9196}\inst{\ref{aff15}}
\and M.~Poncet\inst{\ref{aff8}}
\and L.~A.~Popa\inst{\ref{aff98}}
\and F.~Raison\orcid{0000-0002-7819-6918}\inst{\ref{aff34}}
\and A.~Renzi\orcid{0000-0001-9856-1970}\inst{\ref{aff99},\ref{aff60},\ref{aff4}}
\and J.~Rhodes\orcid{0000-0002-4485-8549}\inst{\ref{aff69}}
\and G.~Riccio\inst{\ref{aff1}}
\and I.~Risso\orcid{0000-0003-2525-7761}\inst{\ref{aff36},\ref{aff37},\ref{aff26}}
\and F.~Rizzo\orcid{0000-0002-9407-585X}\inst{\ref{aff29}}
\and E.~Romelli\orcid{0000-0003-3069-9222}\inst{\ref{aff29}}
\and M.~Roncarelli\orcid{0000-0001-9587-7822}\inst{\ref{aff4}}
\and B.~Rusholme\orcid{0000-0001-7648-4142}\inst{\ref{aff68}}
\and R.~Saglia\orcid{0000-0003-0378-7032}\inst{\ref{aff35},\ref{aff34}}
\and Z.~Sakr\orcid{0000-0002-4823-3757}\inst{\ref{aff100},\ref{aff101},\ref{aff102}}
\and D.~Sapone\orcid{0000-0001-7089-4503}\inst{\ref{aff103}}
\and B.~Sartoris\orcid{0000-0003-1337-5269}\inst{\ref{aff104}}
\and M.~Schirmer\orcid{0000-0003-2568-9994}\inst{\ref{aff5}}
\and P.~Schneider\orcid{0000-0001-8561-2679}\inst{\ref{aff83}}
\and A.~Secroun\orcid{0000-0003-0505-3710}\inst{\ref{aff61}}
\and E.~Sihvola\orcid{0000-0003-1804-7715}\inst{\ref{aff105}}
\and P.~Simon\inst{\ref{aff83}}
\and C.~Sirignano\orcid{0000-0002-0995-7146}\inst{\ref{aff99},\ref{aff60}}
\and G.~Sirri\orcid{0000-0003-2626-2853}\inst{\ref{aff33}}
\and J.~Skottfelt\orcid{0000-0003-1310-8283}\inst{\ref{aff106}}
\and L.~Stanco\orcid{0000-0002-9706-5104}\inst{\ref{aff60}}
\and P.~Tallada-Cresp\'{i}\orcid{0000-0002-1336-8328}\inst{\ref{aff46},\ref{aff47}}
\and A.~N.~Taylor\inst{\ref{aff6}}
\and I.~Tereno\orcid{0000-0002-4537-6218}\inst{\ref{aff107},\ref{aff108}}
\and N.~Tessore\orcid{0000-0002-9696-7931}\inst{\ref{aff27}}
\and S.~Toft\orcid{0000-0003-3631-7176}\inst{\ref{aff109},\ref{aff110}}
\and R.~Toledo-Moreo\orcid{0000-0002-2997-4859}\inst{\ref{aff111},\ref{aff112}}
\and F.~Torradeflot\orcid{0000-0003-1160-1517}\inst{\ref{aff47},\ref{aff46}}
\and A.~Tsyganov\inst{\ref{aff113}}
\and I.~Tutusaus\orcid{0000-0002-3199-0399}\inst{\ref{aff114},\ref{aff115},\ref{aff101}}
\and J.~Valiviita\orcid{0000-0001-6225-3693}\inst{\ref{aff78},\ref{aff79}}
\and T.~Vassallo\orcid{0000-0001-6512-6358}\inst{\ref{aff29},\ref{aff64}}
\and G.~Verdoes~Kleijn\orcid{0000-0001-5803-2580}\inst{\ref{aff3}}
\and Y.~Wang\orcid{0000-0002-4749-2984}\inst{\ref{aff68}}
\and J.~Weller\orcid{0000-0002-8282-2010}\inst{\ref{aff35},\ref{aff34}}
\and A.~Zacchei\orcid{0000-0003-0396-1192}\inst{\ref{aff29},\ref{aff28}}
\and G.~Zamorani\orcid{0000-0002-2318-301X}\inst{\ref{aff4}}
\and F.~M.~Zerbi\orcid{0000-0002-9996-973X}\inst{\ref{aff26}}
\and I.~A.~Zinchenko\orcid{0000-0002-2944-2449}\inst{\ref{aff116}}
\and E.~Zucca\orcid{0000-0002-5845-8132}\inst{\ref{aff4}}
\and V.~Allevato\orcid{0000-0001-7232-5152}\inst{\ref{aff1}}
\and I.~T.~Andika\orcid{0000-0001-6102-9526}\inst{\ref{aff35}}
\and M.~Ballardini\orcid{0000-0003-4481-3559}\inst{\ref{aff117},\ref{aff118},\ref{aff4}}
\and P.~Bergamini\orcid{0000-0003-1383-9414}\inst{\ref{aff4}}
\and S.~Borgani\orcid{0000-0001-6151-6439}\inst{\ref{aff90},\ref{aff28},\ref{aff29},\ref{aff30},\ref{aff119}}
\and E.~Bozzo\orcid{0000-0002-8201-1525}\inst{\ref{aff58}}
\and C.~Burigana\orcid{0000-0002-3005-5796}\inst{\ref{aff120},\ref{aff121}}
\and R.~Cabanac\orcid{0000-0001-6679-2600}\inst{\ref{aff101}}
\and A.~Cappi\inst{\ref{aff104},\ref{aff4}}
\and F.~Caro\orcid{0009-0003-1053-0507}\inst{\ref{aff14}}
\and T.~Castro\orcid{0000-0002-6292-3228}\inst{\ref{aff122},\ref{aff29},\ref{aff28}}
\and D.~Di~Ferdinando\inst{\ref{aff33}}
\and J.~A.~Escartin~Vigo\inst{\ref{aff34}}
\and J.~Garc\'ia-Bellido\orcid{0000-0002-9370-8360}\inst{\ref{aff100}}
\and T.~Gasparetto\orcid{0000-0002-7913-4866}\inst{\ref{aff14}}
\and V.~Gautard\inst{\ref{aff123}}
\and C.~M.~Gutierrez\orcid{0000-0001-7854-783X}\inst{\ref{aff11},\ref{aff12}}
\and A.~Hall\orcid{0000-0002-3139-8651}\inst{\ref{aff6}}
\and Y.~Kang\orcid{0009-0000-8588-7250}\inst{\ref{aff18}}
\and A.~Loureiro\orcid{0000-0002-4371-0876}\inst{\ref{aff124},\ref{aff125}}
\and J.~Macias-Perez\orcid{0000-0002-5385-2763}\inst{\ref{aff126}}
\and R.~Maoli\orcid{0000-0002-6065-3025}\inst{\ref{aff127},\ref{aff14}}
\and R.~B.~Metcalf\orcid{0000-0003-3167-2574}\inst{\ref{aff85},\ref{aff4}}
\and A.~A.~Nucita\inst{\ref{aff128},\ref{aff129},\ref{aff130}}
\and M.~P\"ontinen\orcid{0000-0001-5442-2530}\inst{\ref{aff78}}
\and E.~Sarpa\orcid{0000-0002-1256-655X}\inst{\ref{aff29}}
\and V.~Scottez\orcid{0009-0008-3864-940X}\inst{\ref{aff131},\ref{aff132}}
\and M.~Sereno\orcid{0000-0003-0302-0325}\inst{\ref{aff4},\ref{aff33}}
\and K.~Tanidis\orcid{0000-0001-9843-5130}\inst{\ref{aff133}}
\and M.~Tenti\orcid{0000-0002-4254-5901}\inst{\ref{aff33}}
\and M.~Tucci\inst{\ref{aff58}}
\and M.~Wiesmann\orcid{0009-0000-8199-5860}\inst{\ref{aff67}}
\and J.~A.~Acevedo~Barroso\orcid{0000-0002-9654-1711}\inst{\ref{aff69}}
\and Y.~Akrami\orcid{0000-0002-2407-7956}\inst{\ref{aff100},\ref{aff134}}
\and G.~Angora\orcid{0000-0002-0316-6562}\inst{\ref{aff1},\ref{aff117}}
\and S.~Anselmi\orcid{0000-0002-3579-9583}\inst{\ref{aff60},\ref{aff99},\ref{aff135}}
\and M.~Archidiacono\orcid{0000-0003-4952-9012}\inst{\ref{aff65},\ref{aff66}}
\and G.~Aric\`o\orcid{0000-0002-2802-2928}\inst{\ref{aff33}}
\and F.~Atrio-Barandela\orcid{0000-0002-2130-2513}\inst{\ref{aff136}}
\and M.~Baes\orcid{0000-0002-3930-2757}\inst{\ref{aff137}}
\and L.~Bazzanini\orcid{0000-0003-0727-0137}\inst{\ref{aff117},\ref{aff4}}
\and J.~R.~Bermejo-Climent\inst{\ref{aff138},\ref{aff139}}
\and D.~Bertacca\orcid{0000-0002-2490-7139}\inst{\ref{aff99},\ref{aff7},\ref{aff60}}
\and M.~Bethermin\orcid{0000-0002-3915-2015}\inst{\ref{aff2}}
\and F.~Beutler\orcid{0000-0003-0467-5438}\inst{\ref{aff6}}
\and H.~B\"ohringer\orcid{0000-0001-8241-4204}\inst{\ref{aff34},\ref{aff64},\ref{aff140}}
\and F.~Bouch\`e\orcid{0000-0002-4663-1786}\inst{\ref{aff141},\ref{aff22}}
\and M.~L.~Brown\orcid{0000-0002-0370-8077}\inst{\ref{aff50}}
\and S.~Bruton\orcid{0000-0002-6503-5218}\inst{\ref{aff142}}
\and A.~Calabro\orcid{0000-0003-2536-1614}\inst{\ref{aff14}}
\and B.~Camacho~Quevedo\orcid{0000-0002-8789-4232}\inst{\ref{aff28},\ref{aff31},\ref{aff29}}
\and C.~S.~Carvalho\inst{\ref{aff143}}
\and F.~Cogato\orcid{0000-0003-4632-6113}\inst{\ref{aff4}}
\and T.~E.~Collett\orcid{0000-0001-5564-3140}\inst{\ref{aff144}}
\and S.~Conseil\orcid{0000-0002-3657-4191}\inst{\ref{aff145},\ref{aff52}}
\and A.~R.~Cooray\orcid{0000-0002-3892-0190}\inst{\ref{aff146}}
\and P.~Corcho-Caballero\orcid{0000-0001-6327-7080}\inst{\ref{aff3}}
\and B.~Csizi\orcid{0000-0003-3227-6581}\inst{\ref{aff18}}
\and O.~Cucciati\orcid{0000-0002-9336-7551}\inst{\ref{aff4}}
\and S.~Davini\orcid{0000-0003-3269-1718}\inst{\ref{aff37}}
\and T.~de~Boer\orcid{0000-0001-5486-2747}\inst{\ref{aff48}}
\and F.~De~Paolis\orcid{0000-0001-6460-7563}\inst{\ref{aff128},\ref{aff129},\ref{aff130}}
\and G.~Desprez\orcid{0000-0001-8325-1742}\inst{\ref{aff3}}
\and A.~D\'iaz-S\'anchez\orcid{0000-0003-0748-4768}\inst{\ref{aff147}}
\and S.~Di~Domizio\orcid{0000-0003-2863-5895}\inst{\ref{aff36},\ref{aff37}}
\and J.~M.~Diego\orcid{0000-0001-9065-3926}\inst{\ref{aff148}}
\and V.~Duret\orcid{0009-0009-0383-4960}\inst{\ref{aff61}}
\and A.~Enia\orcid{0000-0002-0200-2857}\inst{\ref{aff123},\ref{aff4}}
\and Y.~Fang\orcid{0000-0002-0334-6950}\inst{\ref{aff35}}
\and A.~Farina\orcid{0009-0000-3420-929X}\inst{\ref{aff26},\ref{aff37}}
\and A.~Finoguenov\orcid{0000-0002-4606-5403}\inst{\ref{aff78}}
\and A.~Franco\orcid{0000-0002-4761-366X}\inst{\ref{aff128},\ref{aff129},\ref{aff130}}
\and A.~Fumagalli\orcid{0009-0004-0300-2535}\inst{\ref{aff29}}
\and Y.~Fu\orcid{0000-0002-0759-0504}\inst{\ref{aff9},\ref{aff3}}
\and K.~Ganga\orcid{0000-0001-8159-8208}\inst{\ref{aff87}}
\and R.~Gavazzi\orcid{0000-0002-5540-6935}\inst{\ref{aff53},\ref{aff149}}
\and E.~Gaztanaga\orcid{0000-0001-9632-0815}\inst{\ref{aff114},\ref{aff115},\ref{aff144}}
\and Z.~Ghaffari\orcid{0000-0002-6467-8078}\inst{\ref{aff29},\ref{aff28}}
\and F.~Giacomini\orcid{0000-0002-3129-2814}\inst{\ref{aff33}}
\and F.~Gianotti\orcid{0000-0003-4666-119X}\inst{\ref{aff4}}
\and G.~Gozaliasl\orcid{0000-0002-0236-919X}\inst{\ref{aff150},\ref{aff78}}
\and A.~Gruppuso\orcid{0000-0001-9272-5292}\inst{\ref{aff4},\ref{aff33}}
\and M.~Guidi\orcid{0000-0001-9408-1101}\inst{\ref{aff32},\ref{aff4}}
\and N.~A.~Hatch\orcid{0000-0001-5600-0534}\inst{\ref{aff151}}
\and C.~Hern\'andez-Monteagudo\orcid{0000-0001-5471-9166}\inst{\ref{aff12},\ref{aff11}}
\and H.~Hildebrandt\orcid{0000-0002-9814-3338}\inst{\ref{aff152}}
\and J.~Hjorth\orcid{0000-0002-4571-2306}\inst{\ref{aff153}}
\and M.~Jauzac\orcid{0000-0003-1974-8732}\inst{\ref{aff154},\ref{aff86},\ref{aff155},\ref{aff156},\ref{aff101}}
\and J.~J.~E.~Kajava\orcid{0000-0002-3010-8333}\inst{\ref{aff157},\ref{aff158},\ref{aff159}}
\and V.~Kansal\orcid{0000-0002-4008-6078}\inst{\ref{aff160},\ref{aff161}}
\and D.~Karagiannis\orcid{0000-0002-4927-0816}\inst{\ref{aff117},\ref{aff162},\ref{aff163}}
\and J.~Kim\orcid{0000-0003-2776-2761}\inst{\ref{aff63}}
\and C.~C.~Kirkpatrick\inst{\ref{aff105}}
\and A.~Kov\'acs\orcid{0000-0002-5825-579X}\inst{\ref{aff139},\ref{aff138}}
\and I.~Kova{\v{c}}i{\'{c}}\orcid{0000-0001-6751-3263}\inst{\ref{aff137}}
\and K.~Koyama\orcid{0000-0001-6727-6915}\inst{\ref{aff144}}
\and S.~Kruk\orcid{0000-0001-8010-8879}\inst{\ref{aff25}}
\and M.~C.~Lam\orcid{0000-0002-9347-2298}\inst{\ref{aff6}}
\and M.~Lattanzi\orcid{0000-0003-1059-2532}\inst{\ref{aff118}}
\and S.~Lee\orcid{0000-0002-8289-740X}\inst{\ref{aff164},\ref{aff69}}
\and M.~D.~Lepinzan\orcid{0000-0003-1287-9801}\inst{\ref{aff29}}
\and G.~Leroy\orcid{0009-0004-2523-4425}\inst{\ref{aff154},\ref{aff86}}
\and J.~Lesgourgues\orcid{0000-0001-7627-353X}\inst{\ref{aff165}}
\and L.~Linke\orcid{0000-0002-2622-8113}\inst{\ref{aff18}}
\and S.~J.~Liu\orcid{0000-0001-7680-2139}\inst{\ref{aff62}}
\and G.~Maggio\orcid{0000-0003-4020-4836}\inst{\ref{aff29}}
\and F.~Mannucci\orcid{0000-0002-4803-2381}\inst{\ref{aff166}}
\and C.~J.~A.~P.~Martins\orcid{0000-0002-4886-9261}\inst{\ref{aff167},\ref{aff39}}
\and M.~Miluzio\inst{\ref{aff25},\ref{aff168}}
\and G.~Morgante\inst{\ref{aff4}}
\and K.~Naidoo\orcid{0000-0002-9182-1802}\inst{\ref{aff144},\ref{aff5}}
\and P.~Natoli\orcid{0000-0003-0126-9100}\inst{\ref{aff117},\ref{aff118}}
\and A.~Navarro-Alsina\orcid{0000-0002-3173-2592}\inst{\ref{aff83}}
\and S.~Nesseris\orcid{0000-0002-0567-0324}\inst{\ref{aff100}}
\and L.~Nicastro\orcid{0000-0001-8534-6788}\inst{\ref{aff4}}
\and F.~Oppizzi\orcid{0000-0003-3904-8370}\inst{\ref{aff37},\ref{aff60},\ref{aff99}}
\and F.~Pace\orcid{0000-0001-8039-0480}\inst{\ref{aff42},\ref{aff43},\ref{aff44},\ref{aff108}}
\and D.~Paoletti\orcid{0000-0003-4761-6147}\inst{\ref{aff4},\ref{aff121}}
\and F.~Passalacqua\orcid{0000-0002-8606-4093}\inst{\ref{aff60}}
\and K.~Paterson\orcid{0000-0001-8340-3486}\inst{\ref{aff5}}
\and L.~Patrizii\inst{\ref{aff33}}
\and C.~Pattison\orcid{0000-0003-3272-2617}\inst{\ref{aff144}}
\and A.~Pisani\orcid{0000-0002-6146-4437}\inst{\ref{aff61}}
\and L.~Pizzuti\orcid{0000-0001-5654-7580}\inst{\ref{aff169}}
\and D.~Potter\orcid{0000-0002-0757-5195}\inst{\ref{aff170}}
\and M.~Poulain\orcid{0000-0002-7664-4510}\inst{\ref{aff171}}
\and G.~W.~Pratt\inst{\ref{aff97}}
\and S.~Quai\orcid{0000-0002-0449-8163}\inst{\ref{aff85},\ref{aff4}}
\and M.~Radovich\orcid{0000-0002-3585-866X}\inst{\ref{aff7}}
\and G.~Rodighiero\orcid{0000-0002-9415-2296}\inst{\ref{aff99},\ref{aff7}}
\and W.~Roster\orcid{0000-0002-9149-6528}\inst{\ref{aff34}}
\and M.~Sahl\'en\orcid{0000-0003-0973-4804}\inst{\ref{aff172}}
\and D.~B.~Sanders\orcid{0000-0002-1233-9998}\inst{\ref{aff48}}
\and J.~Schaye\orcid{0000-0002-0668-5560}\inst{\ref{aff9}}
\and A.~Schneider\orcid{0000-0001-7055-8104}\inst{\ref{aff170}}
\and D.~Sciotti\orcid{0009-0008-4519-2620}\inst{\ref{aff14},\ref{aff84}}
\and E.~Sellentin\orcid{0009-0002-2655-3458}\inst{\ref{aff173},\ref{aff9}}
\and S.~Serjeant\orcid{0000-0002-0517-7943}\inst{\ref{aff174}}
\and F.~Shankar\orcid{0000-0001-8973-5051}\inst{\ref{aff175}}
\and L.~C.~Smith\orcid{0000-0002-3259-2771}\inst{\ref{aff176}}
\and J.~G.~Sorce\orcid{0000-0002-2307-2432}\inst{\ref{aff177},\ref{aff59}}
\and I.~Szapudi\orcid{0000-0003-2274-0301}\inst{\ref{aff48}}
\and F.~Tarsitano\orcid{0000-0002-5919-0238}\inst{\ref{aff178},\ref{aff179}}
\and G.~Testera\orcid{0000-0003-2970-766X}\inst{\ref{aff37}}
\and R.~Teyssier\orcid{0000-0001-7689-0933}\inst{\ref{aff180}}
\and S.~Tosi\orcid{0000-0002-7275-9193}\inst{\ref{aff36},\ref{aff26},\ref{aff37}}
\and A.~Troja\orcid{0000-0003-0239-4595}\inst{\ref{aff29}}
\and C.~Uhlemann\orcid{0000-0001-7831-1579}\inst{\ref{aff181},\ref{aff182}}
\and M.~Urbano\orcid{0000-0001-5640-0650}\inst{\ref{aff151},\ref{aff2}}
\and C.~Valieri\inst{\ref{aff33}}
\and A.~Venhola\orcid{0000-0001-6071-4564}\inst{\ref{aff171}}
\and D.~Vergani\orcid{0000-0003-0898-2216}\inst{\ref{aff4}}
\and P.~Vielzeuf\orcid{0000-0003-2035-9339}\inst{\ref{aff61}}
\and E.~Vilenius\orcid{0000-0002-6184-7681}\inst{\ref{aff27}}
\and S.~Vinciguerra\orcid{0009-0005-4018-3184}\inst{\ref{aff53}}
\and M.~von~Wietersheim-Kramsta\orcid{0000-0003-4986-5091}\inst{\ref{aff86},\ref{aff154}}
\and N.~A.~Walton\orcid{0000-0003-3983-8778}\inst{\ref{aff176}}
\and A.~H.~Wright\orcid{0000-0001-7363-7932}\inst{\ref{aff152}}
\and H.~W.~Yeung\orcid{0000-0002-4993-9014}\inst{\ref{aff6}}
\and D.~Scott\orcid{0000-0002-6878-9840}\inst{\ref{aff183}}}
										   
%%%% please do not edit the affiliation list -- contact ECEB Bureau for changes
\institute{INAF-Osservatorio Astronomico di Capodimonte, Via Moiariello 16, 80131 Napoli, Italy\label{aff1}
\and
Universit\'e de Strasbourg, CNRS, Observatoire astronomique de Strasbourg, UMR 7550, 67000 Strasbourg, France\label{aff2}
\and
Kapteyn Astronomical Institute, University of Groningen, PO Box 800, 9700 AV Groningen, The Netherlands\label{aff3}
\and
INAF-Osservatorio di Astrofisica e Scienza dello Spazio di Bologna, Via Piero Gobetti 93/3, 40129 Bologna, Italy\label{aff4}
\and
Max-Planck-Institut f\"ur Astronomie, K\"onigstuhl 17, 69117 Heidelberg, Germany\label{aff5}
\and
Institute for Astronomy, University of Edinburgh, Royal Observatory, Blackford Hill, Edinburgh EH9 3HJ, UK\label{aff6}
\and
INAF-Osservatorio Astronomico di Padova, Via dell'Osservatorio 5, 35122 Padova, Italy\label{aff7}
\and
Centre National d'Etudes Spatiales -- Centre spatial de Toulouse, 18 avenue Edouard Belin, 31401 Toulouse Cedex 9, France\label{aff8}
\and
Leiden Observatory, Leiden University, Einsteinweg 55, 2333 CC Leiden, The Netherlands\label{aff9}
\and
Space Telescope Science Institute, 3700 San Martin Dr, Baltimore, MD 21218, USA\label{aff10}
\and
Instituto de Astrof\'{\i}sica de Canarias, E-38205 La Laguna, Tenerife, Spain\label{aff11}
\and
Universidad de La Laguna, Dpto. Astrof\'\i sica, E-38206 La Laguna, Tenerife, Spain\label{aff12}
\and
School of Mathematics and Physics, University of Surrey, Guildford, Surrey, GU2 7XH, UK\label{aff13}
\and
INAF-Osservatorio Astronomico di Roma, Via Frascati 33, 00078 Monteporzio Catone, Italy\label{aff14}
\and
Space Science Data Center, Italian Space Agency, via del Politecnico snc, 00133 Roma, Italy\label{aff15}
\and
Institute of Physics, Laboratory of Astrophysics, Ecole Polytechnique F\'ed\'erale de Lausanne (EPFL), Observatoire de Sauverny, 1290 Versoix, Switzerland\label{aff16}
\and
Leibniz-Institut f\"{u}r Astrophysik (AIP), An der Sternwarte 16, 14482 Potsdam, Germany\label{aff17}
\and
Universit\"at Innsbruck, Institut f\"ur Astro- und Teilchenphysik, Technikerstr. 25/8, 6020 Innsbruck, Austria\label{aff18}
\and
Centro de Estudios de F\'isica del Cosmos de Arag\'on (CEFCA), Plaza San Juan, 1, planta 2, 44001, Teruel, Spain\label{aff19}
\and
ARAID Foundation, Avda. de Ranillas, 1-D, E-50018 Zaragoza, Spain\label{aff20}
\and
Dipartimento di Fisica "E. Pancini", Universita degli Studi di Napoli Federico II, Via Cinthia 6, 80126, Napoli, Italy\label{aff21}
\and
INFN -- Sezione di Napoli, Via Cinthia 6, 80126, Napoli, Italy\label{aff22}
\and
Instituto de Astrof\'isica de Andaluc\'ia, CSIC, Glorieta de la Astronom\'\i a, 18080, Granada, Spain\label{aff23}
\and
Department of Physics and Astronomy, University of Victoria, Victoria, BC V8P 5C2, Canada\label{aff24}
\and
ESAC/ESA, Camino Bajo del Castillo, s/n., Urb. Villafranca del Castillo, 28692 Villanueva de la Ca\~nada, Madrid, Spain\label{aff25}
\and
INAF-Osservatorio Astronomico di Brera, Via Brera 28, 20122 Milano, Italy\label{aff26}
\and
Mullard Space Science Laboratory, University College London, Holmbury St Mary, Dorking, Surrey RH5 6NT, UK\label{aff27}
\and
IFPU, Institute for Fundamental Physics of the Universe, via Beirut 2, 34151 Trieste, Italy\label{aff28}
\and
INAF-Osservatorio Astronomico di Trieste, Via G. B. Tiepolo 11, 34143 Trieste, Italy\label{aff29}
\and
INFN, Sezione di Trieste, Via Valerio 2, 34127 Trieste TS, Italy\label{aff30}
\and
SISSA, International School for Advanced Studies, Via Bonomea 265, 34136 Trieste TS, Italy\label{aff31}
\and
Dipartimento di Fisica e Astronomia, Universit\`a di Bologna, Via Gobetti 93/2, 40129 Bologna, Italy\label{aff32}
\and
INFN-Sezione di Bologna, Viale Berti Pichat 6/2, 40127 Bologna, Italy\label{aff33}
\and
Max Planck Institute for Extraterrestrial Physics, Giessenbachstr. 1, 85748 Garching, Germany\label{aff34}
\and
Universit\"ats-Sternwarte M\"unchen, Fakult\"at f\"ur Physik, Ludwig-Maximilians-Universit\"at M\"unchen, Scheinerstr.~1, 81679 M\"unchen, Germany\label{aff35}
\and
Dipartimento di Fisica, Universit\`a di Genova, Via Dodecaneso 33, 16146, Genova, Italy\label{aff36}
\and
INFN-Sezione di Genova, Via Dodecaneso 33, 16146, Genova, Italy\label{aff37}
\and
Department of Physics "E. Pancini", University Federico II, Via Cinthia 6, 80126, Napoli, Italy\label{aff38}
\and
Instituto de Astrof\'isica e Ci\^encias do Espa\c{c}o, Universidade do Porto, CAUP, Rua das Estrelas, PT4150-762 Porto, Portugal\label{aff39}
\and
Faculdade de Ci\^encias da Universidade do Porto, Rua do Campo de Alegre, 4150-007 Porto, Portugal\label{aff40}
\and
European Southern Observatory, Karl-Schwarzschild-Str.~2, 85748 Garching, Germany\label{aff41}
\and
Dipartimento di Fisica, Universit\`a degli Studi di Torino, Via P. Giuria 1, 10125 Torino, Italy\label{aff42}
\and
INFN-Sezione di Torino, Via P. Giuria 1, 10125 Torino, Italy\label{aff43}
\and
INAF-Osservatorio Astrofisico di Torino, Via Osservatorio 20, 10025 Pino Torinese (TO), Italy\label{aff44}
\and
INAF-IASF Milano, Via Alfonso Corti 12, 20133 Milano, Italy\label{aff45}
\and
Centro de Investigaciones Energ\'eticas, Medioambientales y Tecnol\'ogicas (CIEMAT), Avenida Complutense 40, 28040 Madrid, Spain\label{aff46}
\and
Port d'Informaci\'{o} Cient\'{i}fica, Campus UAB, C. Albareda s/n, 08193 Bellaterra (Barcelona), Spain\label{aff47}
\and
Institute for Astronomy, University of Hawaii, 2680 Woodlawn Drive, Honolulu, HI 96822, USA\label{aff48}
\and
Dipartimento di Fisica e Astronomia "Augusto Righi" - Alma Mater Studiorum Universit\`a di Bologna, Viale Berti Pichat 6/2, 40127 Bologna, Italy\label{aff49}
\and
Jodrell Bank Centre for Astrophysics, Department of Physics and Astronomy, University of Manchester, Oxford Road, Manchester M13 9PL, UK\label{aff50}
\and
European Space Agency/ESRIN, Largo Galileo Galilei 1, 00044 Frascati, Roma, Italy\label{aff51}
\and
Universit\'e Lyon 1, CNRS/IN2P3, IP2I Lyon, Villeurbanne, France\label{aff52}
\and
Aix-Marseille Universit\'e, CNRS, CNES, LAM, Marseille, France\label{aff53}
\and
Institut de Ci\`{e}ncies del Cosmos (ICCUB), Universitat de Barcelona (IEEC-UB), Mart\'{i} i Franqu\`{e}s 1, 08028 Barcelona, Spain\label{aff54}
\and
Instituci\'o Catalana de Recerca i Estudis Avan\c{c}ats (ICREA), Passeig de Llu\'{\i}s Companys 23, 08010 Barcelona, Spain\label{aff55}
\and
Institut de Ciencies de l'Espai (IEEC-CSIC), Campus UAB, Carrer de Can Magrans, s/n Cerdanyola del Vall\'es, 08193 Barcelona, Spain\label{aff56}
\and
Universit\'e Lyon 1, CNRS/IN2P3, IUF, IP2I Lyon, Villeurbanne, France \label{aff57}
\and
Department of Astronomy, University of Geneva, ch. d'Ecogia 16, 1290 Versoix, Switzerland\label{aff58}
\and
Universit\'e Paris-Saclay, CNRS, Institut d'astrophysique spatiale, 91405, Orsay, France\label{aff59}
\and
INFN-Padova, Via Marzolo 8, 35131 Padova, Italy\label{aff60}
\and
Aix-Marseille Universit\'e, CNRS/IN2P3, CPPM, Marseille, France\label{aff61}
\and
INAF-Istituto di Astrofisica e Planetologia Spaziali, via del Fosso del Cavaliere, 100, 00100 Roma, Italy\label{aff62}
\and
Department of Physics, University of Oxford, Keble Road, Oxford OX1 3RH, UK\label{aff63}
\and
University Observatory, LMU Faculty of Physics, Scheinerstr.~1, 81679 Munich, Germany\label{aff64}
\and
Dipartimento di Fisica "Aldo Pontremoli", Universit\`a degli Studi di Milano, Via Celoria 16, 20133 Milano, Italy\label{aff65}
\and
INFN-Sezione di Milano, Via Celoria 16, 20133 Milano, Italy\label{aff66}
\and
Institute of Theoretical Astrophysics, University of Oslo, P.O. Box 1029 Blindern, 0315 Oslo, Norway\label{aff67}
\and
Caltech/IPAC, 1200 E. California Blvd., Pasadena, CA 91125, USA\label{aff68}
\and
Jet Propulsion Laboratory, California Institute of Technology, 4800 Oak Grove Drive, Pasadena, CA, 91109, USA\label{aff69}
\and
School of Physics and Astronomy, Lancaster University, Lancaster, LA1 4YB, UK\label{aff70}
\and
Felix Hormuth Engineering, Goethestr. 17, 69181 Leimen, Germany\label{aff71}
\and
Technical University of Denmark, Elektrovej 327, 2800 Kgs. Lyngby, Denmark\label{aff72}
\and
Cosmic Dawn Center (DAWN), Denmark\label{aff73}
\and
Universit\'e PSL, Observatoire de Paris, Sorbonne Universit\'e, CNRS, LERMA, 75014, Paris, France\label{aff74}
\and
Universit\'e Paris-Cit\'e, 5 Rue Thomas Mann, 75013, Paris, France\label{aff75}
\and
NASA Goddard Space Flight Center, Greenbelt, MD 20771, USA\label{aff76}
\and
Universit\'e de Gen\`eve, D\'epartement de Physique Th\'eorique and Centre for Astroparticle Physics, 24 quai Ernest-Ansermet, CH-1211 Gen\`eve 4, Switzerland\label{aff77}
\and
Department of Physics, P.O. Box 64, University of Helsinki, 00014 Helsinki, Finland\label{aff78}
\and
Helsinki Institute of Physics, Gustaf H{\"a}llstr{\"o}min katu 2, University of Helsinki, 00014 Helsinki, Finland\label{aff79}
\and
Laboratoire d'etude de l'Univers et des phenomenes eXtremes, Observatoire de Paris, Universit\'e PSL, Sorbonne Universit\'e, CNRS, 92190 Meudon, France\label{aff80}
\and
SKAO, Jodrell Bank, Lower Withington, Macclesfield SK11 9FT, UK\label{aff81}
\and
Centre de Calcul de l'IN2P3/CNRS, 21 avenue Pierre de Coubertin 69627 Villeurbanne Cedex, France\label{aff82}
\and
Universit\"at Bonn, Argelander-Institut f\"ur Astronomie, Auf dem H\"ugel 71, 53121 Bonn, Germany\label{aff83}
\and
INFN-Sezione di Roma, Piazzale Aldo Moro, 2 - c/o Dipartimento di Fisica, Edificio G. Marconi, 00185 Roma, Italy\label{aff84}
\and
Dipartimento di Fisica e Astronomia "Augusto Righi" - Alma Mater Studiorum Universit\`a di Bologna, via Piero Gobetti 93/2, 40129 Bologna, Italy\label{aff85}
\and
Department of Physics, Institute for Computational Cosmology, Durham University, South Road, Durham, DH1 3LE, UK\label{aff86}
\and
Universit\'e Paris Cit\'e, CNRS, Astroparticule et Cosmologie, 75013 Paris, France\label{aff87}
\and
CNRS-UCB International Research Laboratory, Centre Pierre Bin\'etruy, IRL2007, CPB-IN2P3, Berkeley, USA\label{aff88}
\and
University of Applied Sciences and Arts of Northwestern Switzerland, School of Engineering, 5210 Windisch, Switzerland\label{aff89}
\and
Dipartimento di Fisica - Sezione di Astronomia, Universit\`a di Trieste, Via Tiepolo 11, 34131 Trieste, Italy\label{aff90}
\and
Telespazio UK S.L. for European Space Agency (ESA), Camino bajo del Castillo, s/n, Urbanizacion Villafranca del Castillo, Villanueva de la Ca\~nada, 28692 Madrid, Spain\label{aff91}
\and
Institut de F\'{i}sica d'Altes Energies (IFAE), The Barcelona Institute of Science and Technology, Campus UAB, 08193 Bellaterra (Barcelona), Spain\label{aff92}
\and
European Space Agency/ESTEC, Keplerlaan 1, 2201 AZ Noordwijk, The Netherlands\label{aff93}
\and
Waterloo Centre for Astrophysics, University of Waterloo, Waterloo, Ontario N2L 3G1, Canada\label{aff94}
\and
Department of Physics and Astronomy, University of Waterloo, Waterloo, Ontario N2L 3G1, Canada\label{aff95}
\and
Perimeter Institute for Theoretical Physics, Waterloo, Ontario N2L 2Y5, Canada\label{aff96}
\and
Universit\'e Paris-Saclay, Universit\'e Paris Cit\'e, CEA, CNRS, AIM, 91191, Gif-sur-Yvette, France\label{aff97}
\and
Institute of Space Science, Str. Atomistilor, nr. 409 M\u{a}gurele, Ilfov, 077125, Romania\label{aff98}
\and
Dipartimento di Fisica e Astronomia "G. Galilei", Universit\`a di Padova, Via Marzolo 8, 35131 Padova, Italy\label{aff99}
\and
Instituto de F\'isica Te\'orica UAM-CSIC, Campus de Cantoblanco, 28049 Madrid, Spain\label{aff100}
\and
Institut de Recherche en Astrophysique et Plan\'etologie (IRAP), Universit\'e de Toulouse, CNRS, UPS, CNES, 14 Av. Edouard Belin, 31400 Toulouse, France\label{aff101}
\and
Universit\'e St Joseph; Faculty of Sciences, Beirut, Lebanon\label{aff102}
\and
Departamento de F\'isica, FCFM, Universidad de Chile, Blanco Encalada 2008, Santiago, Chile\label{aff103}
\and
Universit\'e C\^{o}te d'Azur, Observatoire de la C\^{o}te d'Azur, CNRS, Laboratoire Lagrange, Bd de l'Observatoire, CS 34229, 06304 Nice cedex 4, France\label{aff104}
\and
Department of Physics and Helsinki Institute of Physics, Gustaf H\"allstr\"omin katu 2, University of Helsinki, 00014 Helsinki, Finland\label{aff105}
\and
Centre for Electronic Imaging, Open University, Walton Hall, Milton Keynes, MK7~6AA, UK\label{aff106}
\and
Departamento de F\'isica, Faculdade de Ci\^encias, Universidade de Lisboa, Edif\'icio C8, Campo Grande, PT1749-016 Lisboa, Portugal\label{aff107}
\and
Instituto de Astrof\'isica e Ci\^encias do Espa\c{c}o, Faculdade de Ci\^encias, Universidade de Lisboa, Campo Grande, 1749-016 Lisboa, Portugal\label{aff108}
\and
Cosmic Dawn Center (DAWN)\label{aff109}
\and
Niels Bohr Institute, University of Copenhagen, Jagtvej 128, 2200 Copenhagen, Denmark\label{aff110}
\and
Universidad Polit\'ecnica de Cartagena, Departamento de Electr\'onica y Tecnolog\'ia de Computadoras,  Plaza del Hospital 1, 30202 Cartagena, Spain\label{aff111}
\and
European University of Technology EUt+, European Union\label{aff112}
\and
Centre for Information Technology, University of Groningen, P.O. Box 11044, 9700 CA Groningen, The Netherlands\label{aff113}
\and
Institute of Space Sciences (ICE, CSIC), Campus UAB, Carrer de Can Magrans, s/n, 08193 Barcelona, Spain\label{aff114}
\and
Institut d'Estudis Espacials de Catalunya (IEEC),  Edifici RDIT, Campus UPC, 08860 Castelldefels, Barcelona, Spain\label{aff115}
\and
Astronomisches Rechen-Institut, Zentrum f\"ur Astronomie der Universit\"at Heidelberg, M\"onchhofstr. 12-14, 69120 Heidelberg, Germany\label{aff116}
\and
Dipartimento di Fisica e Scienze della Terra, Universit\`a degli Studi di Ferrara, Via Giuseppe Saragat 1, 44122 Ferrara, Italy\label{aff117}
\and
Istituto Nazionale di Fisica Nucleare, Sezione di Ferrara, Via Giuseppe Saragat 1, 44122 Ferrara, Italy\label{aff118}
\and
ICSC - Centro Nazionale di Ricerca in High Performance Computing, Big Data e Quantum Computing, Via Magnanelli 2, Bologna, Italy\label{aff119}
\and
INAF, Istituto di Radioastronomia, Via Piero Gobetti 101, 40129 Bologna, Italy\label{aff120}
\and
INFN-Bologna, Via Irnerio 46, 40126 Bologna, Italy\label{aff121}
\and
Department of Mathematical Physics, Institute of Physics, University of S\~ao Paulo, R. do Mat\~ao 1371, 05508-090, S\~ao Paulo, SP, Brazil\label{aff122}
\and
CEA Saclay, DFR/IRFU, Service d'Astrophysique, Bat. 709, 91191 Gif-sur-Yvette, France\label{aff123}
\and
Oskar Klein Centre for Cosmoparticle Physics, Department of Physics, Stockholm University, Stockholm, SE-106 91, Sweden\label{aff124}
\and
Astrophysics Group, Blackett Laboratory, Imperial College London, London SW7 2AZ, UK\label{aff125}
\and
Univ. Grenoble Alpes, CNRS, Grenoble INP, LPSC-IN2P3, 53, Avenue des Martyrs, 38000 Grenoble, France\label{aff126}
\and
Dipartimento di Fisica, Sapienza Universit\`a di Roma, Piazzale Aldo Moro 2, 00185 Roma, Italy\label{aff127}
\and
Department of Mathematics and Physics E. De Giorgi, University of Salento, Via per Arnesano, CP-I93, 73100, Lecce, Italy\label{aff128}
\and
INFN, Sezione di Lecce, Via per Arnesano, CP-193, 73100, Lecce, Italy\label{aff129}
\and
INAF-Sezione di Lecce, c/o Dipartimento Matematica e Fisica, Via per Arnesano, 73100, Lecce, Italy\label{aff130}
\and
Institut d'Astrophysique de Paris, 98bis Boulevard Arago, 75014, Paris, France\label{aff131}
\and
ICL, Junia, Universit\'e Catholique de Lille, LITL, 59000 Lille, France\label{aff132}
\and
Center for Astrophysics and Cosmology, University of Nova Gorica, Nova Gorica, Slovenia\label{aff133}
\and
CERCA/ISO, Department of Physics, Case Western Reserve University, 10900 Euclid Avenue, Cleveland, OH 44106, USA\label{aff134}
\and
Laboratoire Univers et Th\'eorie, Observatoire de Paris, Universit\'e PSL, Universit\'e Paris Cit\'e, CNRS, 92190 Meudon, France\label{aff135}
\and
Departamento de F{\'\i}sica Fundamental. Universidad de Salamanca. Plaza de la Merced s/n. 37008 Salamanca, Spain\label{aff136}
\and
Universiteit Gent, Department of Physics and Astronomy, Proeftuinstraat 86 N3, 9000 Ghent, Belgium
\label{aff137}
\and
Konkoly Observatory, HUN-REN CSFK, MTA Centre of Excellence, Budapest, Konkoly Thege Mikl\'os {\'u}t 15-17. H-1121, Hungary\label{aff138}
\and
MTA-CSFK Lend\"ulet Large-Scale Structure Research Group, Konkoly-Thege Mikl\'os \'ut 15-17, H-1121 Budapest, Hungary\label{aff139}
\and
Max-Planck-Institut f\"ur Physik, Boltzmannstr. 8, 85748 Garching, Germany\label{aff140}
\and
Scuola Superiore Meridionale, Via Mezzocannone 4, 80138, Napoli, Italy\label{aff141}
\and
California Institute of Technology, 1200 E California Blvd, Pasadena, CA 91125, USA\label{aff142}
\and
Instituto de Astrof\'isica e Ci\^encias do Espa\c{c}o, Faculdade de Ci\^encias, Universidade de Lisboa, Tapada da Ajuda, 1349-018 Lisboa, Portugal\label{aff143}
\and
Institute of Cosmology and Gravitation, University of Portsmouth, Portsmouth PO1 3FX, UK\label{aff144}
\and
Centre de Recherche Astrophysique de Lyon, UMR5574, CNRS, Universit\'e Claude Bernard Lyon 1, ENS de Lyon, 69230, Saint-Genis-Laval, France\label{aff145}
\and
Department of Physics \& Astronomy, University of California Irvine, Irvine CA 92697, USA\label{aff146}
\and
Departamento F\'isica Aplicada, Universidad Polit\'ecnica de Cartagena, Campus Muralla del Mar, 30202 Cartagena, Murcia, Spain\label{aff147}
\and
Instituto de F\'isica de Cantabria, Edificio Juan Jord\'a, Avenida de los Castros, 39005 Santander, Spain\label{aff148}
\and
Institut d'Astrophysique de Paris, UMR 7095, CNRS, and Sorbonne Universit\'e, 98 bis boulevard Arago, 75014 Paris, France\label{aff149}
\and
Department of Computer Science, Aalto University, PO Box 15400, Espoo, FI-00 076, Finland\label{aff150}
\and
School of Physics and Astronomy, University of Nottingham, University Park, Nottingham NG7 2RD, UK\label{aff151}
\and
Ruhr University Bochum, Faculty of Physics and Astronomy, Astronomical Institute (AIRUB), German Centre for Cosmological Lensing (GCCL), 44780 Bochum, Germany\label{aff152}
\and
DARK, Niels Bohr Institute, University of Copenhagen, Jagtvej 155, 2200 Copenhagen, Denmark\label{aff153}
\and
Department of Physics, Centre for Extragalactic Astronomy, Durham University, South Road, Durham, DH1 3LE, UK\label{aff154}
\and
Astrophysics Research Centre, University of KwaZulu-Natal, Westville Campus, Durban 4041, South Africa\label{aff155}
\and
School of Mathematics, Statistics \& Computer Science, University of KwaZulu-Natal, Westville Campus, Durban 4041, South Africa\label{aff156}
\and
Department of Physics and Astronomy, Vesilinnantie 5, University of Turku, 20014 Turku, Finland\label{aff157}
\and
Finnish Centre for Astronomy with ESO (FINCA), Quantum, Vesilinnantie 5, University of Turku, 20014 Turku, Finland\label{aff158}
\and
Serco for European Space Agency (ESA), Camino bajo del Castillo, s/n, Urbanizacion Villafranca del Castillo, Villanueva de la Ca\~nada, 28692 Madrid, Spain\label{aff159}
\and
ARC Centre of Excellence for Dark Matter Particle Physics, Melbourne, Australia\label{aff160}
\and
Centre for Astrophysics \& Supercomputing, Swinburne University of Technology,  Hawthorn, Victoria 3122, Australia\label{aff161}
\and
Department of Physics and Astronomy, University of the Western Cape, Bellville, Cape Town, 7535, South Africa\label{aff162}
\and
Van Swinderen Institute, University of Groningen, Nijenborgh 3, 9747 AG Groningen, The Netherlands\label{aff163}
\and
Ohio University, Physics \& Astronomy Department,1 Ohio University, Athens, OH 45701, USA\label{aff164}
\and
Institute for Theoretical Particle Physics and Cosmology (TTK), RWTH Aachen University, 52056 Aachen, Germany\label{aff165}
\and
INAF-Osservatorio Astrofisico di Arcetri, Largo E. Fermi 5, 50125, Firenze, Italy\label{aff166}
\and
Centro de Astrof\'{\i}sica da Universidade do Porto, Rua das Estrelas, 4150-762 Porto, Portugal\label{aff167}
\and
HE Space for European Space Agency (ESA), Camino bajo del Castillo, s/n, Urbanizacion Villafranca del Castillo, Villanueva de la Ca\~nada, 28692 Madrid, Spain\label{aff168}
\and
Dipartimento di Fisica ``G. Occhialini", Universit\`a degli Studi di Milano Bicocca, Piazza della Scienza 3, 20126 Milano, Italy\label{aff169}
\and
Department of Astrophysics, University of Zurich, Winterthurerstrasse 190, 8057 Zurich, Switzerland\label{aff170}
\and
Space physics and astronomy research unit, University of Oulu, Pentti Kaiteran katu 1, FI-90014 Oulu, Finland\label{aff171}
\and
Theoretical astrophysics, Department of Physics and Astronomy, Uppsala University, Box 524, 751 20 Uppsala, Sweden\label{aff172}
\and
Mathematical Institute, University of Leiden, Einsteinweg 55, 2333 CA Leiden, The Netherlands\label{aff173}
\and
School of Physical Sciences, The Open University, Milton Keynes, MK7 6AA, UK\label{aff174}
\and
School of Physics \& Astronomy, University of Southampton, Highfield Campus, Southampton SO17 1BJ, UK\label{aff175}
\and
Institute of Astronomy, University of Cambridge, Madingley Road, Cambridge CB3 0HA, UK\label{aff176}
\and
Univ. Lille, CNRS, Centrale Lille, UMR 9189 CRIStAL, 59000 Lille, France\label{aff177}
\and
Kobayashi-Maskawa Institute for the Origin of Particles and the Universe, Nagoya University, Chikusa-ku, Nagoya, 464-8602, Japan\label{aff178}
\and
Institute for Particle Physics and Astrophysics, Dept. of Physics, ETH Zurich, Wolfgang-Pauli-Strasse 27, 8093 Zurich, Switzerland\label{aff179}
\and
Department of Astrophysical Sciences, Peyton Hall, Princeton University, Princeton, NJ 08544, USA\label{aff180}
\and
Fakult\"at f\"ur Physik, Universit\"at Bielefeld, Postfach 100131, 33501 Bielefeld, Germany\label{aff181}
\and
School of Mathematics, Statistics and Physics, Newcastle University, Herschel Building, Newcastle-upon-Tyne, NE1 7RU, UK\label{aff182}
\and
Department of Physics and Astronomy, University of British Columbia, Vancouver, BC V6T 1Z1, Canada\label{aff183}}      
%%%% please do not edit the author list -- contact ECEB Bureau for changes

%
% For a Key Project paper, please use instead:
%
% \title{\Euclid\/ preparation} % or another appropriate pre-phrase
% \subtitle{Your title}
%
% \author{Euclid Collaboration: F.~Author, .....}
%

% \institute{$^{1}$ Dark Energy Institute\\
% $^{2}$ Center for Large-Scale Structure\\
% $^{3}$ Institute for Fundamental Research \\
% $^{4}$ Institute for Astrophysics and Cosmology\\}

% 
% Put your abstract here:
%
\abstract{
\massi{We searched within the first data release of the Euclid Wide Survey ($\simeq 1900\,\mathrm{deg}^2$) for resolved stellar systems between distances from 10\,kpc and several Mpc, using an isochrone filtering technique.
We detect 15 systems that have never been resolved into stars before. Five are new discoveries, three in the Milky Way halo  and two beyond the Local Group. The remaining ten are dwarf galaxies that were previously catalogued as unresolved sources in shallower ground-based surveys and lack reliable distance measures.
We measured distance moduli for all 15 systems, and half-light radii, ellipticities, position angles, absolute magnitudes, and stellar masses for 14 of them.}
In the Milky Way halo, \UMi{}, at $\simeq 20$\,kpc, has $M_V \simeq +2.0$ and a stellar mass of $\simeq 40\,M_{\sun}$, and is one of the least luminous stellar systems known.
\Fornax{} is located near the tidal limit of the Fornax dwarf spheroidal galaxy and may be physically associated with it.
\Eridanus{} lies at $\simeq 165$\,kpc and has a stellar mass of $\simeq 550\,M_{\sun}$, which places it in the region of the size--luminosity plane where globular clusters and ultra-faint dwarf galaxies overlap.
Beyond the Local Group, the 12 dwarf galaxies lie at distances of 3--5.5\,Mpc. Three of them reach or exceed a mean surface brightness within the half-light radius of $\mu_V\simeq28$\,\magarc{}, a regime in which almost no galaxies at these distances were previously known.
\massi{Ten of them are candidate members of the M81 group: eight have three-dimensional distances from M81 smaller than or comparable to the radius of its zero-velocity surface, and two lie close to the peripheral member NGC~4236, one of which is a candidate satellite of this galaxy.}
These results demonstrate \Euclid's transformative potential for low-surface-brightness science, enabling the study of an unprecedented range of environments from the MW halo out to the Local Volume within a single, homogeneous survey.}
%
% Provide up to five key words:
%
%    from the list in
%     https://www.aanda.org/for-authors/latex-issues/information-files#pop}
%
% Add short versions of title and author list for page headings
%
   \titlerunning{ Pushing back the borders of the invisible with \Euclid}
   \authorrunning{Euclid Collaboration: M.~Gatto et al.}
   
   \maketitle
%
%-------------------------------------------------------------------
%
%
%   Start the main text of your paper here
%
   
\section{\label{sc:Intro}Introduction}

A complete map of the satellite population of the Milky Way (MW) is a longstanding challenge in Galactic astronomy, with significant cosmological implications. While the first identifications of dwarf galaxies orbiting the MW, specifically the Sculptor and Fornax dwarfs, date back nearly a century \citep{Shapley1938}, the field has experienced a profound renaissance over the last two decades. The advent of wide-field digital sky surveys, coupled with the development of automated detection algorithms for mining deep datasets in search of over-densities of resolved stars, has led to a rapid surge in the census of known MW satellites \citep[e.g.,][]{Willman-2005a, Willman-2005b, Belokurov-2007, Walsh-2009, Cerny-2023}.
%\massi{as well as of dwarfs resolved into individual stars in the Local Group (LG) and in the Local Volume \citep[LV, see, e.g.,][and references therein]{mutlu21,mutlu24,jones23_pavo,jones24_corvus,kara24,kara25,amandine26}.}\par
Large-scale photometric surveys, such as the Sloan Digital Sky Survey \citep[SDSS;][]{York-2000}, Panoramic Survey Telescope \& Rapid Response System \citep[Pan-STARRS1;][]{Chambers-2016}, and the Dark Energy Survey \citep[DES;][]{DES-Abbott-2016}, have been responsible for the majority of these recent discoveries \citep{Belokurov-2007,Laevens-2015a,Laevens-2015b,Bechtol-2015,Koposov-2015}. These surveys have allowed us to probe the faintest end of the galactic luminosity function, uncovering the realm of ultra-faint dwarf galaxies \citep[UFDs;][]{Simon2019}\footnote{We follow the definition of \citet{Simon2019} that denoted UFDs as those galaxies with a total luminosity $L<10^5\,L_\odot$ or $M_V > -7.7$.}.\par
\massi{In the current $\Lambda$ cold dark matter ($\Lambda$CDM) cosmological paradigm \citep[see][and references therein]{bertone18,condon18,peebles25}, local ($z \simeq 0$) dwarf galaxies are considered as the living relics of the hierarchical assembly process leading to the formation of large galaxies.
Since the faintest among them appear as those whose mass budget is most dominated by dark matter \citep{mateo98,mc12,Simon2019}, they serve as unique laboratories for addressing current small-scale challenges to the $\Lambda$CDM model \citep{Bullock&Boylan-Kolchin2017}.}\par
The current census reaches approximately 50 confirmed dwarf satellites of the MW and about 550 dwarfs in the Local Volume (LV), within 6~Mpc from the Sun \citep{Pace2025}. However, many works suggest that the total population of MW satellites remains significantly underestimated \citep[e.g.][]{Jethwa-2018, Newton-2018, Drlica-Wagner-2020, Santos-Santos-2026, Tan-2026}, implying that the most elusive, diffuse, and distant systems have yet to be detected. The primary barrier to discovery is their extremely low surface brightness, which keeps a vast population of UFDs below the detection limits of current ground-based facilities. %Characterizing these distant, sparse, or ultra-diffuse systems is pivotal for a comprehensive understanding of the formation of the MW halo.\par
\massi{Beyond the MW halo, our knowledge of isolated dwarfs or of the satellite populations around different hosts in the LV is even more incomplete, given the higher distances involved, creating a significant observational bias.} A telling example is offered by the Andromeda galaxy (M31), whose satellite system has been progressively extended to ever fainter dwarf members \citep{Martin-2006, Martin-2013, Richardson-2011,Arias-2025,Smith-2025,amandine26,Sakowska-2026}, yet down to a brighter limit than in the MW, owing to M31's greater distance \citep{Doliva-Dolinsky-2023}. \massi{In particular, the faintest known dwarf satellite of M31 is more than four magnitudes brighter than the faintest MW satellite \citep[in $M_V$,][]{Pace2025}.
This illustrates how the accessible faint end of the satellite luminosity function becomes increasingly truncated as host distance grows, leaving the faint dwarf populations in the LV largely uncharted.}\par
Extending the study of UFDs and classical dwarfs beyond the immediate vicinity of the MW is therefore crucial to testing $\Lambda$CDM predictions across a diverse range of host masses, morphologies, and environmental conditions.
\massi{Beyond the Local Group (LG), considerable effort has been devoted in recent years to building censuses of satellites and isolated dwarfs around nearby hosts. Deep resolved-star surveys of individual groups, such as Centaurus~A \citep{Crnojevic-2016} and M81 \citep{Okamoto-2019, Gozman-2024}, and of nearby MW-mass or low-mass spirals \citep[e.g.,][]{Carlin-2016, Carlin-2024, Smercina-2018,mutlu24}, have detected satellites close to the transition between classical dwarfs and UFDs, but each of them covers only the virial volume of one or a few hosts. 
In parallel, a growing number of isolated dwarfs in the LV has been discovered and resolved into stars \citep[see, e.g.,][and references therein]{jones23_pavo,jones24_corvus,kara24,kara25}. 
As a result, the faint end of the dwarf population beyond the LG is known only around a handful of hosts, and no homogeneous search based on resolved stars at space-based resolution has yet been carried out over thousands of square degrees.}\par
\massi{Deep, wide-field space-based imaging, combining low-surface-brightness sensitivity with stable and homogeneous image quality, therefore offers a major opportunity to extend satellite censuses to fainter luminosities and around hosts well beyond the MW.}
In this context, the \Euclid mission \citep{Mellier-2025} represents a transformative step forward for local low-surface brightness astronomy.
\Euclid{}’s unique combination of a vast survey area ($\simeq$\num{14000}\,deg$^2$ at the end of the six-year mission), high spatial resolution in the visible (\IE) with the VIS instrument \citep{EuclidSkyVIS}, and deep near-infrared (\YE, \JE, \HE) photometry with the NISP instrument \citep{EuclidSkyNISP}, provides an unprecedented discovery space, thanks to the unrivaled sensitivity of searches based on exploring the surface density field as traced by individual resolved stars down to very faint magnitude limits. 
Unlike previous surveys, \Euclid is uniquely positioned to bridge the gap between different scales: it possesses the sensitivity to detect the most elusive UFDs within the MW halo and the resolution required to identify dwarf galaxies by star counts at much larger distances. Furthermore, it explores large swaths of sky, making it the first optical-IR imaging space telescope operating in survey mode.\par
In this paper, we present the results of a systematic search for resolved low-surface brightness stellar systems in the $\simeq$1900\,deg$^2$ of the Euclid Wide Survey (EWS) in the Euclid Data Release 1 \citep[DR1;][]{DR1cite}, exploiting an isochrone filter technique applied over a wide range of distances, from the MW halo out to the LV. \massi{In this contribution, we present several newly discovered local stellar systems and we resolve into stars for the first time several already catalogued galaxies, thus illustrating the power of this kind of search in the \Euclid environment.
A systematic study of the selection window will be presented elsewhere (\textcolor{blue}{Rusterucci et al. in prep.}). 
In the future the sensitivity will be improved by tuning of the method and by progress of \Euclid surveys and data reduction.}\par
The paper is organized as follows. In Sect.~\ref{sc:catalog} we describe the \Euclid photometric catalogue and the stellar sample selection. In Sect.~\ref{sc:method} we outline the isochrone filter technique used to detect stellar systems. In Sect.~\ref{sc:results} we present the identified systems and their derived structural and photometric properties. In Sect.~\ref{sc:discussion} we discuss the properties of the discovered stellar systems and their environmental context. Finally, in Sect.~\ref{sc:summary} we summarize our main conclusions.

\section{\label{sc:catalog} \Euclid catalogue}

In this analysis, we used data from the DR1 of the EWS, covering an approximate area of $1900$\,deg$^2$, reaching a 5$\sigma$ point-source depth of 26.7, 24.4, 24.4, 24.4 in \IE, \YE, \JE, and \HE, respectively \citep{Mellier-2025}. 
Specifically, we adopt the template-fitting photometric measurements provided in the MERge processing function (MER) catalogues.
\massi{The MERge processing function \citep[MER;][]{DR1-TP009} detects sources in the \Euclid images, measures their properties, and combines them into a single multi-wavelength catalogue.}
%From this catalogue, we adopted the template-fitting photometry.}\par
The search for resolved dwarf galaxies requires a sample of stars that is both complete and pure. We perform %our search on VIS detections only, performing
the initial star/galaxy separation using \texttt{POINT\_LIKE\_PROB > 0.05}. This parameter is inspired by the \texttt{SPREAD\_MODEL} from \texttt{SourceExtractor} and is designed to be highly conservative and biased toward a high-purity of star selection \citep{Romelli-2026}. Therefore, despite this low threshold value, this criterion still yields a stellar sample with purity above 80\% across all magnitudes (\textcolor{blue}{Rusterucci et al. in prep.}).\par
To account for the reddening, all apparent magnitudes are corrected for Galactic extinction prior to further photometric filtering. We convert the {\it Planck} reddening maps \citep{Planck-2014} into stellar extinction values using the absorption coefficients $R_{I_\mathrm{E}} = 2.28$, $R_{Y_\mathrm{E}} = 1.14$, $R_{J_\mathrm{E}} = 0.79$, and $R_{H_\mathrm{E}} = 0.51$. These were derived from the $A_{\lambda}/A_V$ values provided by the PARSEC web interface\footnote{\url{https://stev.oapd.inaf.it/cgi-bin/cmd}}, which adopts the extinction curve of \citet{Cardelli-1989} and \citet{ODonnell-1994}, assuming $R_V=3.1$. Throughout the remainder of this work, all reported magnitudes are implicitly reddening-corrected.\par
\begin{figure}
    \centering
    \includegraphics[width=1.\linewidth]{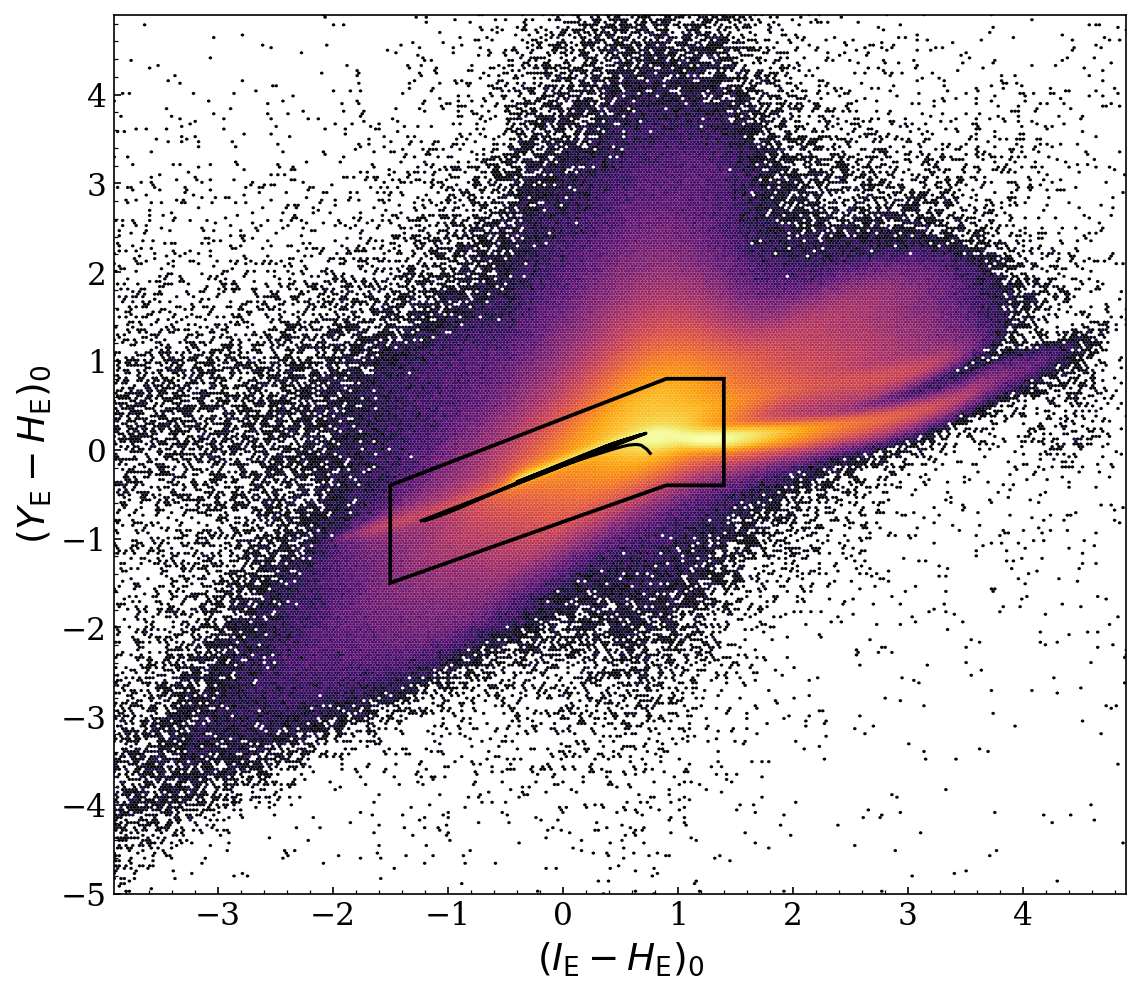}
    \caption{Distribution of sources in the de-reddened ($\IE-\HE$) vs. ($\YE-\HE$) colour--colour plane. The solid black contour defines the polygonal selection region used to isolate the stellar population and minimize contamination from residual background compact galaxies, while the black solid line represents a stellar isochrone with $\logten(t/\rm{yr}) = 10.12$ and \feh{} = $-2.2$ shown as a reference for the expected stellar locus.}
    \label{fig:color_color}
\end{figure}

Since the primary source of contamination in stellar catalogues arises from background compact galaxies, with galaxy groups providing small scale overdensities that can mimic local dwarf galaxies in automated pipelines, we further refine our star/galaxy separation photometrically. This is achieved by exploiting the de-reddened ($\IE-\HE$) vs. ($\YE-\HE$) colour--colour space, adopting a procedure similar to those described in \citet{Hunt-2025} and \citet{Annibali-2026}. 
Figure~\ref{fig:color_color} illustrates the distribution of sources in this colour--colour plane, where a well-defined stellar locus is clearly visible. To highlight this feature, we overplot a PARSEC stellar isochrone \citep{Bressan-2012} with $\logten(t/\text{yr}) = 10.12$ and \feh{} = $-2.2$ (see Sect.~\ref{sc:method} for the justification of these parameters). 
We isolate the stellar population by selecting all sources falling within a polygonal region defined by the vertices: $(-1.5, -1.5)$, $(-1.5, -0.4)$, $(0.9, 0.8)$, $(1.4, 0.8)$, $(1.4, -0.4)$, and $(0.9, -0.4)$, indicated by the solid black contour in the figure. 
\massi{We verified that this region fully encloses main sequence, red giant branch and horizontal branch stars of old isochrones with metallicities up to \feh{} $\simeq-1$.}
This colour--colour cut filters out residual compact galaxies, significantly enhancing the final purity of our stellar catalogue.

\section{\label{sc:method} Isochrone filter technique}

To identify candidate satellite systems, we employed an isochrone filtering technique \citep{Koposov-2007,Walsh-2009}.
This method remains the gold standard for uncovering dwarf galaxies in resolved stellar population catalogues. Its efficiency is well documented, having enabled the discovery of a wide range of stellar systems, including ultra-faint star clusters \citep[e.g.][]{Kim-2015a, Kim-2015b, Kim-2016c, Mau-2020, Cerny-2023}, extremely faint stellar systems whose nature is still debated \citep[e.g.][]{Bechtol-2015, Koposov-2015, Luque-2016, Smith-2024}, and UFDs within the MW or surrounding external hosts \citep[e.g.][]{Laevens-2015a, Laevens-2015b, Homma-2019, Homma-2024, Cerny-2023, Cerny-2025, Gatto-2024, Smith-2025}.\par% 
The core objective of this strategy is to enhance the signal-to-noise ratio of low surface brightness stellar systems against the foreground of MW stars. This is achieved by performing a preliminary selection in the colour--magnitude diagram (CMD) using the \HE vs. ($\IE-\HE$) colour space. We isolate sources that are photometrically consistent with the evolutionary tracks of ancient, metal-poor stellar populations, which are characteristic of the satellite galaxies targeted in this study, \massi{by selecting stars within 0.05~mag in colour from a given isochrone. To account for the increasing photometric uncertainty towards fainter magnitudes, this fixed colour threshold is combined in quadrature with the propagated photometric error on the ($\IE-\HE$) colour at each star's magnitude, so that the effective width of the selection window increases smoothly for fainter sources.}
We utilized the PARSEC isochrone library \citep{Bressan-2012}, adopting an age of $\logten(t/\rm{yr}) = 10.12$ and a metallicity of \feh{} = $-$2.2. 
This choice is representative of the stellar populations expected to inhabit the UFDs \citep{Pace2025}.
\massi{We note that the target isochrone is used only to enhance the contrast of stellar overdensities against the foreground and background contaminants, and not to characterize the detected systems, whose stellar populations are subsequently modeled with a wider range of isochrones (Sect.~\ref{sc:photo_analysis}).}\par
We systematically scanned a wide range of distance moduli, 15 $\leq (m - M)_0 \leq 28$ (corresponding to a distance range from $10$ kpc to approximately $4$ Mpc), using a step size of $0.5$ mag. This process resulted in the generation of 27 distinct filtered catalogues, each optimized for a specific distance interval.
\massi{Owing to the broadening of the selection window at faint magnitudes, systems slightly beyond the upper limit of this range can still be detected in the filtered catalogue with the largest distance modulus.}
For each filtered dataset, we generated stellar density maps with a pixel resolution of $30\arcsecond$, smoothed using a Gaussian kernel with a bandwidth of $1\arcminute$. 
We quantified the significance $S_{\text{pxl}}$ of each pixel as
\begin{equation}
    S_{\text{pxl}} = \frac{N_{\text{stars}} - N_{\text{bkg}}}{\sigma_{\text{bkg}}},
\end{equation}
where $N_{\text{stars}}$ denotes the star count within a given pixel, while $N_{\text{bkg}}$ and $\sigma_{\text{bkg}}$ represent the mean and standard deviation of the estimated local stellar count, respectively. These background statistics were computed by considering all pixels located within an annulus centred on the analyzed pixel, with an inner radius of 6\arcminute~and an outer radius of 15\arcminute.\par 
%  \NFM{comment: What is the S/N used to consider a detection significant?}\par
To validate our detection pipeline, we performed a sanity check to ensure that the algorithm could consistently recover both UFDs and more distant dwarf galaxies within the \Euclid DR1 footprint. 
\massi{Within the DR1 footprint, the \citet{Pace2025} catalogue lists 42 known systems classified as galaxies or ambiguous objects with a distance modulus $(m-M) < 28$~mag \citep{Tully-2009}. Our algorithm recovers 37 of these 42 systems, corresponding to an overall recovery fraction of $\simeq 88\%$. All five missed systems lie at $(m-M) \gtrsim 27.8$~mag, i.e.\ at heliocentric distances $\gtrsim 3.6$~Mpc, indicating that incompleteness sets in only beyond this distance.
Specifically, 25 galaxies lie at $(m-M) \gtrsim 27.8$~mag (i.e.\ beyond $\simeq 3.6$~Mpc), the algorithm recovers 20 of these 25 galaxies, whose absolute $V$-band magnitudes range from $-18.34$ to $-7.26$~mag, while all five missed objects are fainter than $M_V \simeq -11.85$~mag.
Restricting the comparison to this faint end, the \citet{Pace2025} catalogue lists 11 galaxies with $M_V \gtrsim -11.85$~mag beyond 3.6~Mpc, of which the algorithm recovers 6. Notably, the algorithm's sensitivity is not strictly limited by luminosity: it successfully recovers galaxies as faint as $M_V \simeq -8.1$~mag, such as dw0944+69 \citep{Chiboucas-2013}, and systems as distant as $(m-M) = 28.42$~mag ($\simeq 4.8$~Mpc), such as DDO~165.
In summary, the algorithm achieves a 100\% recovery fraction for all known galaxies within $\simeq 3.6$~Mpc, decreasing to $\simeq 55\%$ for galaxies beyond this distance and fainter than $M_V \simeq -11.85$~mag. This drop in the recovery fraction likely reflects the small number of resolved stars that \Euclid\ can detect in such faint, distant systems, which limits the algorithm's ability to identify them as significant stellar overdensities.
A systematic  analysis of the selection function will be presented elsewhere (\textcolor{blue}{Rusterucci et al. in prep.}).}
%The sensitivity and selection function of the isochrone filter method in the context of the \Euclid DR1 dataset will be analyzed and described in detail in a forthcoming paper (\textcolor{blue}{Rusterucci et al.;} in prep.).
%In particular, the algorithm successfully detected Ursa Minor II, the first candidate UFD discovered within the \Euclid consortium (Balbinot et al., in prep.). Furthermore, the algorithm also demonstrated its effectiveness for objects beyond the Local Group by correctly identifying galaxies such as ESO 115-021, which lies at an estimated distance modulus of (m - M) = 28.53 mag \citep{Pace2025}.

\section{\label{sc:results}  Results}
\begin{figure*}[htbp!]
    \centering
    \includegraphics[width=.3\linewidth]{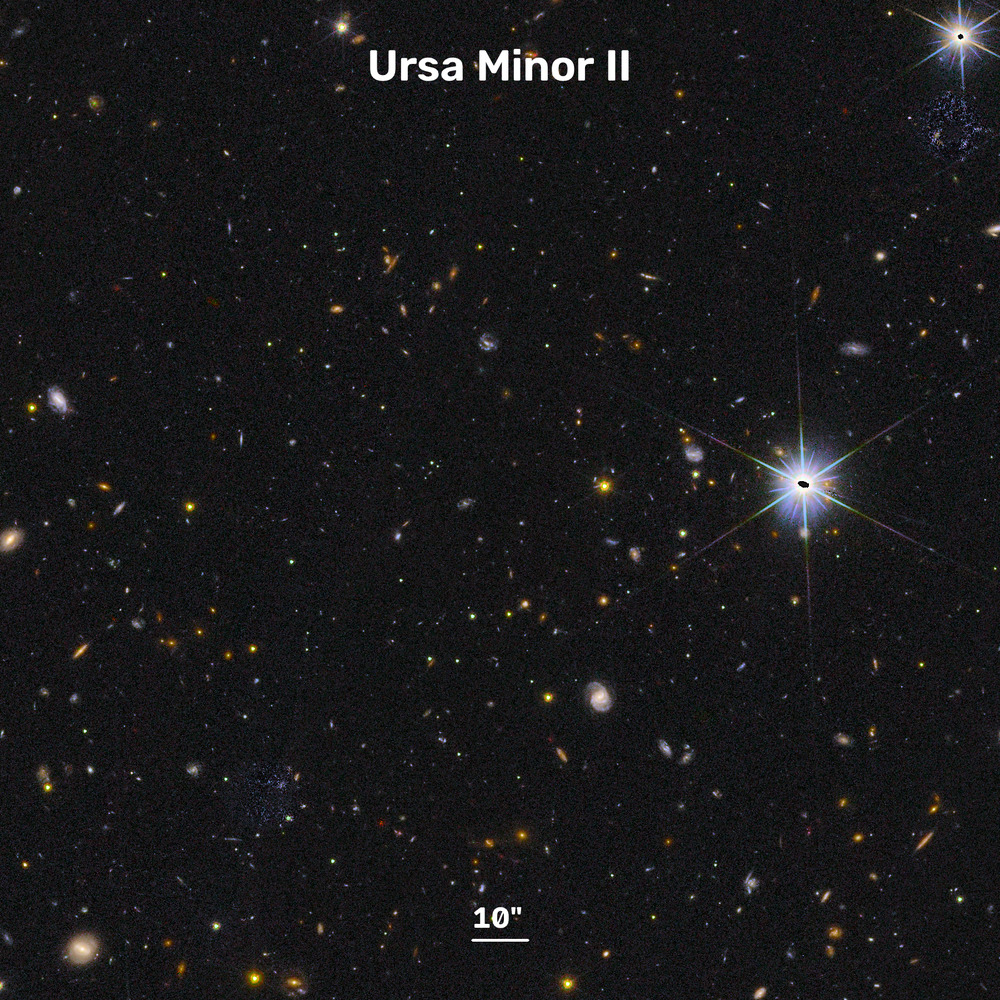}
    \includegraphics[width=.3\linewidth]{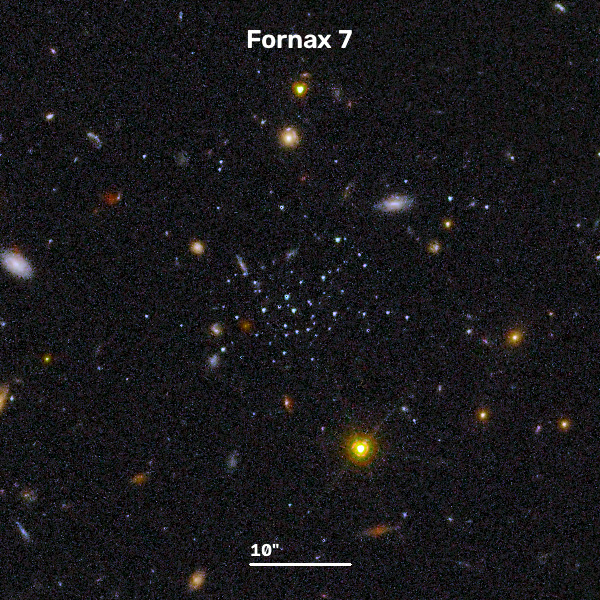}
    \includegraphics[width=.3\linewidth]{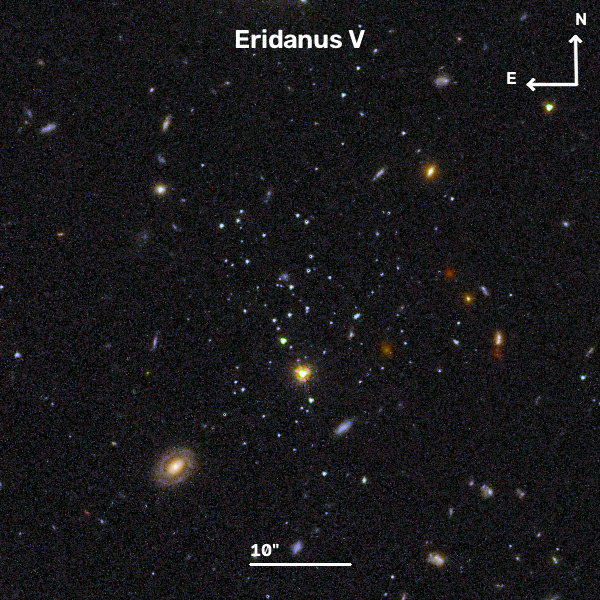}\\%[1ex]
    \includegraphics[width=.3\linewidth]{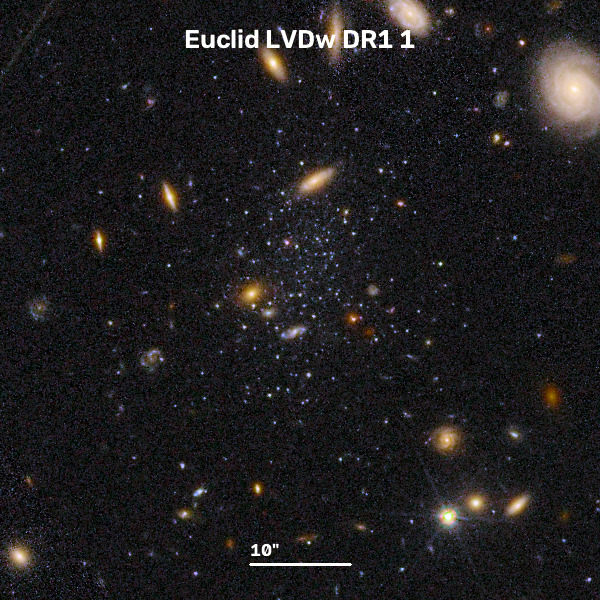}
    \includegraphics[width=.3\linewidth]{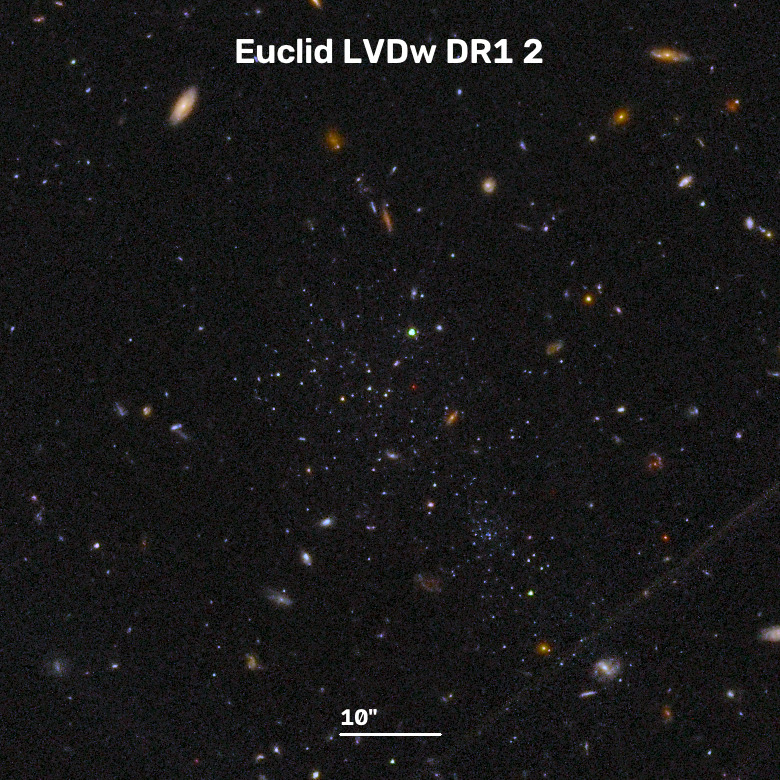}
    \includegraphics[width=.3\linewidth]{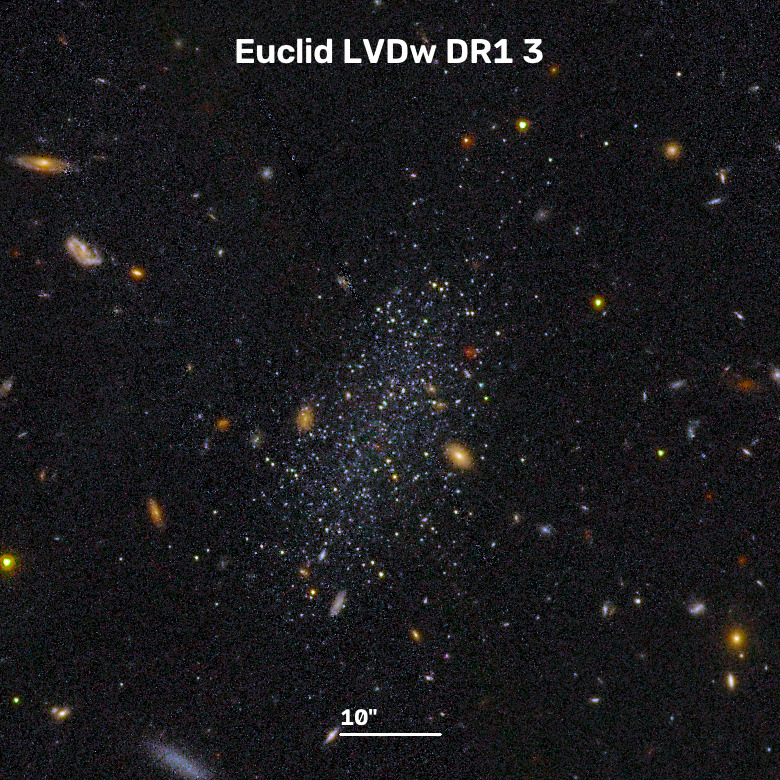}\\%[1ex]
    \includegraphics[width=.3\linewidth]{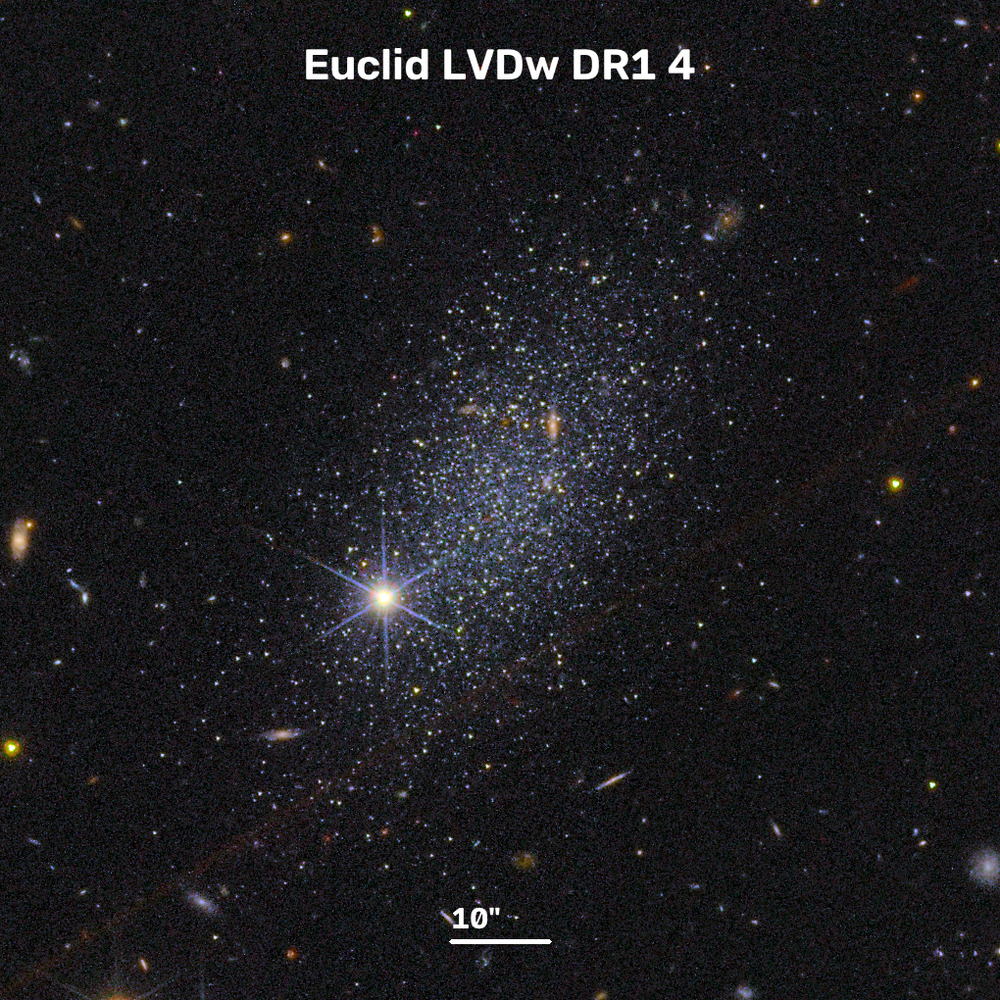}
    \includegraphics[width=.3\linewidth]{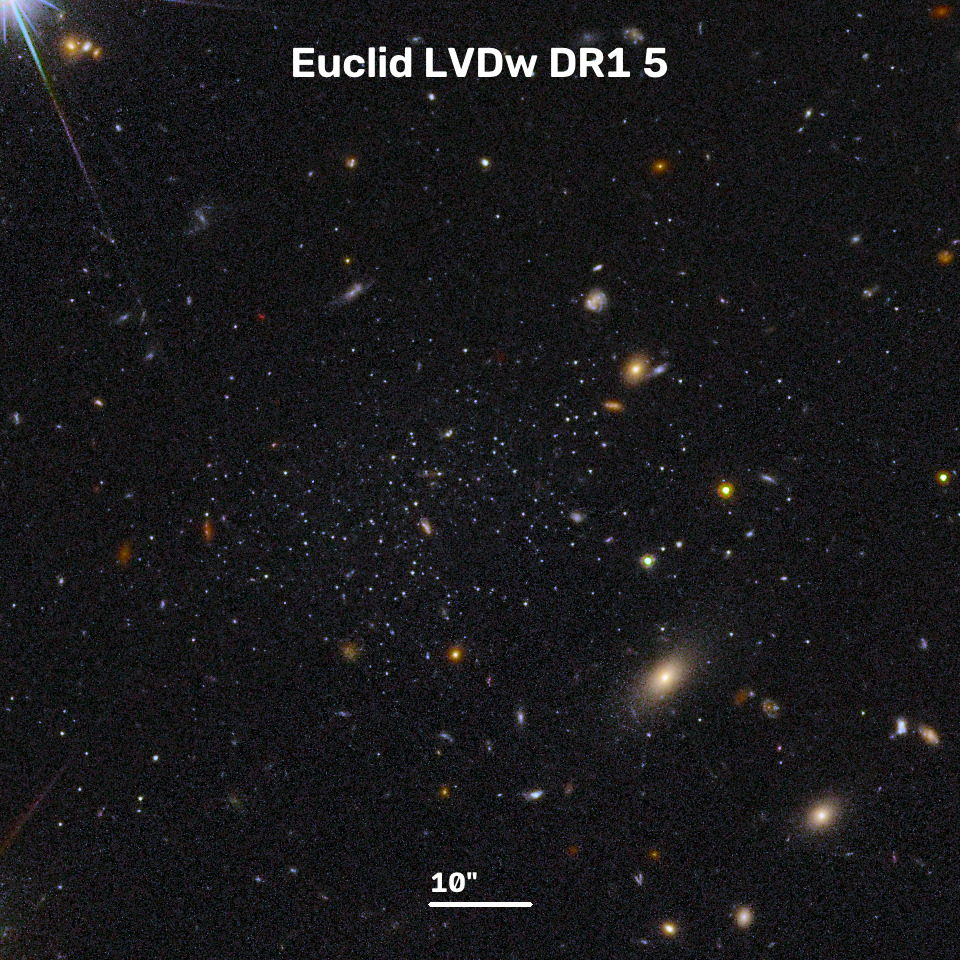}
    \includegraphics[width=.3\linewidth]{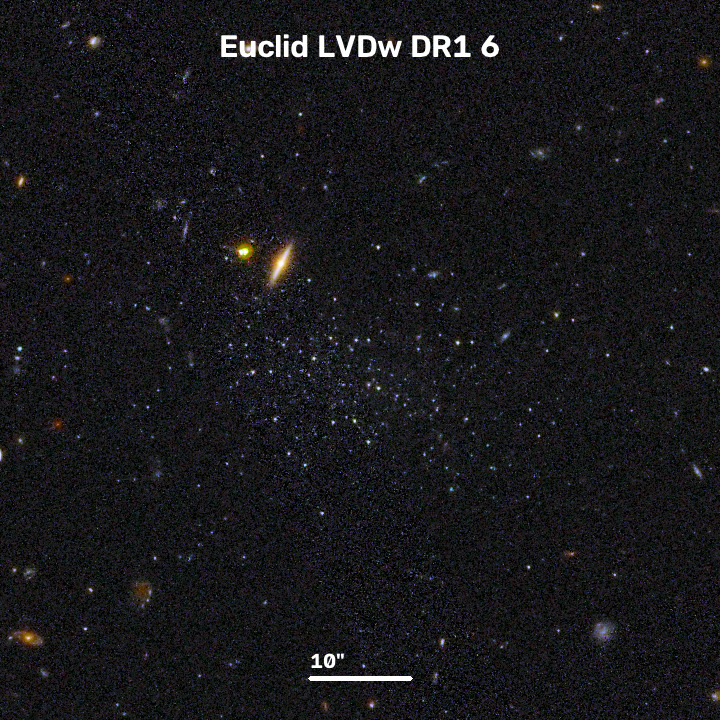}\\%[1ex]
    \includegraphics[width=.3\linewidth]{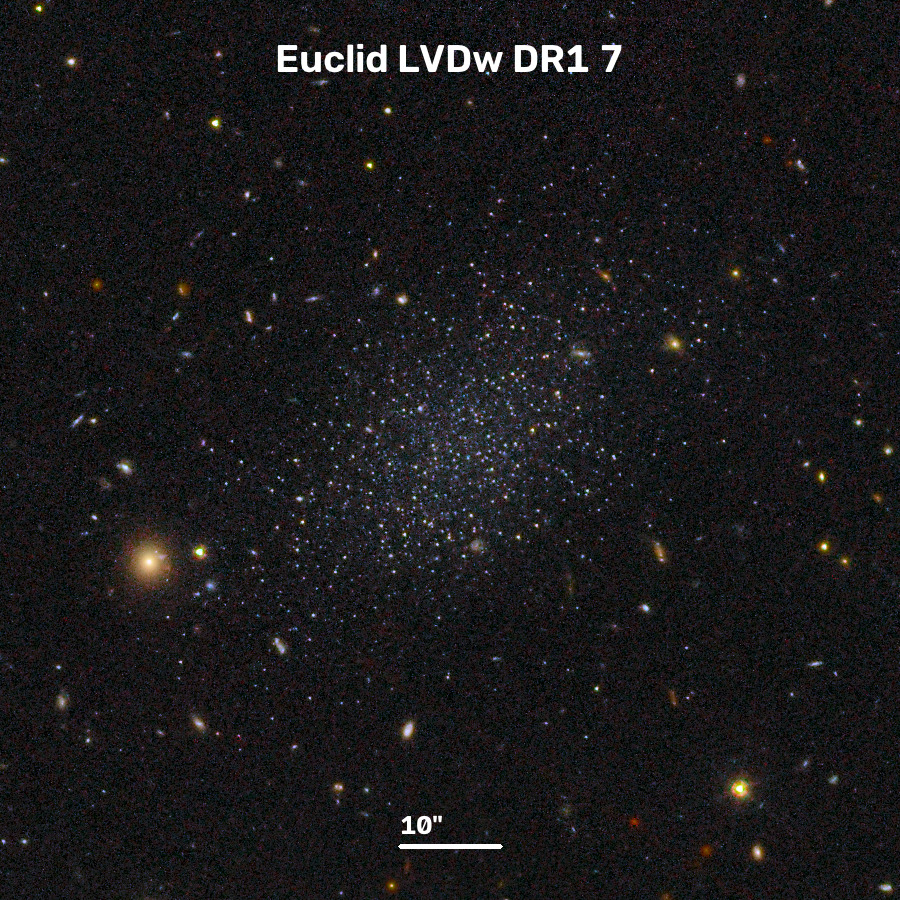}
    \includegraphics[width=.3\linewidth]{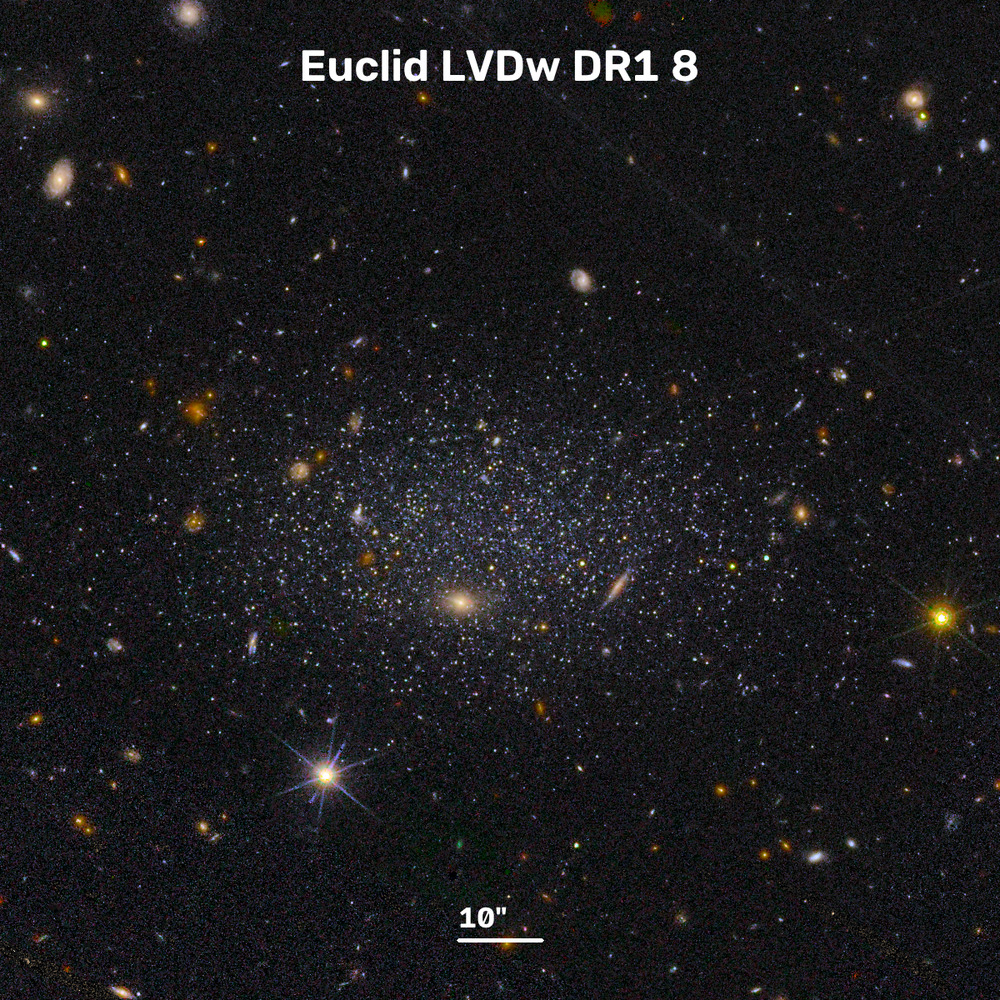}
    \includegraphics[width=.3\linewidth]{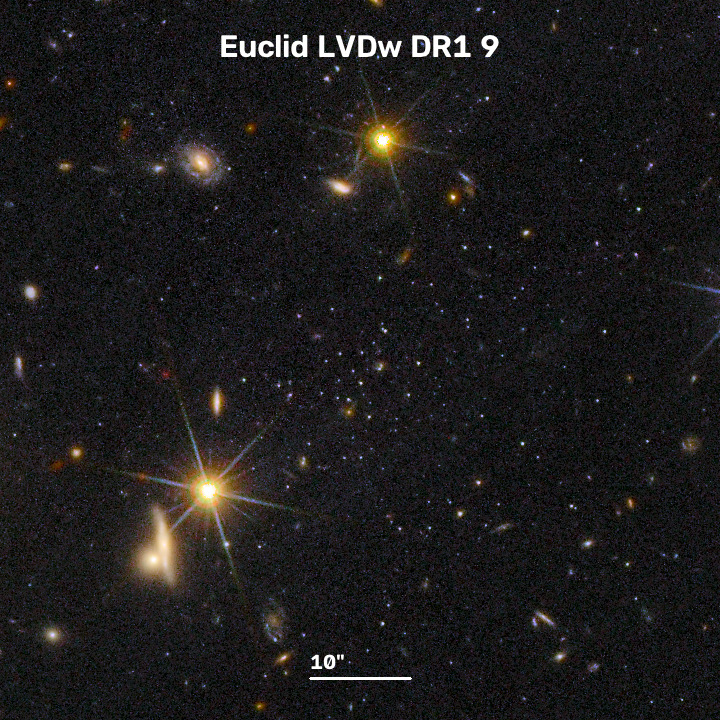}
    \caption{\massi{\Euclid\ composite colour images of the 15 stellar systems presented in this work. Each panel displays the system name at the top and a $10''$ scale bar at the bottom. The north-east orientation is indicated by the white arrows in the top-right panel.}}
    \label{fig:mosaic}
\end{figure*}

% Seconda pagina (continuazione)
\begin{figure*}[htbp]
    \centering
    \ContinuedFloat 
    \includegraphics[width=.3\linewidth]{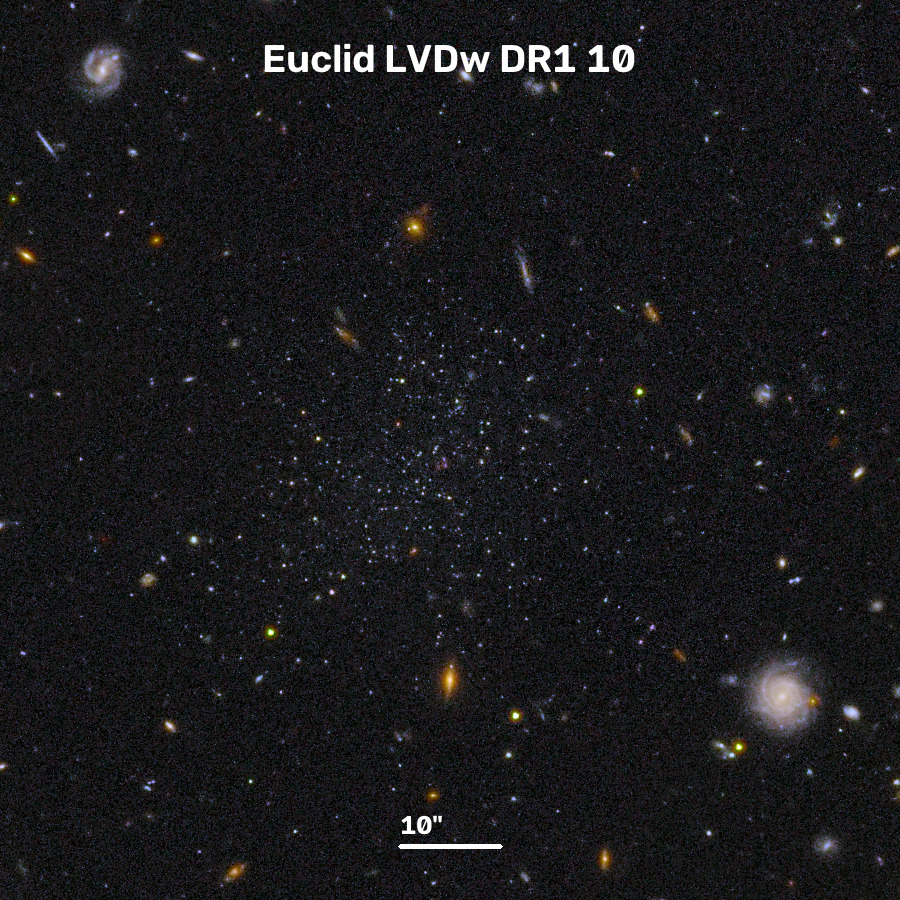}
    \includegraphics[width=.3\linewidth]{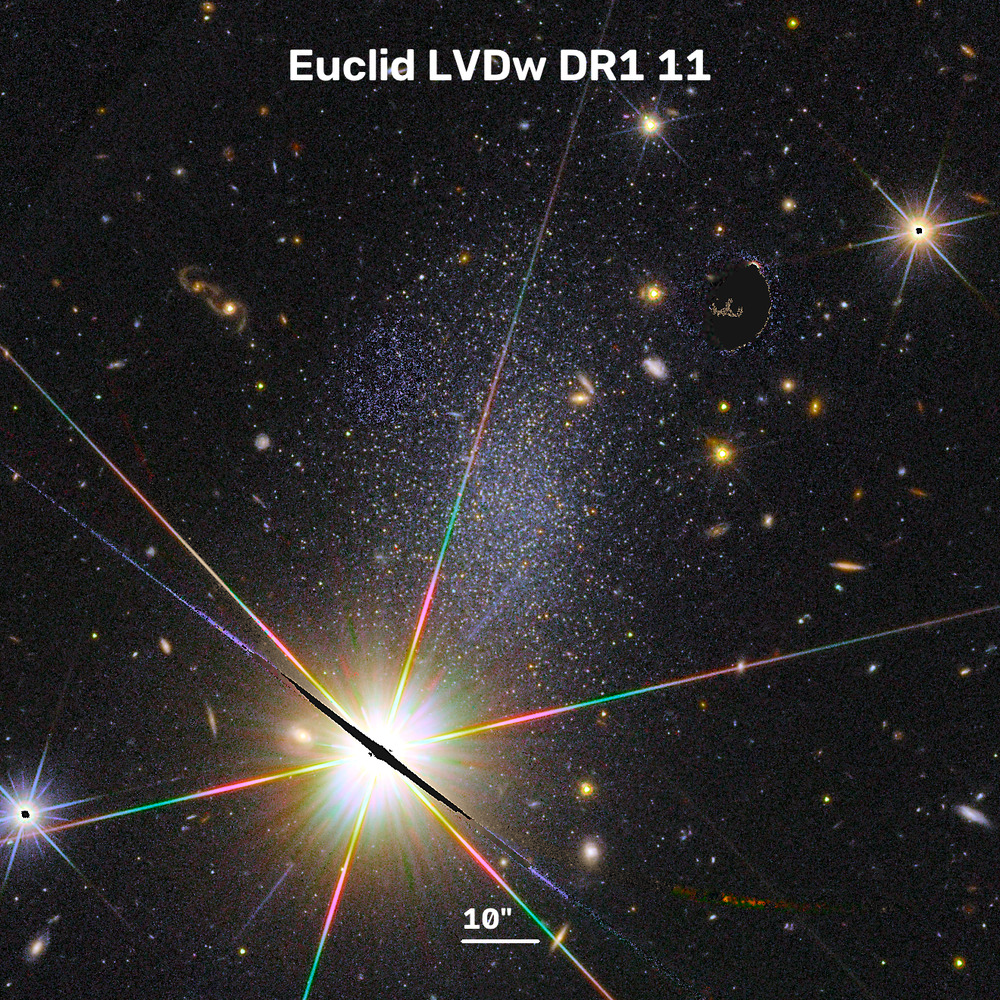}
    \includegraphics[width=.3\linewidth]{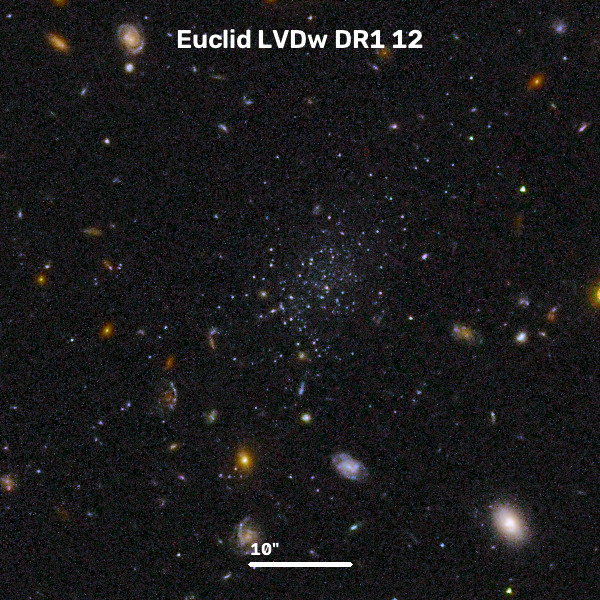}
    \caption{continued.}
\end{figure*}

By applying the isochrone filter technique described in Sect.~\ref{sc:method} to the full DR1 footprint, we identified five previously unknown stellar systems.\footnote{These stellar systems were independently found by applying \texttt{erose}, a search algorithm designed to automatically detect (semi-)resolved dwarf galaxies in wide-field surveys \citep[see][for further details about the algorithm]{Rusterucci-2026}.} 
Our search also recovered ten systems that were already catalogued, but had never been resolved into stars.
Thanks to the angular resolution and depth of \Euclid, we resolved their brightest stars for the first time, which allowed us to constrain their distances directly from their resolved stellar populations.
\massi{Each candidate detection was first visually inspected on the deep \Euclid VIS stacked images to confirm its reality against artifacts or chance projections; only systems confirmed in this way are presented as secure discoveries in this work.}
\massi{We designate the three MW systems, whose nature (globular cluster or UFD) cannot be established from our photometry, with both a constellation-based name and a \Euclid designation as ultra-faint stellar systems (Euclid~UFSS~DR1~1 to DR1~3), and the twelve Local Volume dwarf galaxies as Euclid~LVDw~DR1~1 to DR1~12, the first two being the new discoveries.}
%In addition to the five new galaxies presented in table 1, that are confirmed by the inspection of the image or by spectroscopic follow-up, in App.~C we present a candidate that is detected as a significant over-density and displays a CMD similar to Eri~V but still lack a clearcut confirmation.
%\massi{This step favours compact systems and those with a sufficient number of bright stars; nearby, very extended systems with few bright members, such as \UMi{}, are barely discernible in the images, and overdensities whose nature cannot be established in this way are not included in this work.}
\massi{Figure~\ref{fig:mosaic} shows colour-composite \Euclid images of all 15 systems\footnote{We adopted the \texttt{eummy} python code (\url{https://github.com/schirmermischa/eummy}) to create \Euclid colour images from the four MER stacked images.}. While the search was performed on the MER catalogue, the analysis of each confirmed system is based on the stellar catalogues produced with the PSF-fitting photometry pipeline described in \citet{Libralato-2024}, \citet{Libralato-2026}, and \citet{Griggio-2026},} 
that has demonstrated higher efficiency and precision in stellar photometry, with respect to the MER catalogue, especially in crowded regions as the main body of the galaxies considered here.\par
%Unless otherwise stated, all photometric measurements, figures, and analyses presented hereafter are based on these catalogues.\par
% CHECK: "and" if two tables, "--" (no spaces) if a range
The main properties of these systems, including their equatorial coordinates and estimated distance moduli, are summarized in Tables~\ref{tab:dwarf_distances} and \ref{tab:dwarf_params}.
The range of recovered distance moduli demonstrates the versatility of our approach: using a consistent methodology across the entire \Euclid dataset, we successfully detected both nearby stellar systems within the Milky Way halo and distant dwarf galaxies located well beyond the canonical boundary of the Local Group.\footnote{The zero-velocity surface, which delineates the Local Group from the expanding Hubble flow, is estimated to have a radius of approximately 1~Mpc \citep[e.g.,][]{Karachentsev-2009}.}\par

\subsection{\label{sc:photo_analysis} Distance estimation}

%\begin{figure}[h!]
%    \centering
    %\includegraphics[width=1.\linewidth]{euclid_coloured_images/EriV_image.jpg}\\
%    \includegraphics[width=1.\linewidth]{CMD/Eridanus V_28.8623_-47.7105_0.49_21.10_CMD_paper.png}
%    \caption{\emph{Left panel:} CMD in the \HE vs. (\IE -- \HE) space for all stars located within 2\,$r_{\rm h}$ of the centre. Red (black) points denote stars selected (not selected) by the isochrone filter. The best-fit PARSEC isochrone, namely $\logten({t/{\rm yr}}) = 10.12$, \feh{} $= -2.2$, (m $-$ M)$_0$ = 21.1 mag is shown as a solid black line, while dashed and dash-dotted lines represent the same isochrone shifted by $\pm 0.3$ mag to illustrate the distance uncertainty. \massi{TThe error bars on the right side indicate the mean photometric uncertainties in different magnitude bins.} 
%    \emph{Right panel}: CMD of a nearby control field of equal area, \massi{centered 5 arcminutes away from the stellar system.}}
%    \label{fig:EridanusV}
%\end{figure}

\begin{figure*}[htbp!]
    \centering
    \includegraphics[width=.33\linewidth]{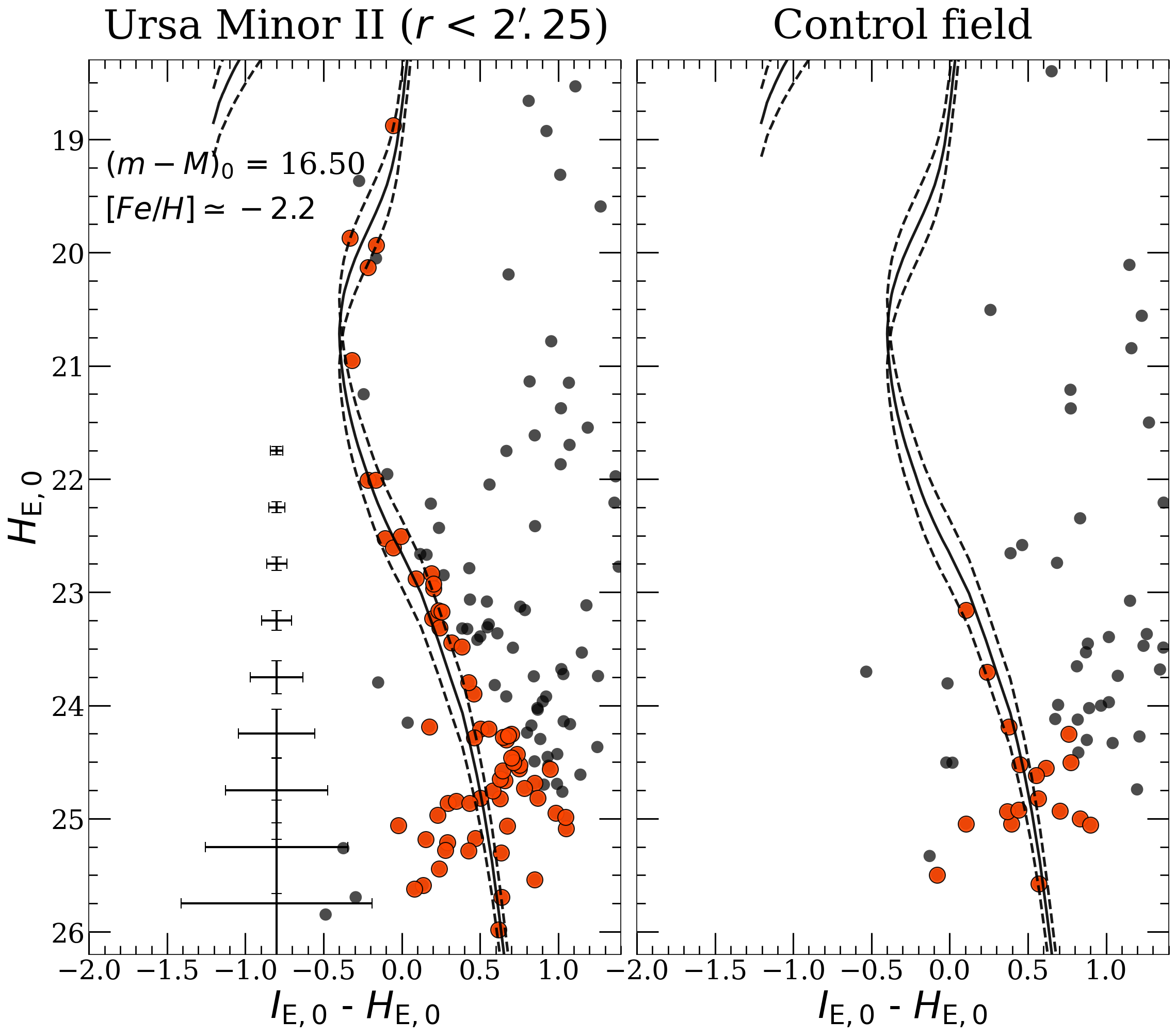}
    \includegraphics[width=.33\linewidth]{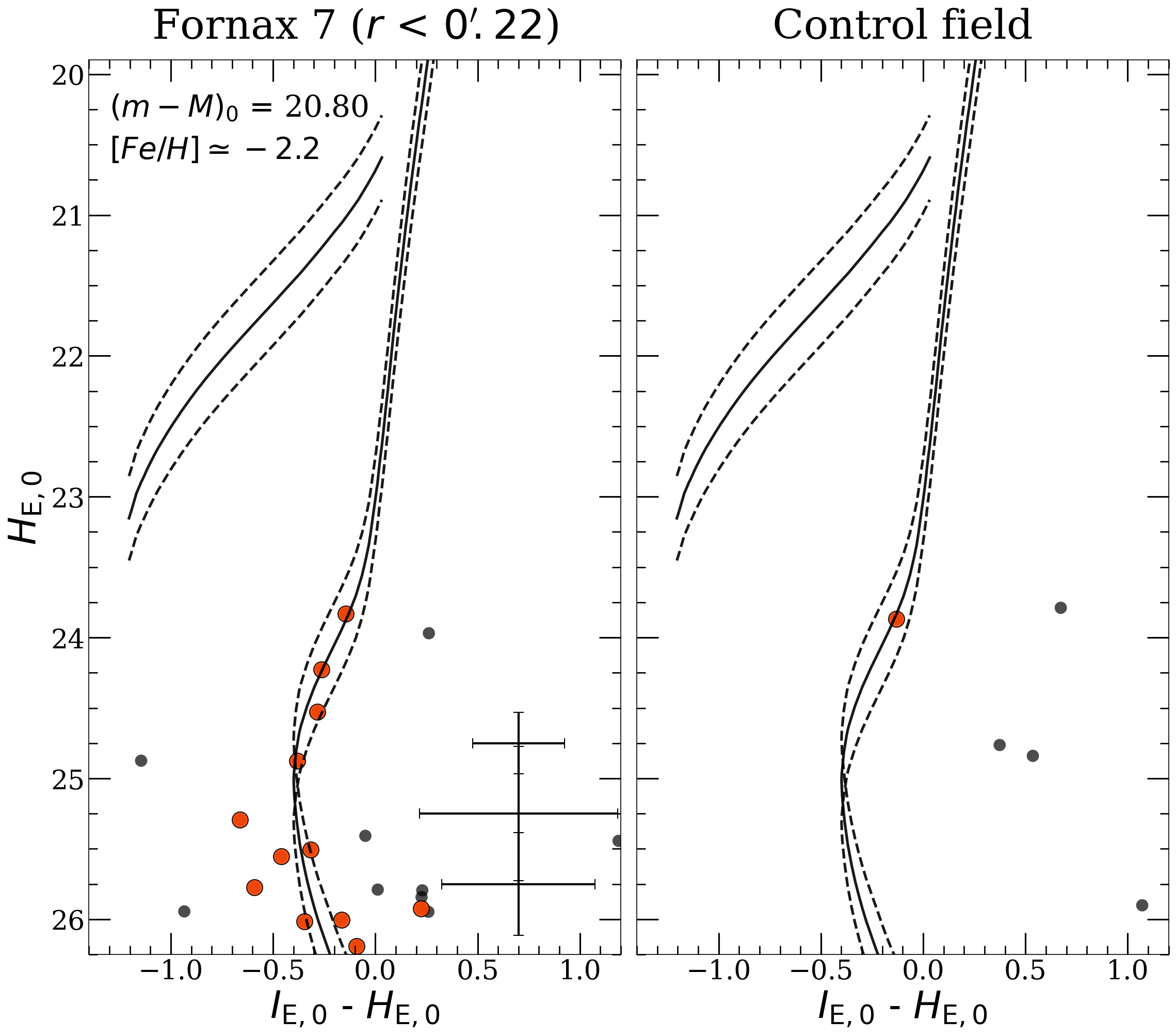}
    \includegraphics[width=.33\linewidth]{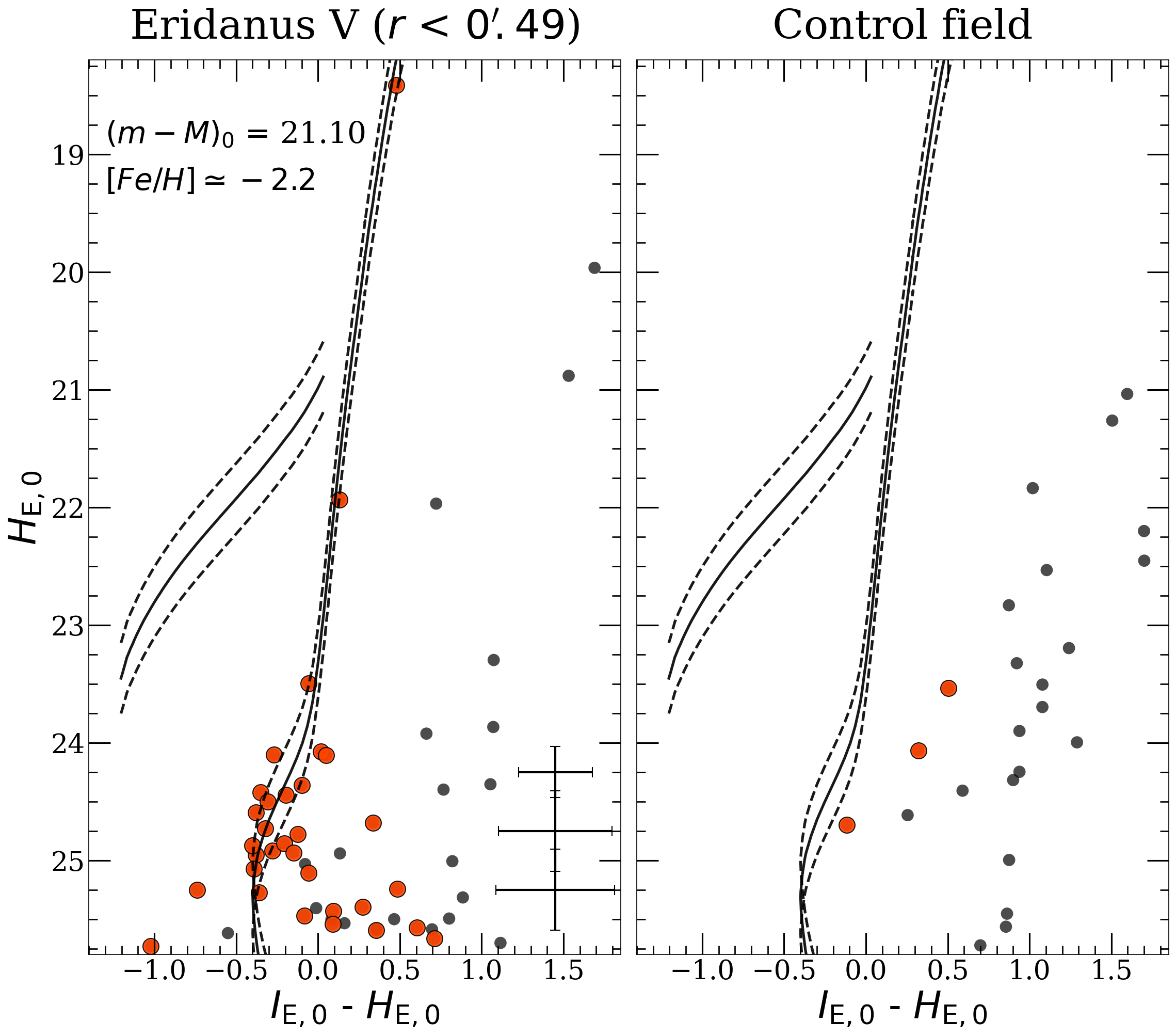}\\
    \includegraphics[width=.33\linewidth]{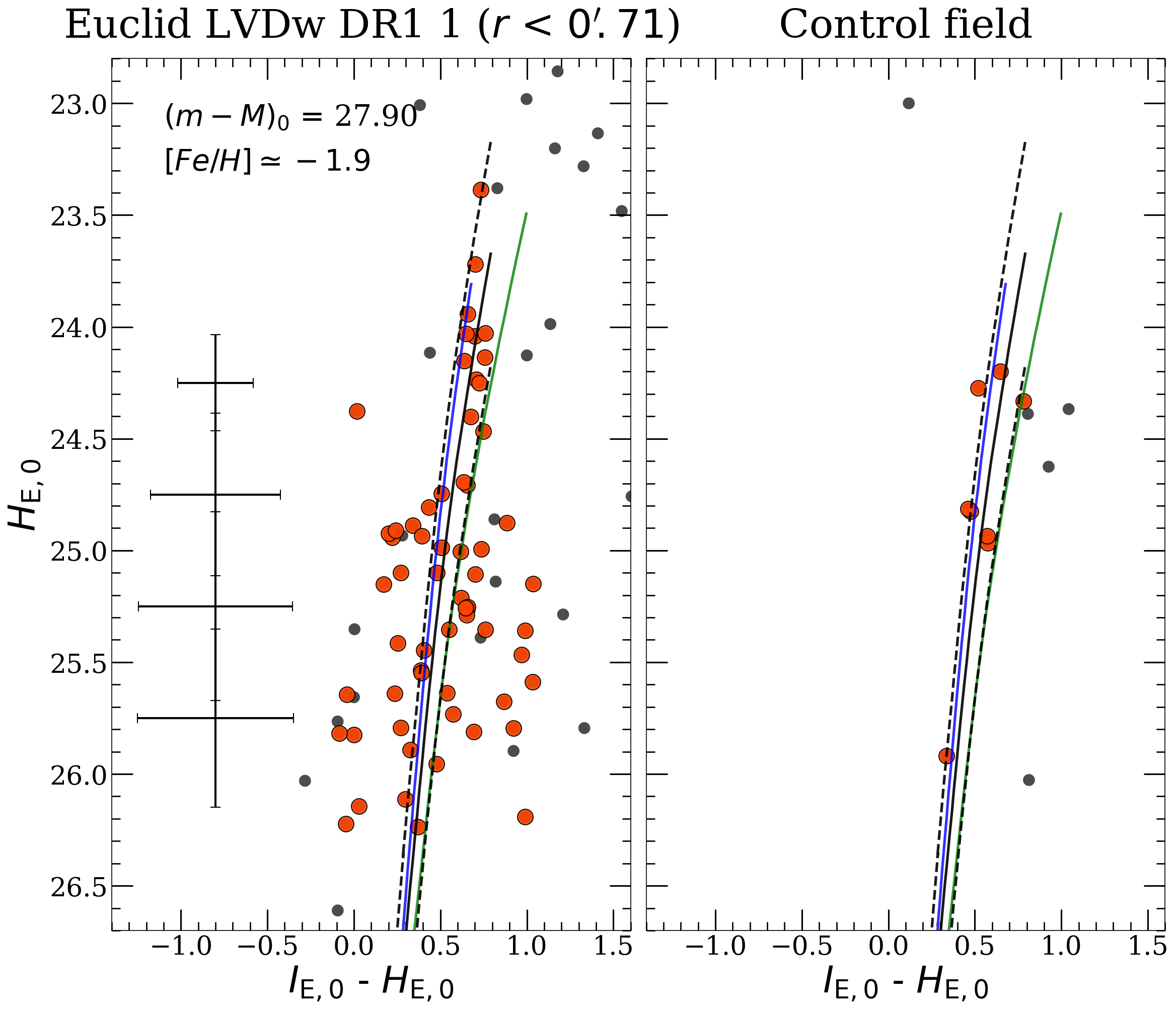}
    \includegraphics[width=.33\linewidth]{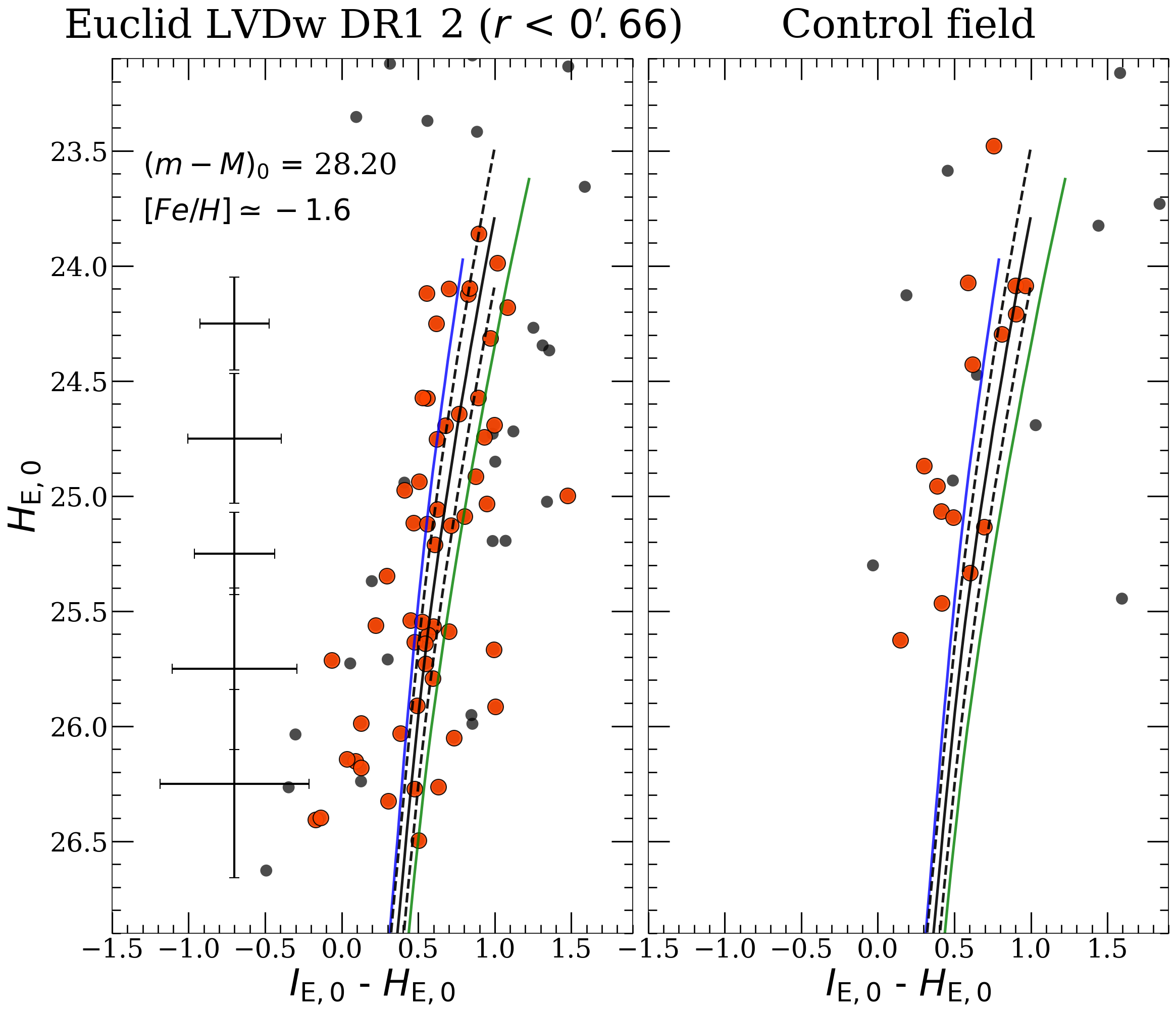}
    \includegraphics[width=.33\linewidth]{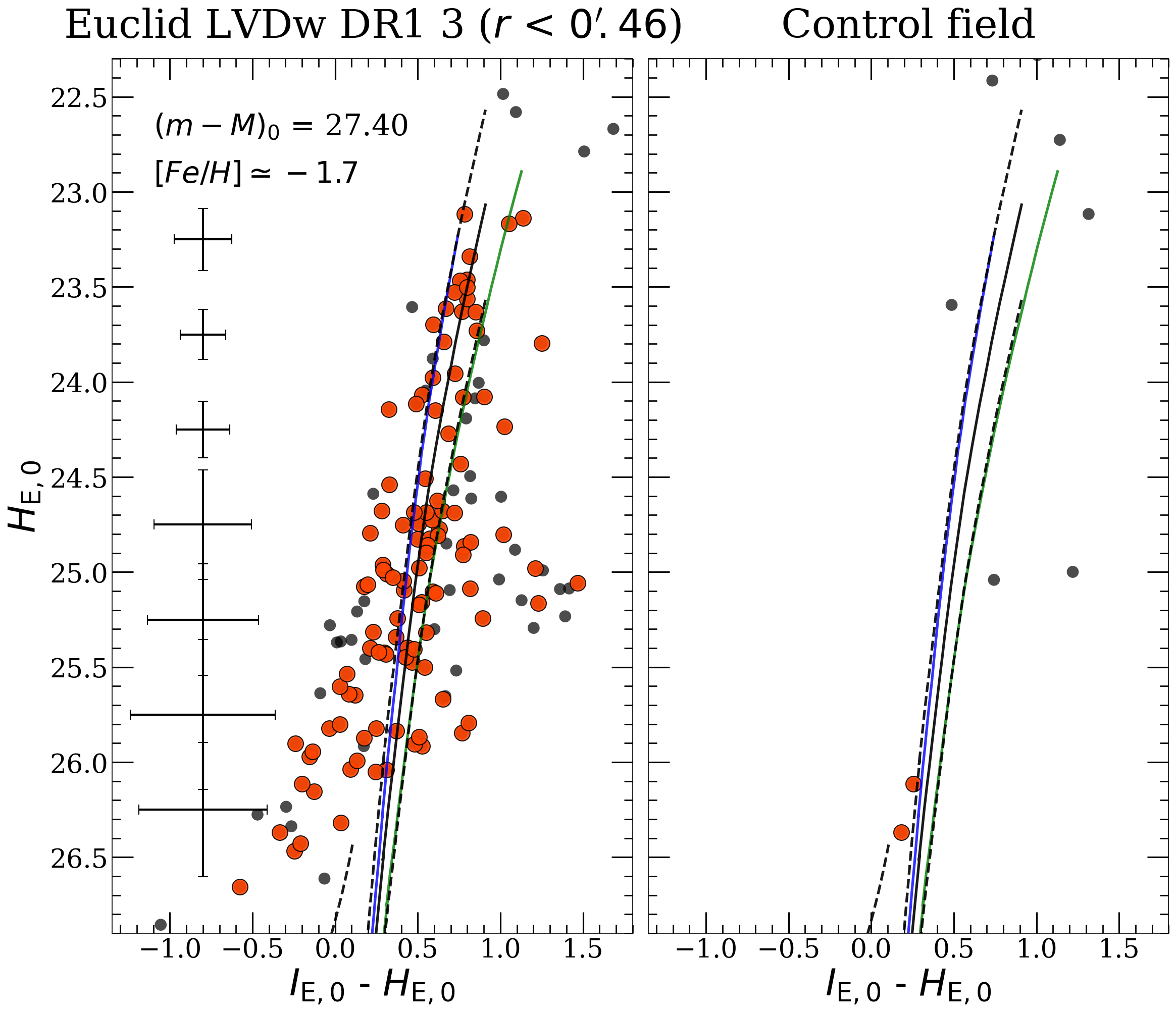}\\
    \includegraphics[width=.33\linewidth]{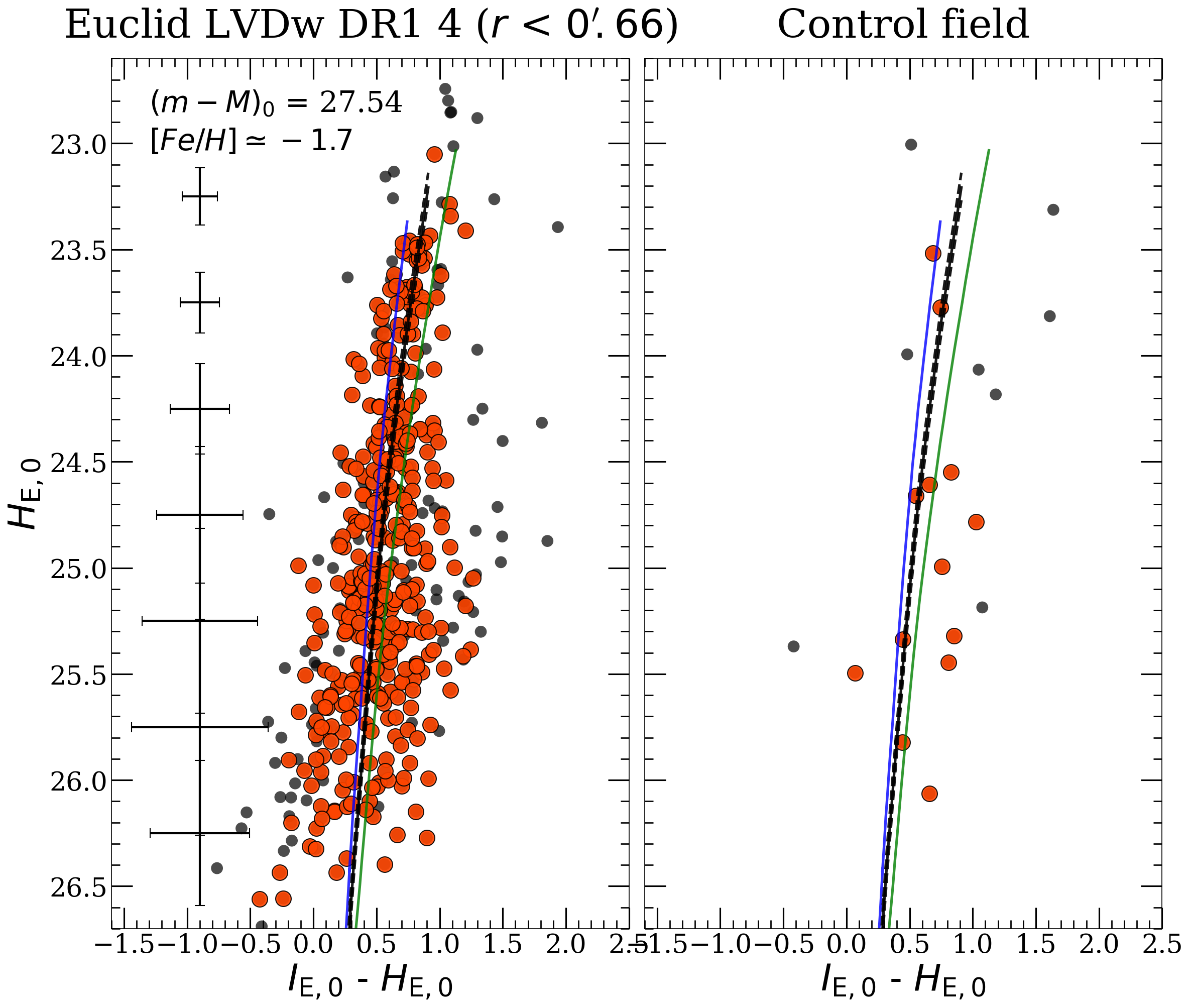}
    \includegraphics[width=.33\linewidth]{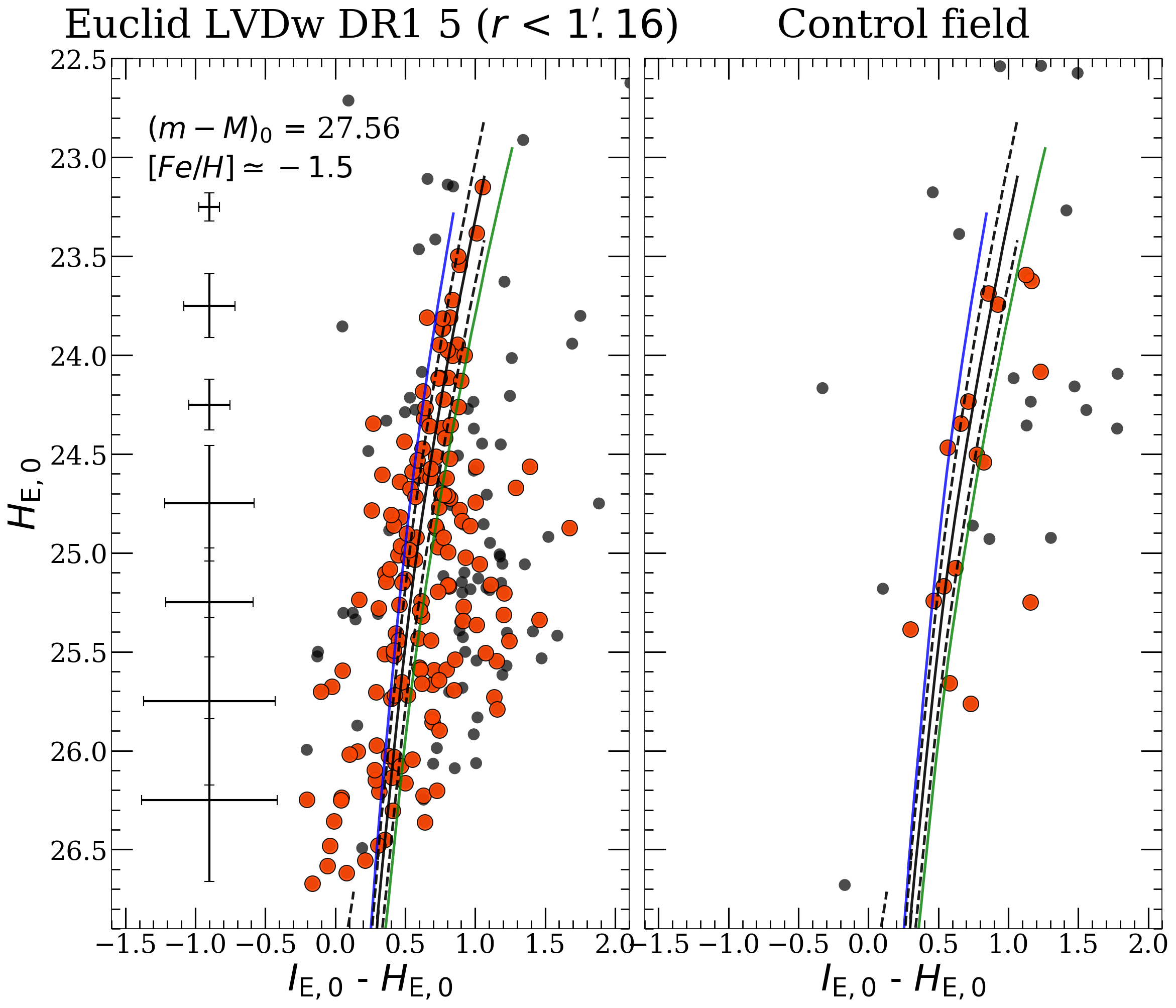}
    \includegraphics[width=.33\linewidth]{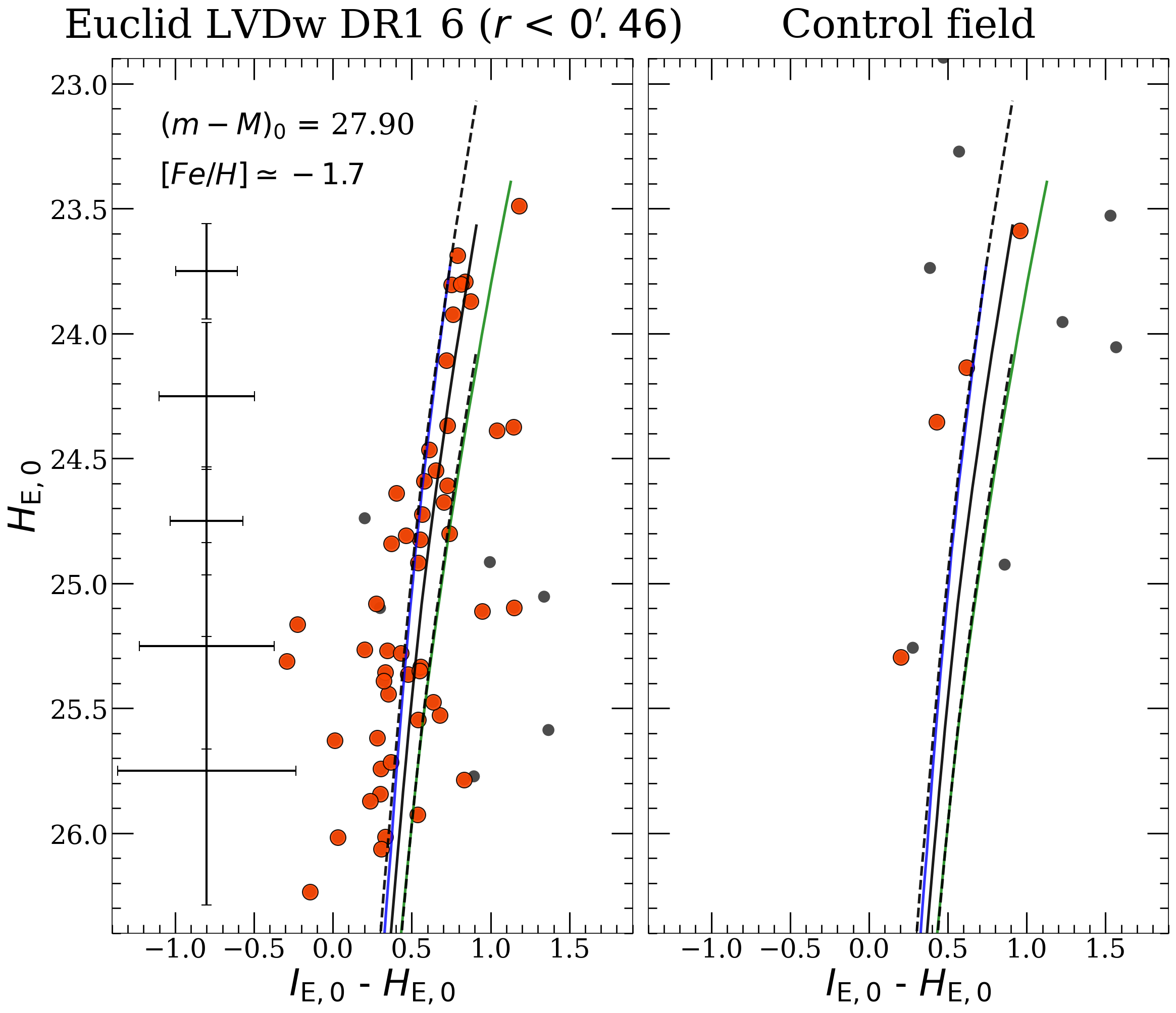}\\
    \includegraphics[width=.33\linewidth]{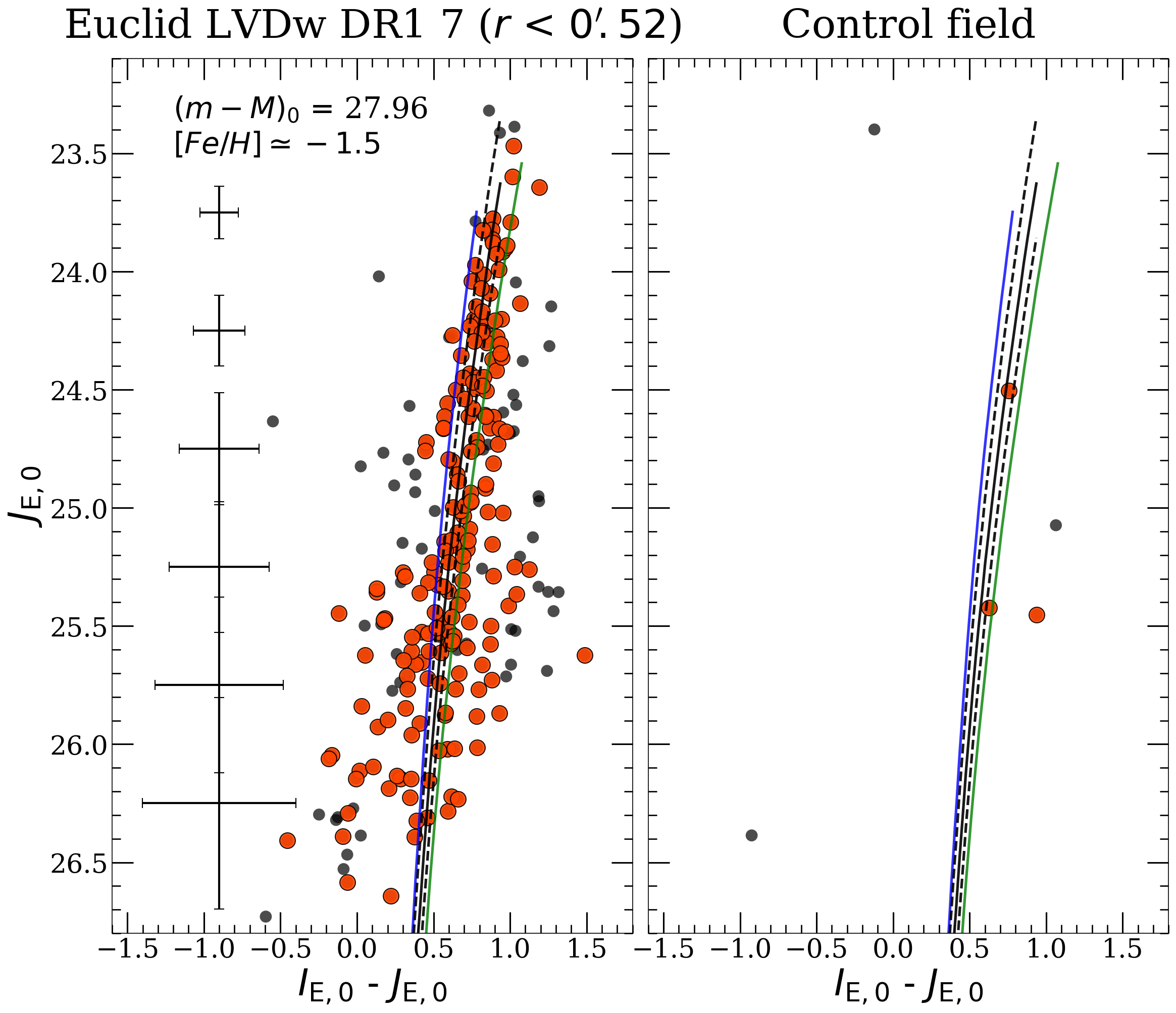}
    \includegraphics[width=.33\linewidth]{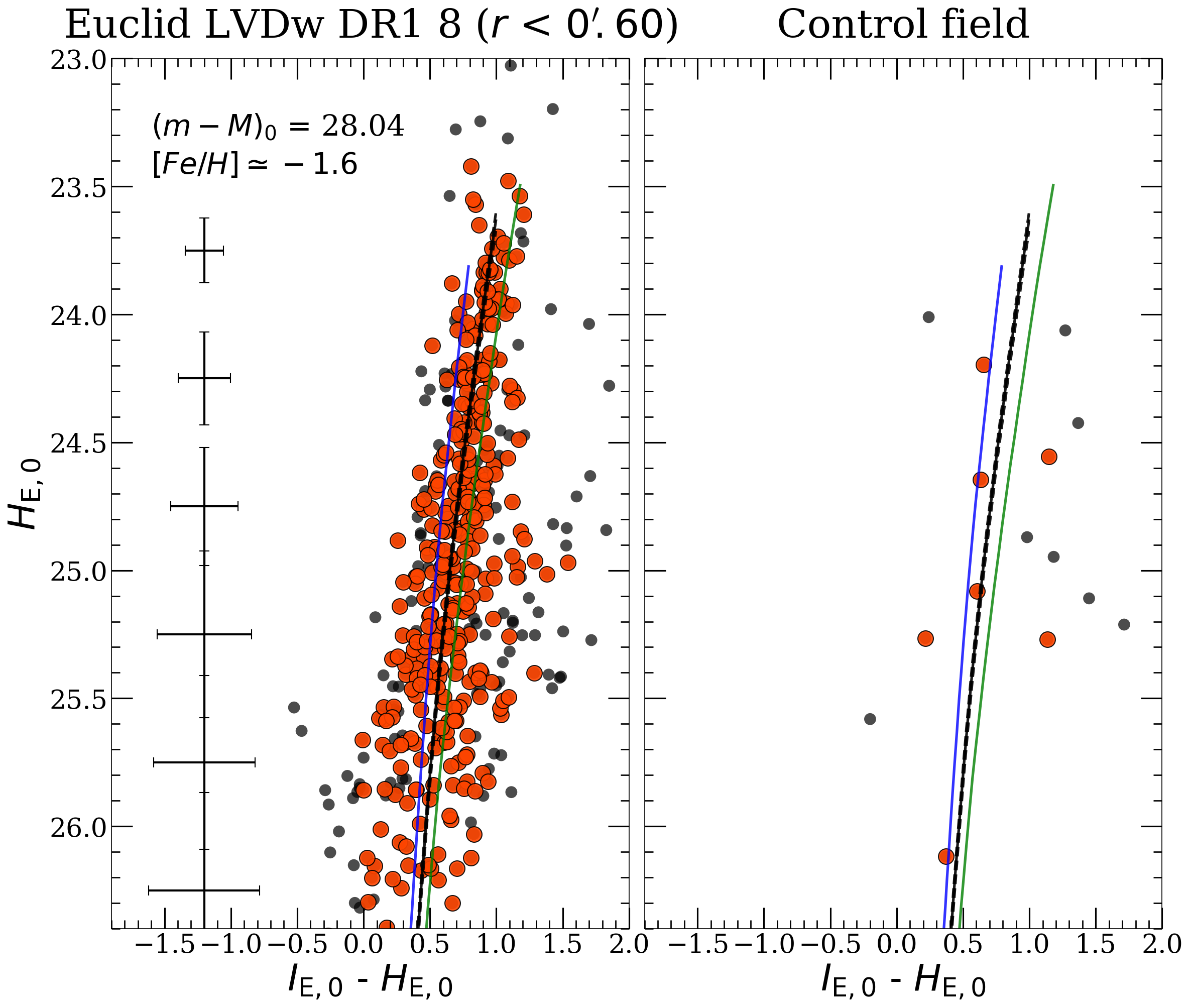}
    \includegraphics[width=.33\linewidth]{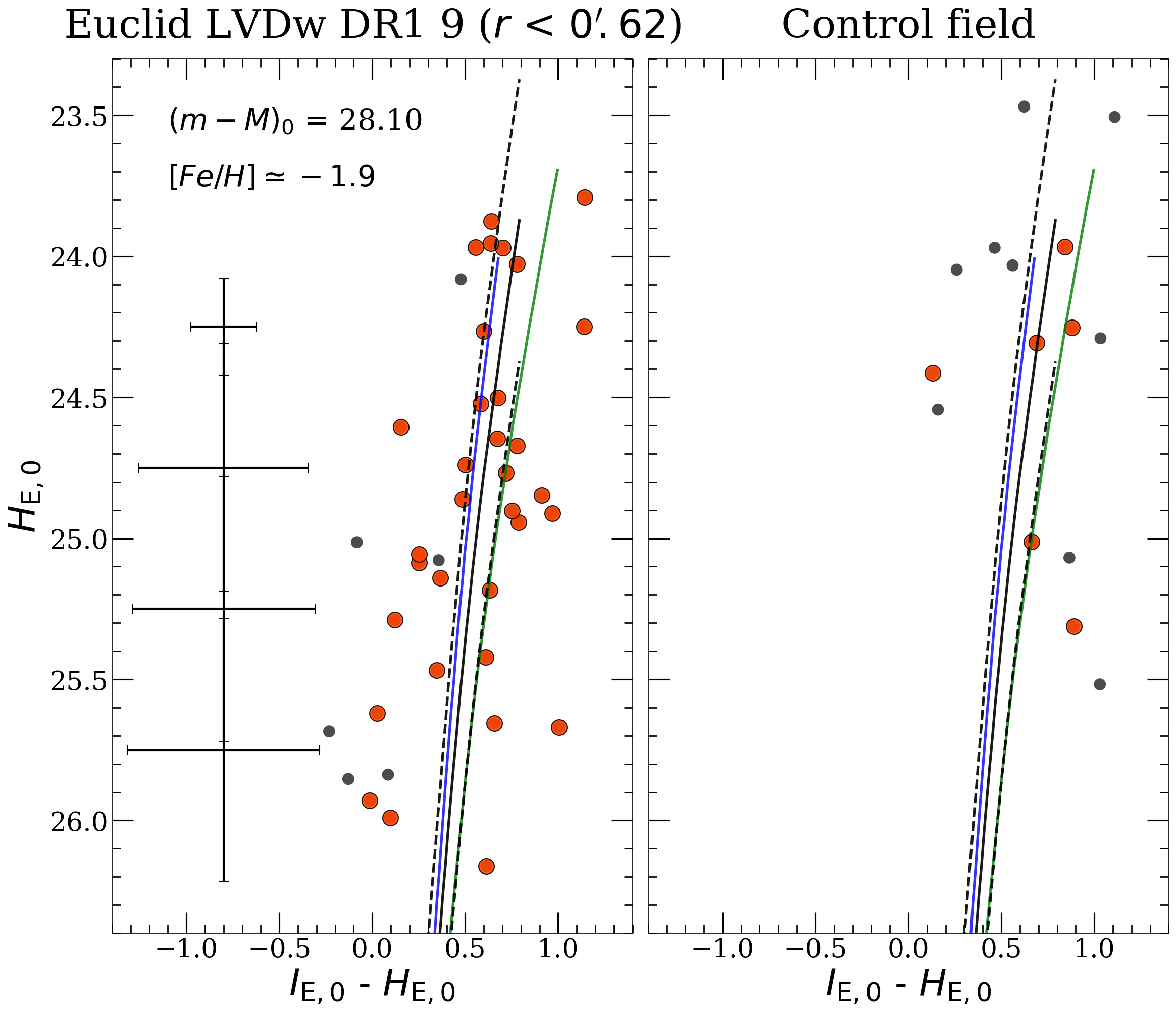}\\
    \caption{\massi{CMD of the 15 stellar systems in the \HE vs. ($\IE-\HE$) space for all stars located within 2\,$r_{\rm h}$ of the centre (left panels). Red (black) points denote stars selected (not selected) by the isochrone filter. The best-fit PARSEC isochrone (whose distance modulus and metallicity are reported in each left panel) is shown as a solid black line, while dashed lines represent the same isochrone shifted by the estimated distance uncertainty. When present, the blue and green solid lines indicate isochrones with the same distance modulus as the best fit, but with metallicities shifted by $-0.3$ and $+0.3$ dex, respectively. The error bars on the side indicate the mean photometric uncertainties in different magnitude bins. The right panels show the CMD of a nearby control field of equal area, centered 5 arcminutes away from the stellar system.}}
    \label{fig:CMD_all}
\end{figure*}

\begin{figure*}[htbp]
    \centering
    \ContinuedFloat 
    \includegraphics[width=.33\linewidth]{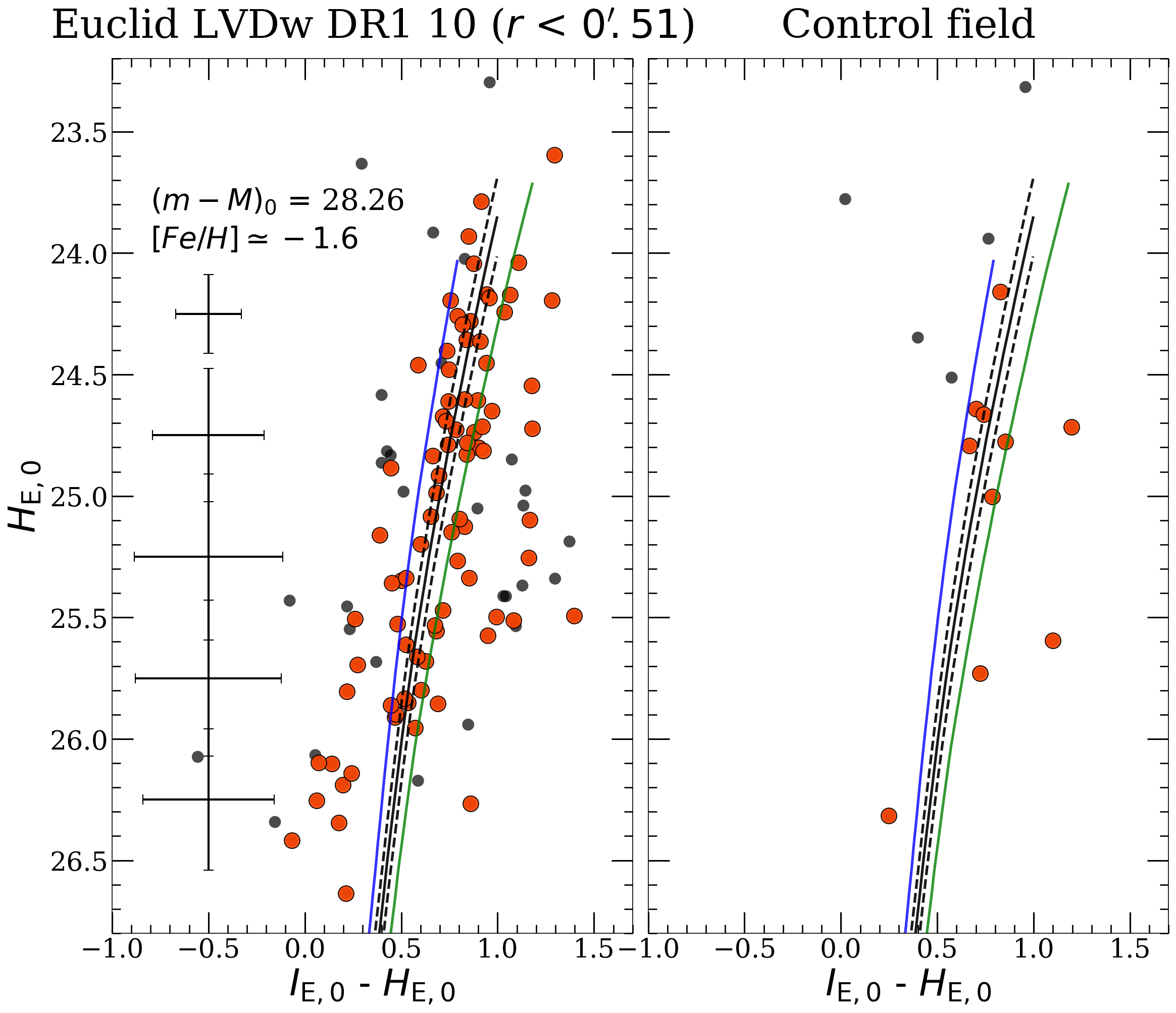}
    \includegraphics[width=.33\linewidth]{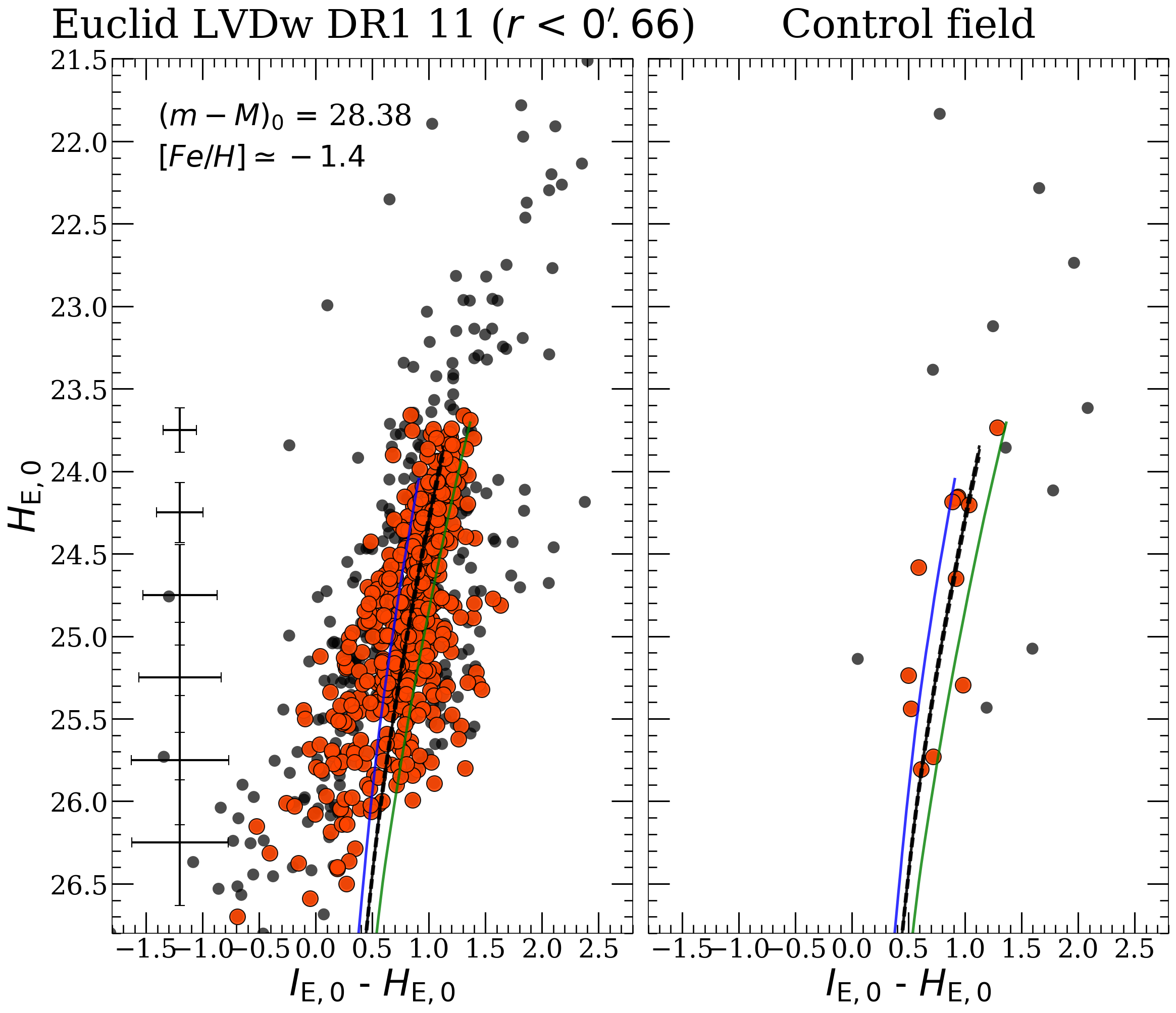}
    \includegraphics[width=.33\linewidth]{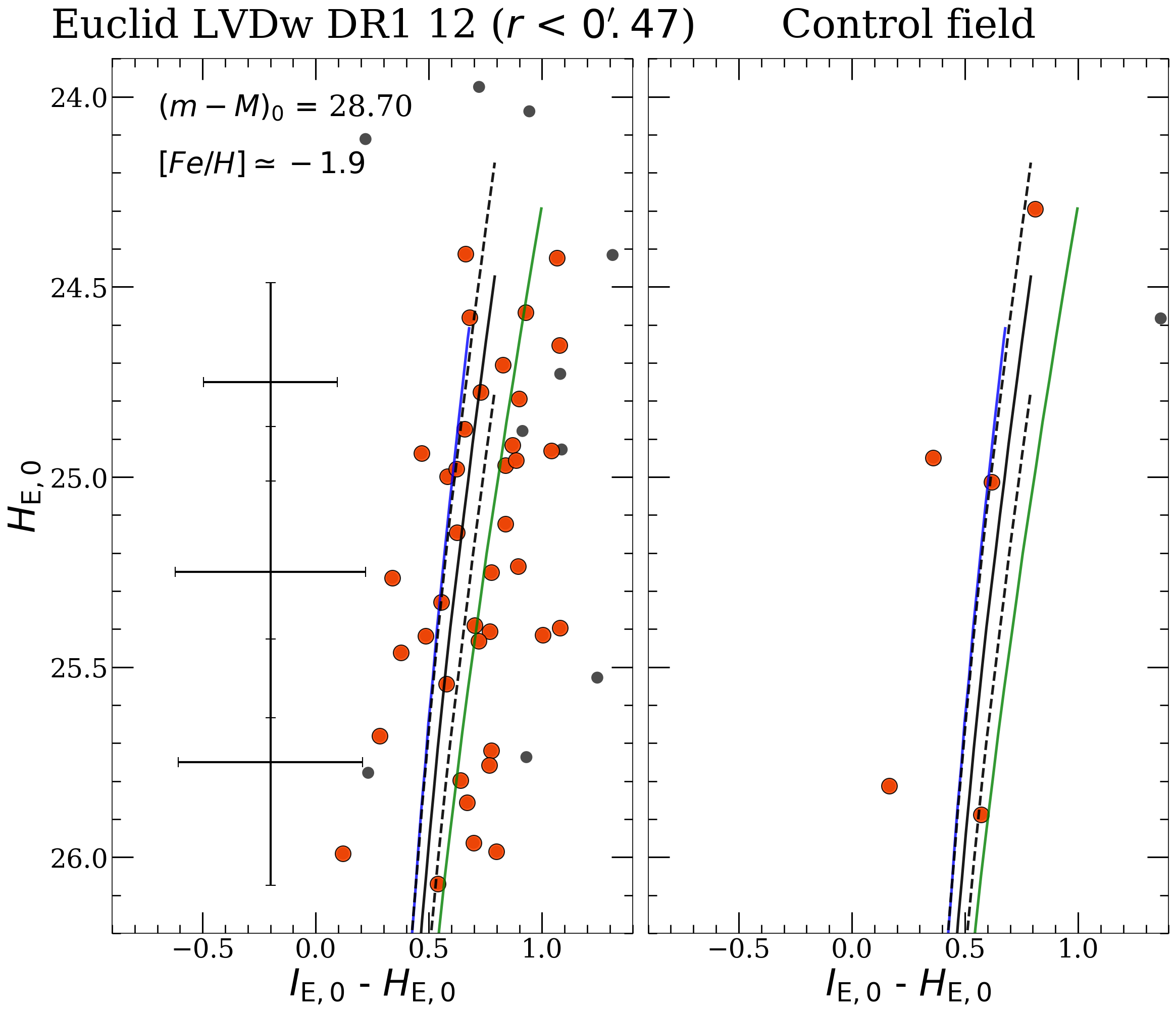}\\
    \caption{continued.}
\end{figure*}

\begin{table*}[]
    \centering
    \small
    \setlength{\tabcolsep}{4pt}
    %\footnotesize
    \caption{Distance moduli, heliocentric distances, and metallicities of the 15 stellar systems.} %\NFM{comment: Add a S/N column.}
    
    \begin{tabular}{l|c|c|c|c|c|c|l|c}
    \hline
    & \multicolumn{3}{c|}{Visual isochrone fit} & \multicolumn{3}{c|}{TRGB} & &\\
    ID & $(m -M)_0$ & $D_\odot$ & \feh{} & $(m -M)_0$ & $D_\odot$ & \feh{} & Other ID & Ref.\\
    & (mag) & (kpc) & & (mag) & (kpc) & & & \\
    \hline
    \UMi{}/Euclid~UFSS~DR1~1 & 16.5 $\pm$ 0.3 & $20 \pm 3$ & $\simeq-2.2$ & & & &\\
     \\[-1em]
    \Fornax{}/Euclid~UFSS~DR1~2 & 20.8 $\pm$ 0.3 & $145_{-19}^{+21}$ & $\simeq-2.2$ & & & &\\
    \\[-1em]
    \Eridanus{}/Euclid~UFSS~DR1~3 & 21.1 $\pm$ 0.3 & $166_{-21}^{+25}$ & $\simeq-2.2$ & & & &\\
    \\[-1em]
    \EuclidII{} & 27.9 $\pm$ 0.5 & $3800_{-780}^{+980}$ & $\simeq-1.9$ & & & &\\
    \\[-1em]
    \EuclidI{} & 28.2 $\pm$ 0.5 & $4400_{-900}^{+1100}$ & $\simeq-1.6$ & & & &\\
    \\[-1em]
    \hline\hline
    \EuclidIV{} & 27.4 $\pm$ 0.5 & $3020_{-620}^{+780}$ & $\simeq-1.7$ & & & & KK2022 Dw0916+6944 & (1) \\
    \\[-1em]
    \EuclidXII{} & 27.7 $\pm$ 0.5 & $3470_{-710}^{+900}$ & $\simeq-1.7$ & $27.54_{-0.07}^{+0.06}$ & $3220_{-100}^{+90}$ & $\simeq-1.7$ & PYY2023 J123423+764330 & (2) \\
    \\[-1em]
    \EuclidVII{} & 28.0 $\pm$ 0.5 & $4000_{-800}^{+1000}$ & $\simeq-1.7$ & $27.56_{-0.29}^{+0.32}$ & $3250_{-410}^{+520}$ & $\simeq-1.5$ & PYY2023 J095637+722617 & (2) \\
    \\[-1em]
    \EuclidVI{} & 27.9 $\pm$ 0.5 & $3800_{-780}^{+980}$ & $\simeq-1.7$ &  & & & SDSS J095022.73+621537.8 & (3) \\
    \\[-1em]
    \EuclidVIII{} & 27.8 $\pm$ 0.5 & $3630_{-750}^{+940}$ & $\simeq-1.5$ & $27.96_{-0.28}^{+0.23}$ & $3910_{-470}^{+440}$ & $\simeq-1.5$ & 
    Dw1155+78 & (4) \\
    \\[-1em]
    \EuclidX{} & 28.0 $\pm$ 0.5 & $4000_{-800}^{+1000}$ & $\simeq-1.5$ & $28.04_{-0.03}^{+0.04}$ & $4055_{-56}^{+75}$ & $\simeq-1.6$ & PYY2023 J121704+750321 & (2) \\
    \\[-1em]
    \EuclidV{} & 28.1 $\pm$ 0.5 & $4200_{-900}^{+1100}$ & $\simeq-1.9$ & & & & SDSS J094132.57+660505.0 & (3) \\
    \\[-1em]
    \EuclidXI{} & 28.1 $\pm$ 0.5 & $4200_{-900}^{+1100}$ & $\simeq-1.3$ & $28.26_{-0.16}^{+0.16}$ & $4490_{-320}^{+340}$ & $\simeq-1.6$ & MAGE1223+6908 & (5) \\
    \\[-1em]
    \EuclidIX{} & 27.8 $\pm$ 0.5 & $3630_{-750}^{+940}$ & $\simeq-1.3$ & $28.38_{-0.03}^{+0.04}$ & $4742_{-65}^{+88}$ & $\simeq-1.4$ & LEDA 166124 & (6) \\
    \\[-1em]
    \EuclidIII{} & 28.7 $\pm$ 0.5 & $5500_{-1100}^{+1400}$ & $\simeq-1.9$ & & & & SMDG J0329164-494928 & (7) \\
    \\[-1em]
    \hline\hline
    \end{tabular}
\tablefoot{The upper section lists the five newly discovered stellar systems, and the lower section the ten galaxies previously catalogued as unresolved sources and resolved into stars here for the first time; for the latter, the last two columns give the designation in the literature and the corresponding reference. Columns 2--4 report the results of the visual isochrone fit. Columns 5--7 report, for the six systems with a sufficiently populated RGB, the results of the TRGB fit, which are the fiducial values adopted for those systems.}
\tablebib{(1)~\citet{Karachentsev&Kaisina2022}; (2)~\citet{Paudel-2023}; (3)~\citet{Adelman-McCarthy-2008}; (4)~\citet{Karachentsev-2023}; (5)~\citet{Hunter-2025}; (6)~\citet{Karachentseva&Karachentsev1998}; (7)~\citet{Zaritsky-2023}.}
    \label{tab:dwarf_distances}
\end{table*}

\massi{Figure~\ref{fig:CMD_all} illustrates the CMD analysis for all 15 stellar systems presented in this work. The left panel of each object} shows the CMD of all stars within $2\,r_{\rm h}$ (that is inferred in Sect.~\ref{sc:structural_parameters}) of the centre of the system, while the right panel shows the CMD of a local field, selected as an equal-area annulus centred on the system at a distance of $5\arcmin$, which characterizes the contaminating population.\par 
%The CMDs of the full sample are provided in Figure~\ref{fig:CMD_all}.\par
We overlaid old ($\simeq$13 Gyr) PARSEC isochrones on the CMD of each system and adjusted the distance modulus $(m-M)_0$ and the metallicity \feh{} until we obtained the best visual agreement with the observed stellar distribution. 
We started from \feh{}~$=-2.2$, typical of ancient ultra-faint stellar systems, and tested progressively more metal-rich values for the most luminous dwarfs in our sample, as expected from the stellar mass--metallicity relation of dwarf galaxies \citep[e.g.][]{Kirby-2013}. The best-fit isochrones are shown as solid black lines in Fig.~\ref{fig:CMD_all}.\par
\massi{The constraints provided by this procedure are necessarily loose, because the systems in our sample are either extremely faint or very distant. The former host very few, if any, post-main-sequence stars, while for the latter the photometry is not deep enough to reach the main-sequence turn-off (MSTO), so that only the upper red giant branch (RGB) and/or asymptotic giant branch (AGB) are resolved. In both cases, the features that usually anchor the distance and metallicity, such as a well-defined horizontal branch (HB), an extended main sequence, or a well-populated RGB, are missing or sparsely sampled.
We estimated the uncertainties on $(m-M)_0$ by shifting the isochrone in distance modulus until it no longer provided an acceptable match to the observed CMD features. Since this process relies on visual judgement, we test the robustness of our measurements below against distances measured from the tip of the red giant branch \citep[TRGB; see][and references therein]{lee93, bellazzini08, madore23}, in the cases in which this was feasible.}\par
\massi{We followed the Bayesian approach of \citet{Bellazzini&Pascale2024}, modelling the bright end of the luminosity function (LF) as the sum of an RGB component, an AGB component above the tip, and a contaminating field population. We adopted their calibration for the \Euclid \IE band, $M_{\rm TRGB}=-3.30\pm0.05$, which is anchored to the metallicity of the Small Magellanic Cloud (SMC).
Although the galaxies analyzed here may be more metal-poor than the SMC this is the most sensible choice, lacking a measure of the TRGB colour in a combination of passbands among those used by \citet{Bellazzini&Pascale2024}. On the other hand, the TRGB method provides an objective, quantitative distance estimate, in contrast to the visual isochrone fit, which is inherently subject to the observer's judgment. 
% TODO: PARSEC test, fill in before circulating
%Since the galaxies analysed here may be more metal-poor than the SMC, we estimated the dependence of $M_{\rm TRGB}$ on metallicity using PARSEC isochrones, finding a variation of [x]\,mag in \IE between [Fe/H]~$=$~[SMC] and $-2.2$, which we [added in quadrature to the TRGB uncertainties / consider negligible compared to the statistical uncertainties]. Despite this caveat, the TRGB method provides an objective, quantitative distance estimate, in contrast to the visual isochrone fit, which is inherently subject to the observer's judgement.
The fit converges reliably only for well-populated RGBs, so that we could measure the TRGB for six systems only. Figure~\ref{fig:trgb_figure} shows the fit for these six dwarf galaxies.}\par
Table~\ref{tab:dwarf_distances} lists, for each system, the distance modulus from the isochrone fit and, where available, from the TRGB, together with the corresponding heliocentric distances.
% CHECK: sigma = combined uncertainty of the two estimates?
\massi{For the six galaxies with both estimates, the two distance moduli agree within about $1\sigma$ in all cases, the largest difference being $\simeq1.2\sigma$ for \EuclidIX{}. This shows that, at least where the RGB is well populated, the visual fit is not significantly biased with respect to a quantitative measurement of the tip, and that its uncertainties are not underestimated (see Appendix~\ref{sc:trgb} for details). We therefore adopted the TRGB distance as fiducial for these six systems, and the isochrone-based distance for the remaining ones.}\par
\massi{The visual isochrone fitting provides a first distance estimate for the full sample, sufficient to unambiguously separate MW satellites from LV dwarfs, while the TRGB method, where applicable, supplies a more accurate and less subjective distance for a well-populated subset. A significant refinement of these parameters will ultimately require deeper photometric or spectroscopic follow-up observations.}

\subsection{\label{sc:structural_parameters} Structural parameters estimation}

We utilized the maximum likelihood estimation framework developed by \citet{Martin-2008,Martin-2016b} to derive the fundamental structural properties of the stellar systems reported in this work. Specifically, the central coordinates (RA, Dec), the projected half-light radius $r_{\rm h}$, ellipticity $\epsilon$, position angle $\theta$, and the number of stars $N_{*}$.
\massi{The ellipticity is defined as $\epsilon = 1 - b/a$, where $a$ and $b$ are the semi-major and semi-minor axes of the elliptical model, respectively.
The position angle of the major axis is measured east of north.}
This method models the projected spatial distribution of stars by combining an exponential radial profile for the stellar system with a constant density term representing the local background.
The posterior probability distributions for the six free parameters $\{{\rm RA}, {\rm Dec}, r_{\rm h}, \epsilon, \theta, N_{*}\}$ were sampled using a Markov chain Monte Carlo (MCMC) approach. 
%For this analysis, we restricted the input catalogue to stars consistent with the best-fit isochrone mask.
For this analysis, we restricted the input catalogue to the stars selected by an isochrone-filter mask (Sect.~\ref{sc:method}) built around the best-fit isochrone of each system.\par
To ensure that the structural modeling was based on reliable detections, we performed a stringent quality cut on the catalogue. Specifically, we retained only sources with \IE $\leq 25.7$. This threshold is one magnitude brighter than the $5\sigma$ point-source depth \citep{Mellier-2025}.
To assess the robustness of our results, we conducted several consistency checks by varying this magnitude threshold down to the full, unfiltered catalogue. We found that the free parameters of the fit remain consistent within $1\sigma$ across these different selection criteria, confirming that our structural estimates are not significantly biased by the inclusion or exclusion of the faintest sources near the detection limit. \EuclidIII{} represents the sole exception; here, a fainter threshold (\IE $\leq 26.2$) was required to retain enough stars for a robust MCMC structural fit.%\par\NFM{Is the data complete at this magnitude limit? Otherwise N* is for the incomplete data and will yield an underestimation of the true luminoisty of the system. I believe Samuel has determined the completeness curves for the data and it's definitely worth a check! If the data is not complete, then I would fold in Samuel's completeness curve into the analysis when doing the Monte Carlo to calculate the PDF on $M_V$.}\par
%\massi{MG: I think the major problem for these galaxies is crowding at their centre, causing a radial incompleteness. I added a comment below.}
We further note that for several galaxies at distances beyond 3~Mpc, the stellar catalogues may be subject to a radial incompleteness effect: the central regions of these systems, where crowding is most severe, are likely to have a lower detection efficiency compared to their outskirts. Correcting for this effect would require dedicated artificial star tests tailored to each individual system, which is beyond the scope of this work. We are nonetheless aware that this incompleteness may introduce a systematic bias in the derived structural parameters: the projected half-light radius $r_{\rm h}$ may be slightly overestimated, as the deficit of detected stars toward the centre effectively could flatten the central density, while the total luminosity and stellar mass might be underestimated, since a fraction of the stellar population in the most crowded regions remains undetected.\par
\begin{figure}[h!]
    \centering
    \includegraphics[width=1.\linewidth]{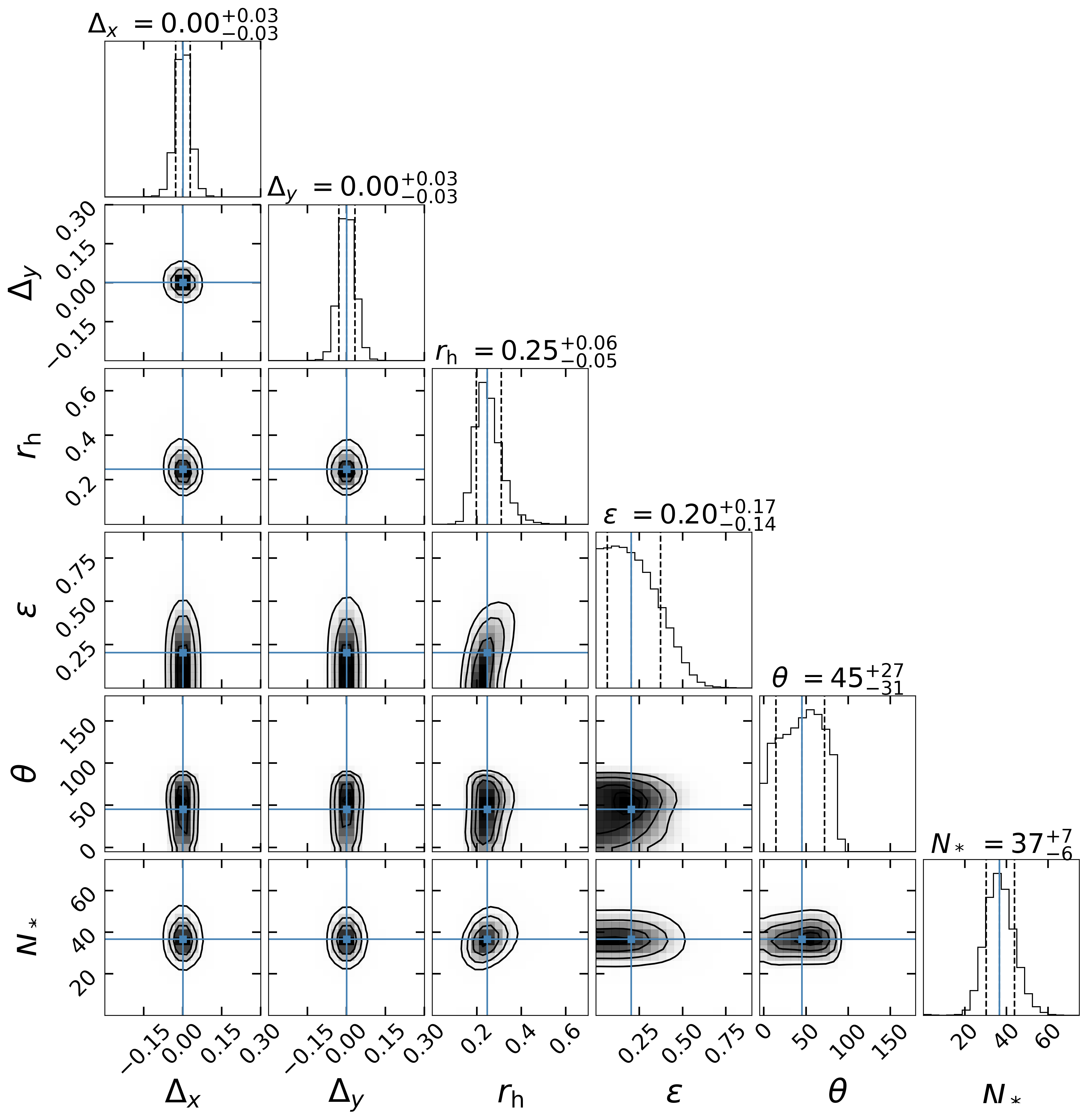}\\
    \includegraphics[width=1.\linewidth]{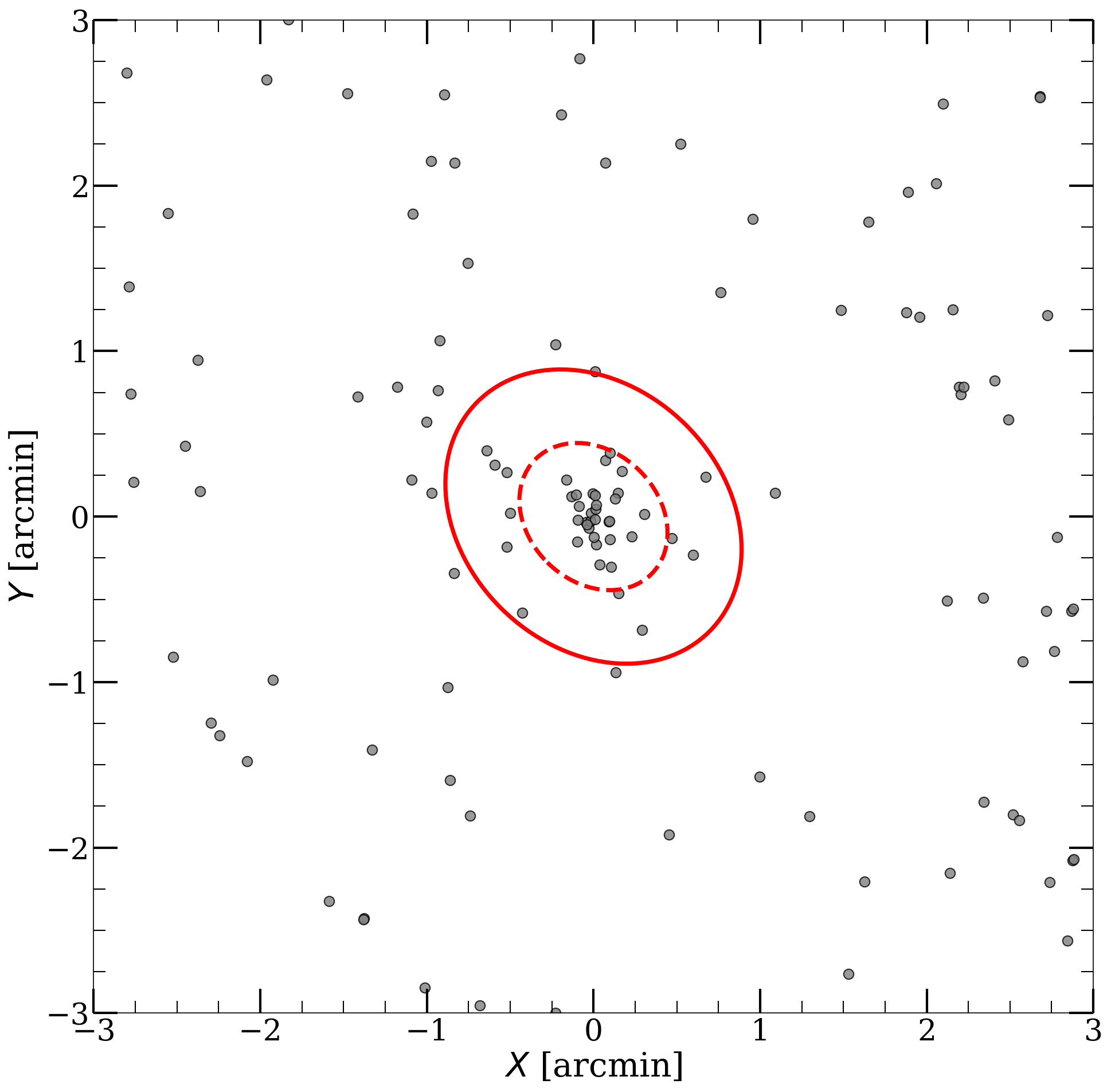}
    \caption{\emph{Top panel:} Posterior probability distributions for the structural parameters of \Eridanus{}, derived from the MCMC analysis. The blue lines indicate the median of the distributions.
    \emph{Bottom panel:} Spatial distribution relative to \Eridanus{} of all stars selected by the best-fit isochrone filter ($\logten{(t/{\rm yr})} = 10.12$, \feh{} $= -2.2$, $(m - M)_0$ = 21.1\,mag) in orthographic projection. The red dashed and solid ellipses represent the best-fit structural model, centred on the coordinates derived from the MCMC analysis, with a semi-major axis equal to 2\,$r_{\rm h}$ and 4\,$r_{\rm h}$, respectively and an orientation defined by the measured position angle $\theta$ and ellipticity $\epsilon$.} 
    %\NFM{comment: the labels of the corner plot need to include units when relevant.} \NFM{comment: invert the X axis so it goes from positive X on the left to negative X on the right.}
    \label{fig:example_mcmc}
\end{figure}

Our MCMC implementation, based on the \texttt{emcee} Python package \citep{emcee}, employed an ensemble of \num{1000} walkers. Each walker performed \num{10000} iterations, with the initial \num{1000} steps discarded as a burn-in phase to ensure the chains had reached convergence.
The final parameter values and their associated $1\sigma$ uncertainties were derived from the 50th, 16th, and 84th percentiles of the marginalized posterior distributions and are listed in Table~\ref{tab:dwarf_params}.
The top panel of Fig.~\ref{fig:example_mcmc} presents the corner plot for \Eridanus{}/Euclid~UFSS~DR1~3 as a representative case. In the bottom panel, we show the spatial distribution of the stars selected through the best-fit isochrone mask and the previously described quality criteria. Overlaid on this stellar distribution is the best-fit model ellipse, centred on the MCMC-derived coordinates and characterized by a semi-major axis corresponding to 2\,$r_{\rm h}$ (dashed ellipse) and to 4\,$r_{\rm h}$ (solid ellipse), respectively.\par
We note, however, that deriving robust structural parameters was not possible for \EuclidX{}, as it is located in close proximity to the edge of the survey footprint, which hampers a reliable application of our MCMC pipeline. %in our sample due to severe observational constraints. Specifically, \EuclidV{} and \EuclidX{} are located in close proximity to the edge of the survey footprint, which hampers a reliable application of our MCMC pipeline. Similarly, a large fraction of \EuclidXI{} is heavily masked by a saturated foreground star. 
Consequently, for this system, we report only its estimated distance modulus.

\begin{table*}[]
    \centering
    \small
    \setlength{\tabcolsep}{4pt}
    \caption{Structural parameters, absolute magnitudes, and stellar masses of the 15 stellar systems.} %\NFM{comment: Add a S/N column.}
    \begin{tabular}{l|c|c|c|c|c|c|c|c|c}
    \hline
    ID & RA & Dec & $r_{\rm h}$ & $r_{\rm h}$ & $\epsilon$ & $\theta$ & $M_V$ & $M_{\rm *}$ & Assoc.\\
    & (J2000) & (J2000) & (arcmin) & (pc) & & (deg) & (mag) & ($M_{\odot}$) & \\
    \hline\hline
    \UMi{} & 196.1542 & 75.5724 & 2.25$_{-0.39}^{+0.40}$ & $13 \pm 3$ & 0.20$_{-0.14}^{+0.17}$ & 162$_{-42}^{+48}$ & +2.0$_{-1.1}^{+0.5}$ & 42$_{-7}^{+9}$ & MW\\\\[-1em]
    \Fornax{} & 38.8244 & $-33.9567$ &  0.11$_{-0.03}^{+0.05}$ & $5_{-1}^{+2}$ & 0.36$_{-0.24}^{+0.26}$ & 106$_{-39}^{+30}$ & +0.51$_{-0.94}^{+0.79}$ & 117$_{-51}^{+71}$ & Fornax dSph\\
    \\[-1em]
    \Eridanus{} & 28.8623 & $-47.7105$ & 0.25$_{-0.05}^{+0.06}$ & $12 \pm 3$ & 0.20$_{-0.14}^{+0.17}$ & 45$_{-31}^{+27}$ & $-1.45_{-0.79}^{+0.65}$  & 550$_{-180}^{+260}$ & MW\\
    \\[-1em]
    \EuclidII{} & 150.0281 & 74.2053 & 0.36$_{-0.11}^{+0.18}$ & $400_{-150}^{+220}$ & 0.63$_{-0.17}^{+0.11}$ & 168$_{-12}^{+11}$ & $-8.06_{-0.83}^{+0.65}$ & $1.9_{-0.9}^{+2.2} \times 10^5$ & M81\\\\[-1em]
    \EuclidI{} & 42.8359 & $-54.3617$ &  0.33$_{-0.12}^{+0.19}$ & $420_{-180}^{+260}$ & 0.35$_{-0.24}^{+0.26}$ & 86$_{-46}^{+44}$ & $-7.9_{-1.2}^{+0.8}$ & $1.3_{-0.9}^{+3.6} \times 10^5$ & \\
    \\[-1em]
    %48.2648 & -53.0490 & \\
    \hline\hline
    \EuclidIV{}  & 139.1833 & 69.7354 &  0.23$\pm$0.04 & $202 \pm 58$ & 0.55$_{-0.11}^{+0.08}$ & 155$_{-11}^{+7}$ & $-8.07_{-0.62}^{+0.53}$ & $1.9_{-0.8}^{+1.5} \times 10^5$ & M81\\\\[-1em]
    \EuclidXII{} & 188.5979 & 76.7248 &  0.33$_{-0.03}^{+0.02}$ & $309_{-30}^{+21}$ & 0.49$_{-0.06}^{+0.04}$ & 146$\pm$4 & $-9.79_{-0.11}^{+0.12}$ & $9.4_{-1.0}^{+1.1} \times 10^5$ & M81\\\\[-1em]
    \EuclidVII{} & 149.1518 & 72.4392 & 0.58$_{-0.11}^{+0.12}$ & $550_{-130}^{+140}$ & 0.53$_{-0.12}^{+0.09}$ & 102$\pm$10 & $-8.58_{-0.55}^{+0.42}$ & $3.4_{-1.1}^{+2.3} \times 10^5$ & M81\\\\[-1em]
    \EuclidVI{} & 147.6009 & 62.2599 & 0.23$_{-0.06}^{+0.10}$ & $250_{-90}^{+130}$ & 0.27$_{-0.18}^{+0.23}$ & 189$_{-40}^{+32}$ & $-7.94_{-0.81}^{+0.66}$ & $1.7_{-0.8}^{+1.9} \times 10^5$ & M81\\\\[-1em]
    \EuclidVIII{} & 178.9786 & 78.0790 & 0.26$\pm$0.03 & $296_{-51}^{+46}$ & 0.42$_{-0.09}^{+0.08}$ & 127$\pm$8 & $-9.62_{-0.51}^{+0.52}$ & $8.7_{-3.3}^{+5.3} \times 10^5$ & M81\\\\[-1em]
    \EuclidX{} & 184.263 & 75.057 & & & & & & & M81\\
    \EuclidV{} & 145.3759 & 66.0859 &  0.31$_{-0.15}^{+0.25}$ & $380_{-200}^{+320}$ & 0.46$_{-0.30}^{+0.26}$ & 66$_{-38}^{+47}$ & $-7.31_{-0.98}^{+0.82}$ & $1.0_{-0.5}^{+1.5} \times 10^5$ & M81\\\\[-1em]
    \EuclidXI{} & 185.7589 & 69.1367 & 0.25$_{-0.05}^{+0.07}$ & $326_{-70}^{+94}$ & 0.37$_{-0.22}^{+0.17}$ & 146$_{-15}^{+23}$ &  $-8.82_{-0.41}^{+0.40}$ & $4.0_{-1.3}^{+1.9} \times 10^5$ & NGC~4236\\\\[-1em]
    \EuclidIX{} & 183.1684 & 68.9277 & 0.33$_{-0.02}^{+0.03}$ & $455_{-28}^{+42}$ & 0.40$_{-0.06}^{+0.05}$ & 171$_{-169}^{+7}$ & $-11.30 \pm 0.11$ & $(4.1 \pm 0.4) \times 10^6$ & NGC~4236\\\\[-1em]
    \EuclidIII{} & 52.3191 & $-49.8249$ & 0.23$_{-0.05}^{+0.07}$ & $370_{-140}^{+120}$ &  0.58$_{-0.17}^{+0.12}$ & 150$\pm$9  & $-8.61_{-0.97}^{+0.77}$ & $3.2_{-1.7}^{+4.7} \times 10^5$ & \\\\[-1em]
    \hline\hline
    \end{tabular}
    \tablefoot{The upper section lists the five newly discovered stellar systems, and the lower section the ten galaxies previously catalogued as unresolved sources and resolved into stars here for the first time. Physical half-light radii, $M_V$, and $M_*$ are computed with the fiducial distances of Table~\ref{tab:dwarf_distances}. No structural parameters are given for \EuclidX{}, which lies close to the edge of the survey footprint. The last column indicates the candidate host or group (see Sect.~\ref{sc:discussion}).}
    \label{tab:dwarf_params}
\end{table*}
 
\subsection{\label{sc:mass_mag} Inferring stellar mass and total magnitude}

We estimated the total luminosity and stellar mass using an approach based on Monte Carlo sampling of synthetic CMDs, following the methodology of \citet{Martin-2008,Martin-2016b}.
For each system, we generated a synthetic stellar population of $10^6$\,$M_\odot$ with the PARSEC library, adopting the age and metallicity of the respective best-fit isochrone and a \citet{Kroupa2002} initial mass function (IMF).
%down to [0.1]\,$M_\odot$. 
From this population, we performed \num{10000} Monte Carlo realizations as follows.
%For each system, we generated \num{10000} synthetic stellar populations using the PARSEC library, adopting the age and metallicity of the respective best-fit isochrones, and a \citet{Kroupa2002} initial mass function (IMF).
For each iteration, the distance modulus $(m - M)_0$ was drawn from a Gaussian distribution centred on the value derived in Sect.~\ref{sc:photo_analysis}, with a width corresponding to its estimated uncertainty.
%We then randomly sampled the luminosity function of the synthetic population until the number of member stars brighter than the \IE = 25.7 (consistent with threshold described in Sect.~\ref{sc:structural_parameters}) reached a target value. 
\massi{We then randomly drew stars from the synthetic population, including those fainter than the detection limit, until the number of stars brighter than $\IE = 25.7$\,mag (the threshold adopted in Sect.~\ref{sc:structural_parameters}) reached a target value. For each iteration, this target was drawn from a Poisson distribution with mean equal to the number of member stars, $N_*$, obtained from our MCMC analysis.}
%This target was independently drawn for each of the \num{10000} iterations from a Poissonian distribution, defined by the value $N_{*}$ as determined from our MCMC analysis.\par 
%We assumed a \citet{Kroupa2002} initial mass function to populate the CMDs. 
%For each realization, we summed the flux and mass of all generated stars to compute the total absolute magnitude in the \Euclid \IE band, and the integrated stellar mass ($M_{\rm *}$), respectively. 
\massi{For each realization, we summed the flux and mass of all generated stars, i.e. both those above and below the detection limit, to compute the total absolute magnitude in the \Euclid \IE band and the integrated stellar mass $M_{\rm *}$, respectively. 
Since only a small fraction of the stellar population lies above the detection limit ($\lesssim$3\% for the MW halo systems, and less for the more distant dwarfs), $M_{\rm *}$ relies on the extrapolation of the mass function (MF) to low masses. 
Adopting the different Chabrier formulations instead of the Kroupa IMF changes the masses by $\lesssim$3\%, while the total magnitudes are, instead, nearly insensitive to the shape of the MF ($\lesssim$0.1\,mag). 
Our estimates therefore assume that the MF below the detection limit follows a canonical IMF: MFs shallower than Kroupa, as reported for some UFDs \citep{Geha-2013, Gennaro-2018}, would result in lower stellar masses.}\par
%{\bf Since only a small fraction of the stellar population lies above the detection limit, the total stellar masses rely on the extrapolation of the MF to low masses. Adopting the different Chabrier formulations instead of the Kroupa IMF changes the masses by $\lesssim$3\%.}
To facilitate comparison with values reported in literature, we transformed these estimates into the Johnson $V$-band absolute magnitude ($M_V$), by adopting the relationship \massi{$M_V$ = $M_{\IE} + 0.3$ expected for an old, metal-poor stellar system \citep{Saifollahi-2025}}.  
The final values reported in Table~\ref{tab:dwarf_params} represent the 50th, the 16th, and the 84th percentiles of the \num{10000} simulated distributions.\par

\section{\label{sc:discussion}  Discussion}

%In the upper panel of Fig.~\ref{fig:MV_rh}, we examine the structural properties of the newly identified stellar systems located within the MW halo by plotting their absolute $V$-band magnitude ($M_{\rm V}$) against their half-light radius ($r_h$). 
%To provide a robust comparative framework, we include confirmed and candidate MW satellites from the  \citet{Pace2025} catalog, as well as Galactic globular clusters from \citet{Pace2025} and \citet{Harris1996} (2010 edition). 
%The lower panel of Fig.~\ref{fig:MV_rh} extends this analysis to the Local Volume, displaying the $M_V - r_{\rm h}$ plane for all galaxies in the \citet{Pace2025} catalogue located at distances between one and five Mpc.
%In the following, we provide a detailed case-by-case description of these systems, highlighting their distribution across the two panels of Fig.~\ref{fig:MV_rh}. 

\begin{figure}
   \centering
   \includegraphics[width=1\linewidth]{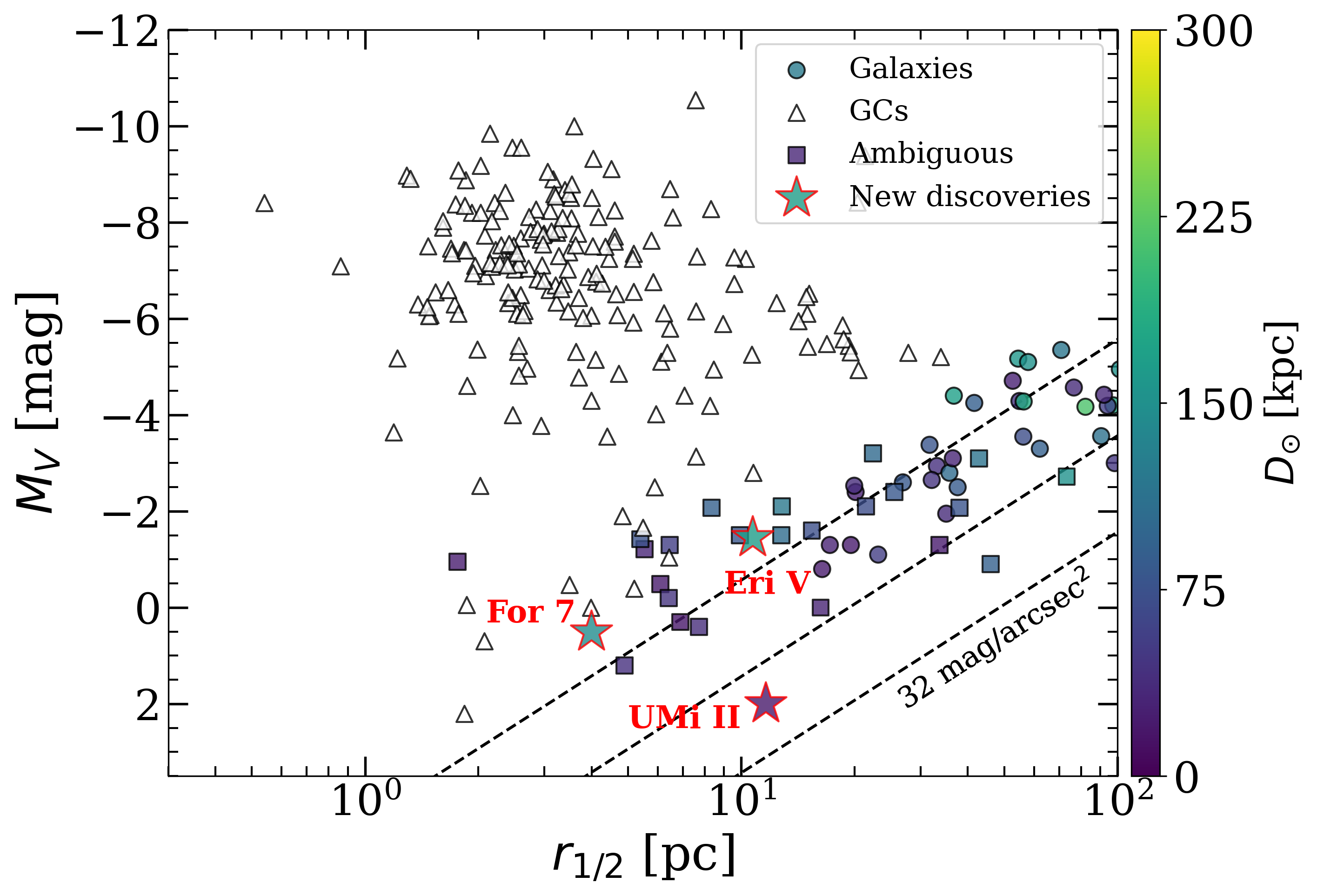}\\
   \includegraphics[width=1\linewidth]{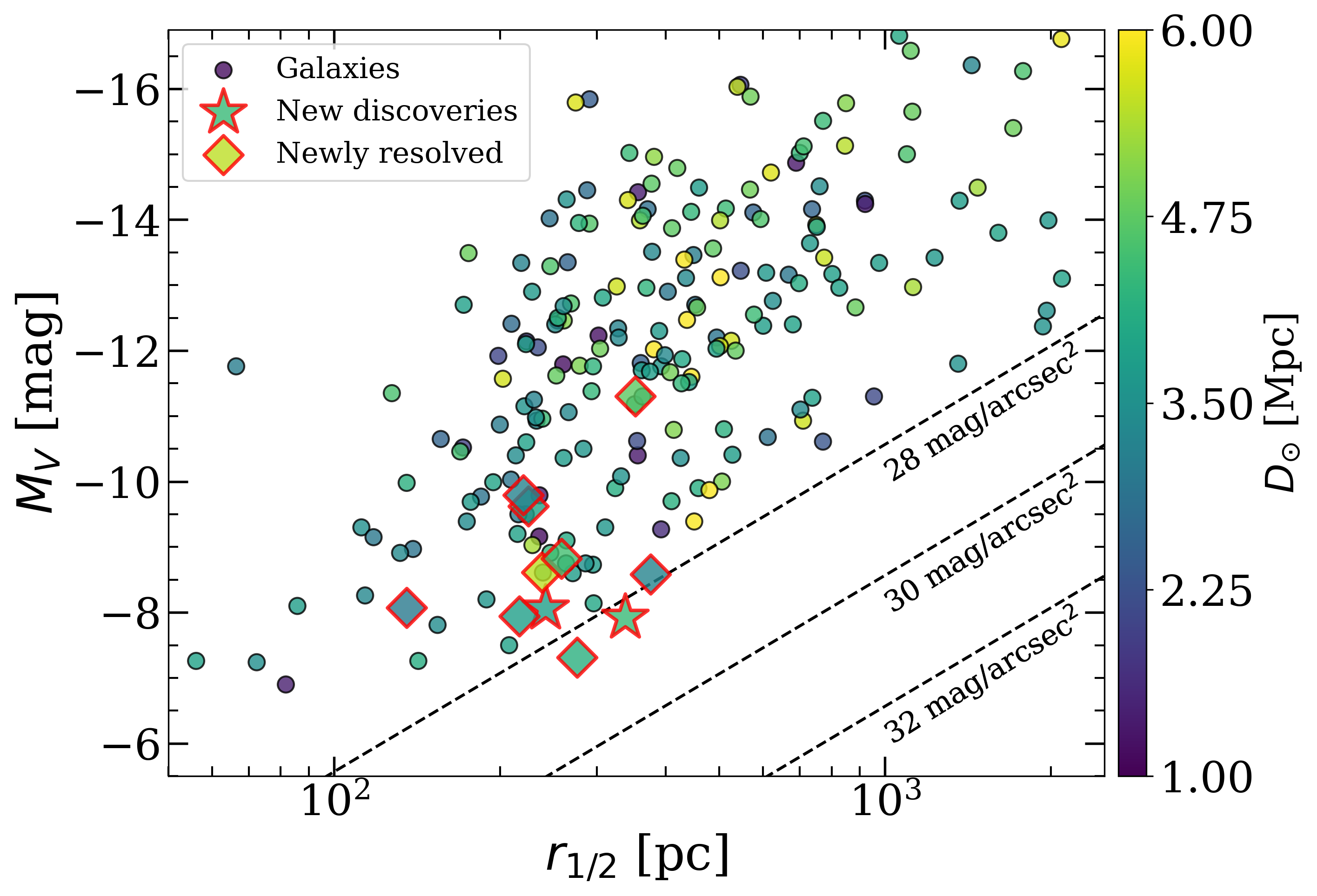}
    \caption{Absolute $V$-band magnitude versus circularised half-light radius, $r_{1/2}=r_{\rm h}\sqrt{1-\epsilon}$. \emph{Upper panel:} systems at $D_\odot\leq300$\,kpc. Circles and squares show galaxies and ambiguous systems from the \citet{Pace2025} catalogue, respectively, colour-coded by heliocentric distance; open triangles show Galactic globular clusters \citep{Harris1996, Pace2025}. Red-edged stars mark \Eridanus{}, \Fornax{}, and \UMi{}. \emph{Lower panel:} galaxies at $1\leq D_\odot/{\rm Mpc}\leq6$ from the \citet{Pace2025} catalogue (circles), colour-coded by heliocentric distance. Red-edged stars mark the two new discoveries, and red-edged diamonds nine of the ten previously catalogued galaxies resolved into stars in this work. In both panels, dashed lines indicate constant mean surface brightness within $r_{1/2}$, $\mu_V=28$, 30, and 32\,\magarc{}.}
    \label{fig:MV_rh}
\end{figure}

\massi{In Fig.~\ref{fig:MV_rh} we put the newly identified stellar systems in context by plotting them in the classical diagnostic plane opposing the absolute $V$-band magnitude $M_V$ to the azimuthally-averaged half-light radius $r_{1/2}$. 
The upper panel is zoomed-in on the faintest and most compact systems, in particular confirmed and candidate MW satellites from the \citet{Pace2025} catalog, as well as Galactic globular clusters listed by \citet[][2010 edition]{Harris1996}, also drawn from \citet{Pace2025}. This panel is best suited for the faint MW satellites.
On the other hand, the lower panel of Fig.~\ref{fig:MV_rh} is extended to include all the galaxies of the \citet{Pace2025} catalogue located between 1~Mpc and 6~Mpc from the Sun, opening the view to the range inhabited by the more distant dwarfs listed in Table~\ref{tab:dwarf_distances}. 
Both panels of Fig.~\ref{fig:MV_rh} show that these systems lie near the low surface brightness end of the distribution of known dwarfs, or even beyond it (\EuclidI{} and \EuclidV{}), demonstrating that searching stellar over-densities with \Euclid is an efficient means of unveiling galaxies in uncharted regions of the parameter space. In the following, we discuss the newly discovered and newly resolved systems on a case by case basis, keeping Fig.~\ref{fig:MV_rh} as our main reference diagram.}

\subsection{\label{sc:mw_satellites} Discovered stellar systems within the Milky Way halo}

\massi{We discovered three new stellar systems that, based on their distance, are most likely new satellites of the MW, albeit one of them may in fact be a satellite of a satellite of the MW \citep{Saifollahi-2026}. \UMi{}/Euclid~UFSS~DR1~1 is the faintest of the three, both in terms of absolute magnitude ($M_V$ = +2.0$_{-1.1}^{+0.5}$) and in surface brightness ($\mu_V \simeq 31~$\magarc{}) and, by far, least massive ($M_* \simeq 40\,M_{\odot}$) and the nearest, as it is located at a heliocentric distance of $\simeq 20$~kpc.}
% In the upper panel of Fig.~\ref{fig:MV_rh}, it lies among the lowest-surface-brightness stellar systems known in the MW halo, comparable only to [name of the ambiguous system at $r_{1/2}\simeq46$\,pc].
Its CMD (Fig.~\ref{fig:CMD_all}, first row and first column) displays a well-defined main sequence down to the photometric limit, about four magnitude below the MSTO. \massi{The extremely low total luminosity points toward a very low mean metallicity, compatible with the isochrone fit presented in Fig.~\ref{fig:CMD_all}.}
A comprehensive analysis of this stellar system, including follow-up spectroscopy acquired with the Gemini-North telescope, will be presented in a dedicated forthcoming paper ({\textcolor{blue}{Balbinot et al., in prep.}}).\par
\Fornax{}/Euclid~UFSS~DR1~2 was identified at the north-west periphery of the Fornax dwarf spheroidal (dSph) galaxy, at a projected angular distance of $\simeq$ 1\fdg1. The system's CMD (Fig.~\ref{fig:CMD_all}, first row, second column) shows a MSTO consistent with a heliocentric distance similar to that of the Fornax dSph ($\simeq 145$~kpc). 
At this distance, \Fornax{} would lie at a projected physical radius of $\simeq 2.8$~kpc from the dSph centre, placing it at the outer edge of its estimated tidal radius \citep[e.g.,][]{Cole-2012}. Consequently, it remains unclear whether \Fornax{} is a gravitationally bound satellite of the Fornax dSph.
Furthermore, the position of \Fornax{} in the $M_V$ vs. $r_{\rm h}$ plane ($r_{\rm 1/2} = 4_{-1}^{+2}$ pc, $M_V = +0.51_{-0.94}^{+0.79}$) does not provide a definitive classification, as it lies in the ambiguity region, where the properties of low-luminosity globular clusters (GCs) and the faint-end of the UFD sequence overlap.
This system has been independently detected within the Euclid Consortium by \citet{Saifollahi-2026}, where a more detailed investigation of its structural properties, its nature and environmental context is provided.
\massi{Although based on different photometric catalogues and analysis techniques, the distance modulus $20.8$, half-light radius ($5.0 \pm 1.0$)\,pc, absolute magnitude $M_V = +0.3 \pm 0.3$, and stellar mass $170_{-62}^{+50}\,M_\odot$ derived in that work are all consistent with ours within the uncertainties.}\par
The CMD of \Eridanus{} (Fig.~\ref{fig:CMD_all}, first row, third column) reveals an over-density near the MSTO reaching \HE $\simeq 25$ at ($\IE-\HE$) $\simeq -0.3$, a few candidate sub-giant branch (SGB) and red giant branch (RGB) stars having no counterpart in the flanking CMD of the nearby control field. The very low number of evolved stars implies a low total stellar mass. 
The sparsity of candidate member stars also introduces challenges in constraining the distance modulus, as the isochrone fitting relies primarily on the magnitude of the main sequence turn-off (MSTO), by definition a nearly vertical feature of the CMD. 
Based on the fit shown in Fig.~\ref{fig:CMD_all}, we estimate a heliocentric distance of approximately 165 kpc.
Our synthetic CMD Monte Carlo analysis confirms the low-mass nature of \Eridanus{} ($M_{\rm *} = 550_{-180}^{+260}\, M_\sun$).\par
In the $M_V$ vs. $r_{\rm h}$ plane (top panel of Fig.~\ref{fig:MV_rh}), \Eridanus{} occupies the transition region between GCs and UFDs $r_{\rm 1/2} = (11 \pm 3$)\,pc, $M_{V} = -1.45_{-0.79}^{+0.65}$. Specifically, its structural parameters are remarkably similar to those of ambiguous systems such as DELVE~6 \citep{Cerny-2023}, Kim~2 \citep{Kim-2015b}.
%, and Eridanus~III \citep{Conn-2018}.%, and also to Koposov~I \citep{Koposov-2007} catalogued as GC.
Given its position in this plane, the ambiguity between a distant GC and a faint UFD cannot be resolved with photometry alone. 
Consequently, follow-up spectroscopic observations are essential to ascertain its nature, by means of its dynamical or chemical properties \citep{ws12, cerny26, bellazz26}.\par
%{\bf Should \Fornax{} and \Eridanus{} be star clusters, both would be dynamically evolved, with evaporation times \citep[][]{Spitzer1987} comparable to (\Eridanus{}, $\sim$9\,Gyr) or much shorter than (\Fornax{}, $\sim$1.5\,Gyr) the age of their stellar populations. In this scenario, \Fornax{} would be the remnant of a substantially more massive progenitor, and the present-day MF of both systems would likely be depleted in low-mass stars. Our fiducial stellar masses should then be regarded as upper limits: removing the low-mass stars by truncating the \citet{Kroupa2002} MF at 0.2, 0.3, and 0.4\,M$_\odot$, as a proxy for evaporation-driven depletion, lowers them by $\sim$20, 40, and 55\%, respectively, while leaving $M_V$ unchanged within $\lesssim$0.1\,mag. As for their classification, only spectroscopic follow-up will be able to discriminate between these two scenarios.\par}

\subsection{Discovered dwarfs beyond the Local Group}

\massi{Our detection pipeline also identified two previously unknown, extremely sparse stellar systems located well beyond the Local Group, at distances of about 4 Mpc (i.e. \EuclidII{} and \EuclidI{}). 
While the Euclid composite images clearly confirm their nature as resolved systems, the CMD includes only a handful of bright RGB stars, insufficient for a proper detection of the RGB tip \citep{madore23}. 
This implies that the inferred distances are especially uncertain (although the comparison with TRGB distances for six other dwarfs, Appendix~\ref{sc:trgb}, indicates that the isochrone-based uncertainties are not underestimated). Their large $r_{\rm h}$ unequivocally classifies them as dwarf galaxies.}
\massi{Based on their total absolute magnitudes, these systems rank among the faintest dwarf galaxies ever discovered at such distances. In the lower panel of Fig.~\ref{fig:MV_rh} \EuclidI{} lies beyond the $\mu_V=28$\,\magarc{} line ($\mu_V\simeq28.3$\,\magarc{}), in a region of the parameter space populated by very few previously known galaxies at 1--6\,Mpc (see Sect.~\ref{sc:comparison_beyondLG}), while \EuclidII{} ($\mu_V\simeq27.4$\,\magarc{}) lies at the low-surface-brightness end of the known population, showcasing the unprecedented capabilities of \Euclid in detecting the most elusive systems in the Local Volume.}
\massi{Indeed, given the extant estimates, they both appear fainter than the classical local dwarf spheroidal galaxies Draco and Ursa Minor while having similar or larger half-light radii. 
Detecting and resolving into stars a Draco-like galaxy at $\simeq$4~Mpc in a blind search over $\simeq$1900\,deg$^2$ is a remarkable achievement.}

\subsection{Confirmed dwarfs beyond the Local Group}

In addition to the new discoveries, our detection pipeline successfully identified ten dwarf galaxies that had been catalogued in previous wide-field surveys, such as the SDSS \citep{Adelman-McCarthy-2008}, DESI legacy imaging surveys \citep{Karachentsev&Kaisina2022, Karachentsev-2023, Paudel-2023}, the Identifying Dwarfs of Magellanic Cloud Analog GalaxiEs \citep[ID-MAGE;][]{Hunter-2025}, and in the Systematically Measuring Ultra-diffuse Galaxies \citep[SMUDGes;][]{Zaritsky-2023}.
%Candidate, Extended Src for SLOAN, ultra-diffuse galaxy (host NGC 3065 at 100 Myr), (host NGC 4589 at 22.3 Mpc)
Due to the shallower photometry and lower spatial resolution of these ground-based datasets, these systems were previously classified only as unresolved objects, extended sources, ultra-diffuse galaxies or candidate galaxies. Consequently, their distances, physical sizes, stellar masses, and other fundamental structural parameters remained entirely unconstrained.
Thanks to the exceptional spatial resolution of \Euclid, we resolve their upper RGB for the first time, enabling us to place the first robust constraints on their primary structural properties.\par 
\massi{Overall, these newly-confirmed systems are located at distances between $3.0$ Mpc (\EuclidIV{}) and $5.5$ Mpc (\EuclidIII{}). Most of them are brighter than our new discoveries in the same distance range. This is not surprising, as they were already bright enough to be detected in shallower surveys. For six of them, the CMD displays an RGB with a population sufficiently rich to get a reliable measure of the distance from the tip of the RGB, broadly confirming the estimates obtained by isochrone fitting (see Sect.~\ref{sc:photo_analysis}).}
\massi{With the only exception of \EuclidIX{}, they are all fainter than $M_V=-10.0$ and have $r_{1/2}\gtrsim130$\,pc, unequivocally identifying them as dwarf galaxies. \EuclidV{} ($\mu_V\simeq28.5$\,\magarc{}) lies beyond the faintest surface brightness of the known galaxies at similar distances in the \citet{Pace2025} compilation, and \EuclidVII{} ($\mu_V\simeq27.9$\,\magarc{}) at its edge, making both cases particularly worthy of follow-up.}\par

\subsection{\label{sc:comparison_beyondLG} Comparison with known stellar systems within the Local Volume}

\begin{figure}
   \centering
   \includegraphics[width=1\linewidth]{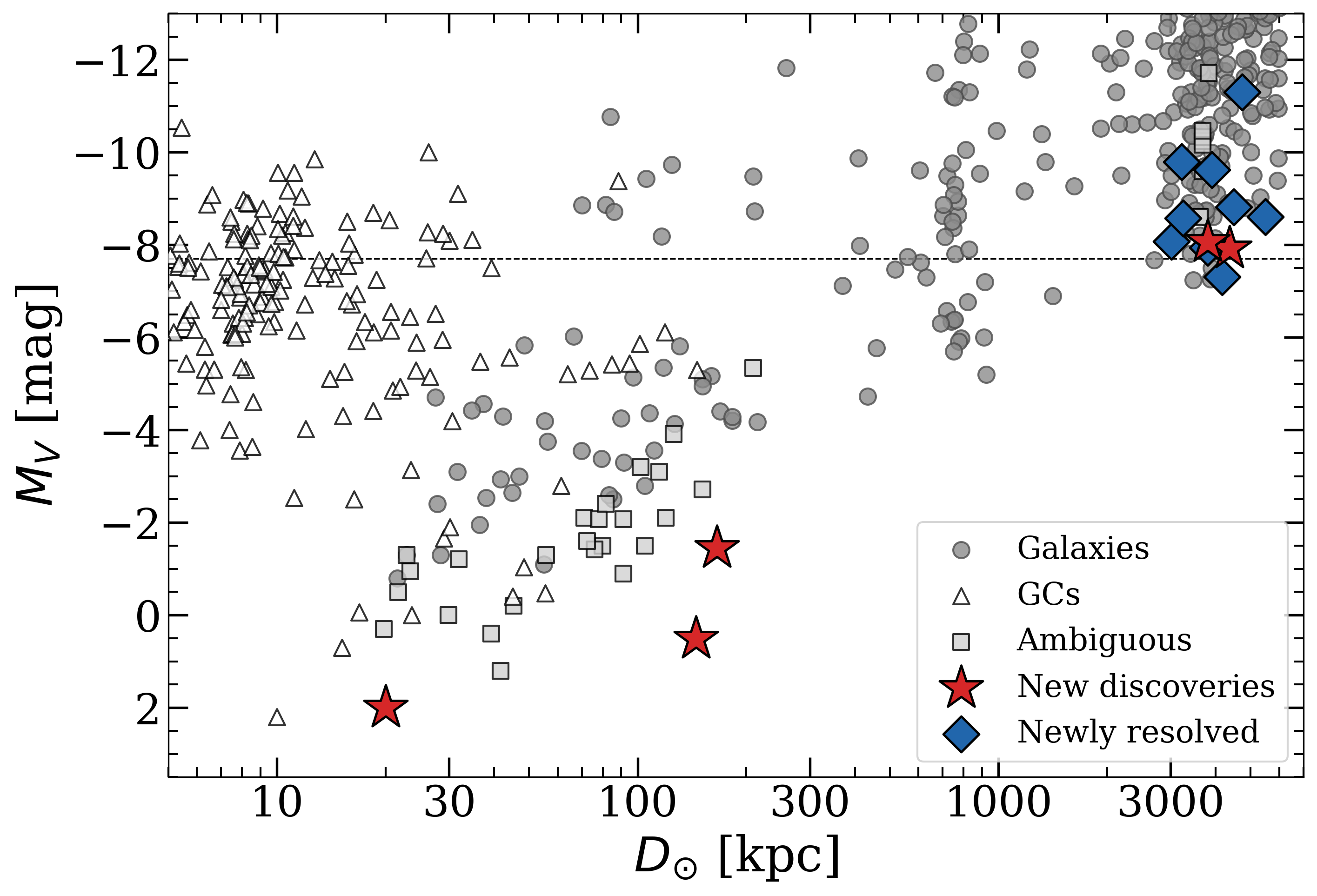}\\
   \caption{Absolute $V$-band magnitude as a function of heliocentric distance. Grey circles, open triangles, and light grey squares show known galaxies, globular clusters, and stellar systems with ambiguous classification \citep{Harris1996, Pace2025}, respectively. Red stars and blue diamonds mark the new discoveries and the previously catalogued galaxies analysed in this work, respectively. The dashed line marks $M_V=-7.7$, the upper luminosity limit adopted for UFDs \citep{Simon2019}.}
    \label{fig:mv_dist}
\end{figure}

\massi{Figure~\ref{fig:mv_dist} places the 14 systems with available $M_V$ in the context of known stellar systems out to $\simeq$6\,Mpc. The faint end of the known population becomes progressively brighter with increasing distance, reflecting the observational limits of previous searches. Our systems lie at or beyond this faint edge over the whole range of distances probed.}
\massi{In the MW halo, \Fornax{} and \Eridanus{}, with $M_V\simeq+0.5$ and $-1.5$, are the faintest stellar systems known beyond $\simeq$100\,kpc, and \UMi{}, at $\simeq$20\,kpc, is among the faintest systems known.}
\massi{Overall, the 11 stellar systems beyond the LG with estimated structural parameters span $-11.3\lesssim M_V\lesssim-7.3$, $135\lesssim r_{1/2}\lesssim380$\,pc, and $25.0\lesssim\mu_V\lesssim28.5$\,\magarc{}, with stellar masses between $1 \times 10^5 M_{\odot}$ and $4 \times 10^6 M_{\odot}$. They therefore populate the regime of the faintest classical dwarfs and of the transition to UFDs ($M_V\simeq-7.7$). 
Eight of them, including the two new discoveries, are fainter than the completeness limit of integrated-light surveys such as ELVES \citep[$M_V\simeq-9$;][]{Carlsten-2022, Carlsten-2026}.
The faint end of our sample (\EuclidV{}, $M_V\simeq-7.3$) is comparable to the faintest confirmed satellites detected by targeted resolved-star surveys around Centaurus~A \citep[$M_V\simeq-7.2$;][]{Crnojevic-2016}, and fainter than the faintest known satellites of M81, whose luminosities have been revised to $M_V\gtrsim-7.7$ by \citet{Gozman-2024}.
If its membership of the M81 group, discussed in Sect.~\ref{sc:environment}, is confirmed, \EuclidV{} would be among the faintest satellites of M81, although the large uncertainty on its luminosity ($M_V = -7.31_{-0.98}^{+0.82}$) does not allow a firmer statement.
Three of our systems reach or exceed $\mu_V=28$\,\magarc{}, a value that no galaxy at 1--6\,Mpc in the \citet{Pace2025} catalogue exceeds. 
Galaxies this diffuse are, however, not unprecedented at these distances: NGC253-SNFC-dw1, a satellite of NGC~253 at 3.6\,Mpc not included in that compilation, has $\mu_V\simeq29.4$\,\magarc{} \citep{Okamoto-2024}.
These results show that a blind search based on resolved stars, so far feasible beyond the LG only with targeted deep imaging of individual hosts, can now reach the transition between classical dwarfs and UFDs out to $\simeq$6\,Mpc over thousands of square degrees. With the full EWS covering about seven times the area of DR1, \Euclid will provide a homogeneous census of these systems across a wide range of environments in the LV.}

\subsection{\label{sc:environment} Environmental context}

Beyond characterizing the intrinsic properties of the newly discovered systems, it is crucial to investigate whether any of these dwarfs are dynamically associated with larger, previously known host galaxies. 
While the possible connection between \Fornax{} and the Fornax dSph has been outlined above and discussed in \citet{Saifollahi-2026}, this section focuses on the environmental context of our more distant LV candidates.\par
%\massi{\EuclidI{} and \EuclidIII{}, located in the southern part of the footprint, have no known massive galaxy in projection at a compatible distance, and are therefore candidate isolated dwarf galaxies.}
\massi{\EuclidI{} and \EuclidIII{}, located in the southern part of the footprint, are not associated with any known group. Each of them, however, lies close in projection to a dwarf galaxy at a similar distance: \EuclidI{} lies $\simeq50'$ from ESO~154-023 [$(m-M)_0 = 28.84 \pm 0.02$, i.e. $D_\odot \simeq 5.9$\,Mpc; \citealt{Tully-2009}], and \EuclidIII{} $\simeq52'$ from IC~1959 [$(m-M)_0 = 28.96 \pm 0.03$, i.e. $D_\odot \simeq 6.2$\,Mpc; \citealt{Tully-2009}].
Both galaxies have stellar masses between those of the SMC and the Large Magellanic Cloud [LMC; $\logten{(M_*/M_\odot)} \simeq 8.9$ and 8.7, respectively; \citealt{Pace2025}], and these angular separations correspond to $\simeq85$ kpc and $\simeq93$\,kpc at their distances. 
The distance moduli of \EuclidI{} and \EuclidIII{} differ from those of ESO~154-023 and IC~1959 by $\simeq1.3\sigma$ and $\simeq0.5\sigma$, respectively. 
However, our distance uncertainties ($\sim1$\,Mpc) are much larger than their virial radii, so that our data can neither confirm nor exclude that \EuclidI{} and \EuclidIII{} are satellites of these galaxies rather than isolated dwarfs.
An independent search for satellites around isolated dwarf galaxies in \Euclid DR1 (\textcolor{blue}{Annibali et al., in prep.}) has also identified \EuclidI{} as a candidate satellite of ESO~154-023, based on the similarity between the CMDs of the two galaxies. 
TRGB distances or radial velocities are required to settle this question.}\par
Several of the newly identified dwarf galaxies lie in the Northern hemisphere, in close spatial proximity and at distance moduli similar to that of the M81 group. The central and most massive galaxy of the group, M81, is located at a distance of ($3.6 \pm 0.1$) Mpc \citep{Tully-2013}, anchoring a filamentary, complex system that contains more than 50 known members, including the LMC-analogue NGC~4236, located at the periphery of the group \citep{Karachentsev-2002}.\par
\begin{figure}
   \centering
   \includegraphics[width=1\linewidth]{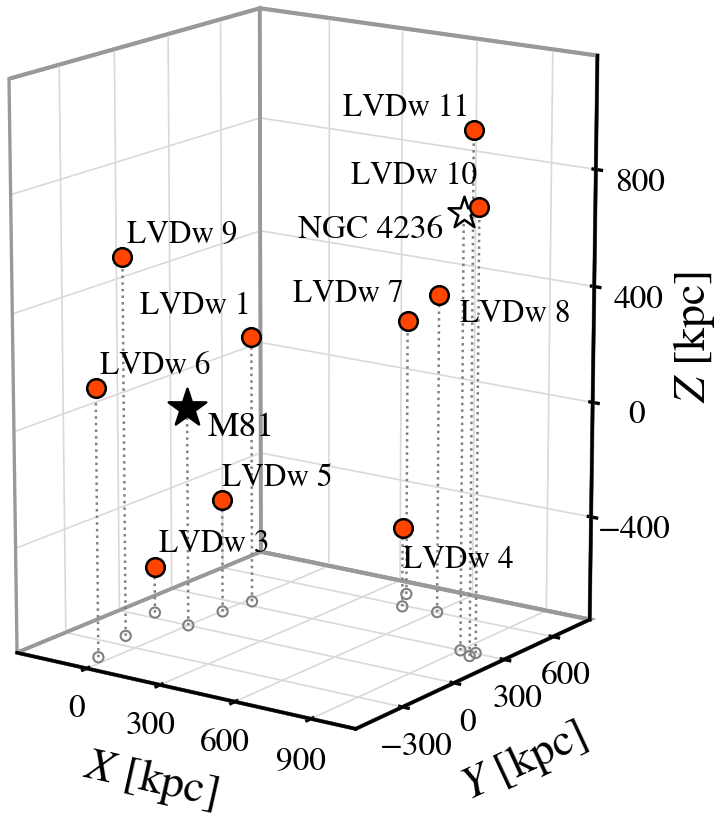}
    \caption{Three-dimensional distribution of the ten systems located in the
direction of the M81 group, in a cartesian frame centred on M81 (filled
black star). The $X$ and $Y$ axes point towards east and north on the sky,
respectively, and the $Z$ axis lies along the line of sight to M81,
increasing with heliocentric distance. Red circles mark the dwarf galaxies and the open star marks NGC~4236. Dotted lines connect each object to its projection (small open circles) on the bottom face of the box.}
    \label{fig:m81_3d}
\end{figure}

 
%  mu = 27.7 ± 0.1 (Kudritzki et al. 2012) (3,467 Mpc - 3,631 Mpc - 3,311 Mpc)
\massi{Figure~\ref{fig:m81_3d} shows the three-dimensional distribution of the
ten systems located in the direction of the M81 group, in a Cartesian
frame centred on M81, together with the position of NGC~4236.}
\massi{First, we note that the previously-reported \EuclidIX{} and \EuclidXI{} are close to NGC~4236, approximately only 0\fdg6 apart from it in projection. At the distance of NGC~4236, $\simeq 4.45$ Mpc \citep{Karachentsev-2002}, this corresponds to $\sim 50$~kpc. These two dwarfs are among those for which we were able to get reliable distances from the RGB tip. According to these measures, \EuclidXI{} lies $\simeq40$\,kpc from NGC~4236 along the line of sight, and is therefore a candidate satellite of this galaxy. \EuclidIX{} lies $\simeq290$\,kpc behind NGC~4236, and therefore its association with NGC~4236 is less secure (see Fig.~\ref{fig:m81_3d}). Radial velocities are required to confirm either association.}\par
The remaining galaxies do not reside in the immediate vicinity of any major, bright member of the M81 complex. Consequently, we assess the probability of their group membership by estimating their three-dimensional physical distances from the centre of the group.
For context, the radius of the zero-velocity surface of the M81 group, marking the boundary where cosmic expansion balances gravitational binding, is estimated to be $R_0 = (0.89 \pm 0.05$) Mpc \citep{Karachentsev&Kashibadze2006}.
To properly account for observational uncertainties, the three-dimensional (3D) distances between our candidates and M81 were computed using a Monte Carlo approach. For each system, we performed \num{10000} stochastic realizations. In each iteration, the heliocentric distance of the candidate satellite was randomly sampled from a Gaussian distribution centred on its estimated distance modulus, with a standard deviation equal to the uncertainty on its fiducial distance modulus (TRGB where available, Table~\ref{tab:dwarf_distances}). The distance of M81 and the value of $R_0$ were also sampled from Gaussian distributions with the uncertainties quoted above.
Assuming M81 marks the gravitational centre of the group, the resulting median values, along with the 16th and 84th percentiles of these Monte Carlo distributions, are reported in Table~\ref{tab:m81_dist}. 
\massi{For each galaxy we also report in Table~\ref{tab:m81_dist} the fraction of the 3D-distance posterior that lies within $R_0$, $P(<R_0)$. This quantity is a geometric measure of how likely a galaxy is to be located inside the zero-velocity surface of the M81 group; it is not a probability of being gravitationally bound, which would require velocity information.} %Because distance uncertainties can only scatter the 3D separation to values larger than the projected one, $P(<R_0)$ is a conservative estimate for genuine members.} %We find $P(<R_0) > [X]$ for [N] galaxies.}
%Based on the values in Table~\ref{tab:m81_dist}, \EuclidIV{}, \EuclidV{}, \EuclidVI{}, \EuclidVII{}, and \EuclidII{} emerge as strong candidates for being physically bound to the M81 group, whereas the membership of \EuclidVIII{}, \EuclidX{}, and \EuclidXII{} is more uncertain. 
%\massi{More accurate distance estimates and, above all, systemic velocity measures are required to settle the issue.}
\massi{Based on the values in Table~\ref{tab:m81_dist}, \EuclidVII{} is the strongest candidate member of the M81 group, with $P(<R_0) = 0.87$. \EuclidII{}, \EuclidIV{}, \EuclidVI{}, \EuclidVIII{}, and \EuclidV{} have $0.56 \leq P(<R_0) \leq 0.66$: they are more likely than not to lie within the zero-velocity surface, but the constraint is limited by their large line-of-sight distance uncertainties.} %and these values do not discriminate among them. 
\massi{\EuclidXII{} and \EuclidX{}, with $P(<R_0) = 0.44$ and $0.27$, have well-constrained 3D distances of $\simeq 0.9$\,Mpc, comparable to $R_0$ itself; they lie at the edge of the group, and their membership cannot be decided on geometric grounds alone. More accurate distances and, above all, systemic velocity measurements are required to establish whether any of these galaxies are gravitationally bound to the group.}
\begin{table}[]
\caption{Three-dimensional distances from M81.}
    \centering
    %\hspace{-1cm}
    %\footnotesize
    \begin{tabular}{l|c|c}
    \hline
    ID & $R$ & $P(<R_0)$\\
    & (kpc) & \\
    \hline\hline
    \EuclidII{} & $669_{-304}^{+638}$ & 0.66\\
    \\[-1em]
    \EuclidIV{} & $731_{-416}^{+551}$ & 0.61\\
    \\[-1em]
    \EuclidXII{} & $896_{-28}^{+37}$ & 0.44\\
    \\[-1em]
    \EuclidVII{} & $496_{-242}^{+366}$ & 0.87\\
    \\[-1em]
    \EuclidVI{} & $719_{-258}^{+634}$ & 0.63\\
    \\[-1em]
    \EuclidVIII{} & $868_{-100}^{+317}$ & 0.56\\
    \\[-1em]
    \EuclidX{} & $923_{-39}^{+45}$ & 0.27\\
    \\[-1em]
    \EuclidV{} & $735_{-447}^{+906}$ & 0.58\\
    \\[-1em]
    \hline\hline
    \end{tabular}
    \label{tab:m81_dist}
    \tablefoot{$P(<R_0)$ is the fraction of the posterior distribution of the 3D distance from M81 that lies within $R_0 = (0.89 \pm 0.05$)\,Mpc.}
\end{table}

\section{\label{sc:summary} Summary}

We conducted a systematic search for resolved stellar systems in the EWS DR1, covering approximately 1900 deg$^2$, using an isochrone filter technique applied over a distance modulus range of $15 \leq (m - M)_{0} \leq 28$ (10 kpc to $\simeq$ 4 Mpc). Systems slightly beyond this limit are still recovered, as discussed in Sect.~\ref{sc:method}. 
Our main results are as follows:
\begin{itemize}
    \item We detected five new stellar systems and recovered ten previously catalogued but unresolved galaxies, resolving their stellar populations for the first time with \Euclid. All 15 systems are presented as secure detections, confirmed by visual inspection of the deep \Euclid VIS stacked images. 
    For all 15 systems, we provide newly derived photometric distance estimates. Additionally, for 14 of them, we derive for the first time structural parameters (half-light radius, ellipticity, and position angle), total absolute magnitudes, and stellar masses from resolved star counts.
    \item Within the MW halo, we identify \Eridanus{}, an ultra-faint stellar system at $\simeq$ 165 kpc with $r_{\rm h}$ = 12 pc and $M_{\rm *} \simeq$ 550 $M_{\odot}$, whose structural properties place it in the ambiguous region between GCs and UFDs. Spectroscopic follow-up is required to determine its true nature.
    \item Also within the Galactic halo, we report the discovery of \UMi{} ($\simeq 20$~kpc), an extremely low-luminosity system with $M_V \simeq +2.0$ and $M_{\rm *} \simeq 40\,M_{\odot}$, and \Fornax{} ($\simeq 145$~kpc), located at the outskirts of the Fornax dSph.
    \item Beyond the Local Group, we report a total of 12 dwarf galaxies at 3–5.5 Mpc. Three of them reach or exceed a mean surface brightness within the half-light radius of $\mu_V\simeq28$\,\magarc{}, which almost no previously known galaxy at 1--6\,Mpc exceed.
    \item \massi{Among these distant targets, ten are candidate members of the M81 group. Eight have 3D distances from M81 smaller than or comparable to the radius of its zero-velocity surface, $R_0 = 0.89 \pm 0.05$\,Mpc. Of these, \EuclidVII{} is the strongest candidate, with a probability of 0.87 of lying within $R_0$; \EuclidII{}, \EuclidIV{}, \EuclidVI{}, \EuclidVIII{}, and \EuclidV{} are more likely than not to lie within it ($0.56 \leq P(<R_0) \leq 0.66$), although their distances are too uncertain to discriminate among them; \EuclidXII{} and \EuclidX{} sit at the edge of the group. The remaining two, \EuclidIX{} and \EuclidXI{}, lie close in projection to NGC~4236, a peripheral member of the group, with \EuclidXI{} being a candidate satellite of that galaxy.} 
    \item \massi{The remaining two, \EuclidI{} and \EuclidIII{}, are not associated with any known group, but lie within $\simeq90$\,kpc in projection of the dwarfs ESO~154-023 and IC~1959, respectively. Given the current distance uncertainties, they may be either satellites of these galaxies or isolated dwarfs.}
    
\end{itemize}
These findings highlight \Euclid's unique capability to bridge the gap between the ultra-faint satellite population of the MW and dwarf galaxies in the broader Local Volume within a single survey, using a uniform data set and a single detection methodology.
\massi{Applying the same technique to the full EWS, we can estimate the number of systems expected by the end of the six-year mission. A simple scaling by the area of the full EWS ($\simeq14\,000$\,deg$^2$) is appropriate only for the systems that are not associated with any known group. It gives $\simeq15$ new systems in the MW halo (excluding \Fornax{}, whose detection is tied to the Fornax dSph) and $\simeq15$ Local Volume dwarfs, each uncertain by about a factor of two given the small numbers involved. Adding the ten systems associated with the M81 group, which belong to a single structure and cannot be scaled by area, we expect at least $\simeq40$ systems by the end of the mission, plus further dwarfs in the vicinity of other nearby groups within the footprint (e.g. the Sculptor group). As in DR1, where 10 of the 12 systems beyond the Local Group were already catalogued as unresolved sources in ground-based surveys, we expect a large fraction of these distant systems to be previously known dwarf-galaxy candidates that \Euclid will resolve into stars for the first time, providing distances, structural parameters, and possibly association with nearby hosts.}

\begin{acknowledgements}
%\AckERO  
\AckEC  
\AckDRone
\\
M.G. acknowledges "Partecipazione LSST - Large Synoptic Survey Telescope" (Ob. Fu.: 1.05.03.06).
ES acknowledges funding through VIDI grant "Pushing Galactic Archaeology to its limits" (with project number VI.Vidi.193.093) which is funded by the Dutch Research Council (NWO). ES acknowledges funding from the European Research Council (ERC, EARLYMW, Grant number 101170507). Views and opinions expressed are those of the author(s) only and do not necessarily reflect those of the European Union or the European Research Council. Neither the European Union nor the granting authority can be held responsible for them.
%\cite{DR1cite}
\end{acknowledgements}

%
% Here comes the reference list, generated via bibtex from
% your bibfile my.bib and Euclid.bib. Please make sure that
% the same paper is not referenced twice, one from your my.bib
% file, and once from Euclid.bib.
%

\bibliography{Euclid, DR1, mybib} 
%
% You can have different .bib files here, e.g.
% \bibliography{Euclid,myown}, but please use the bibentries
% in Euclid.bib for all EC publications to ensure that these
% references are up-to-date and in correct format.
%

%
% Now you can add appendices.
% In this example, the appendices are in one column mode.
% If that is not requires, comment out \onecolumn
% Note that appendices in A\&A come {\it after\/} the references.

\begin{appendix}
  %\onecolumn %If you don't want single column for the Appendix, please
             %comment this out
  
%\section{\label{sc:supplementary_plots} Supplementary diagnostic plots}

%In this section we display the diagnostics plots for all 15 stellar systems reported in this work.

%\section{\label{sc:candidate} A high-confident candidate}
%Here we discuss the possibility that an other overdensity detected by the algorithm might be an actual stellar system.
%Alike the other 15 stellar systems presented in the work, its larger projected half-light radius makes it much more diffuse on the sky, hampering to reveal in the composite \Euclid images a clear overdensity of stars with respect to the average background, despite some stars are visible in the centre of the image (see Figure~\ref{fig:cand_image})

%\begin{figure}
%    \centering
%    \includegraphics[width=\linewidth]{euclid_coloured_images/Cand.jpg}
%    \caption{}
%    \label{fig:cand_image}
%\end{figure}

%\begin{figure}
%    \centering
%    \includegraphics[width=\linewidth]{CMD/Candidate_65.0983_-32.8324_1.73_21.40_CMD_paper.png}
%    \caption{}
%    \label{fig:cand_CMD}
%\end{figure}

\section{\label{sc:trgb} Tip red giant branch approach}
\begin{figure*}
    \centering
    \includegraphics[width=.4\linewidth]{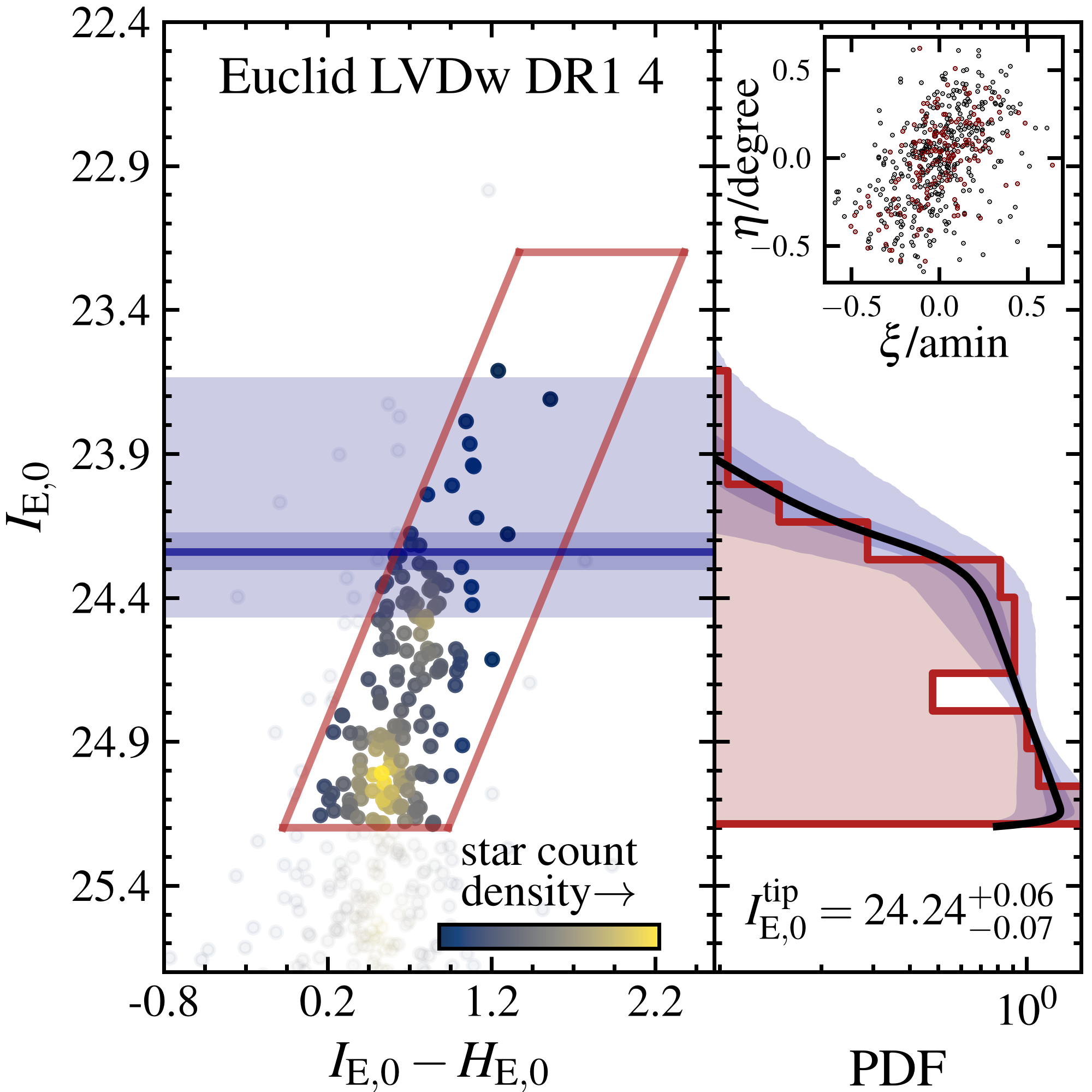}
    \includegraphics[width=.4\linewidth]{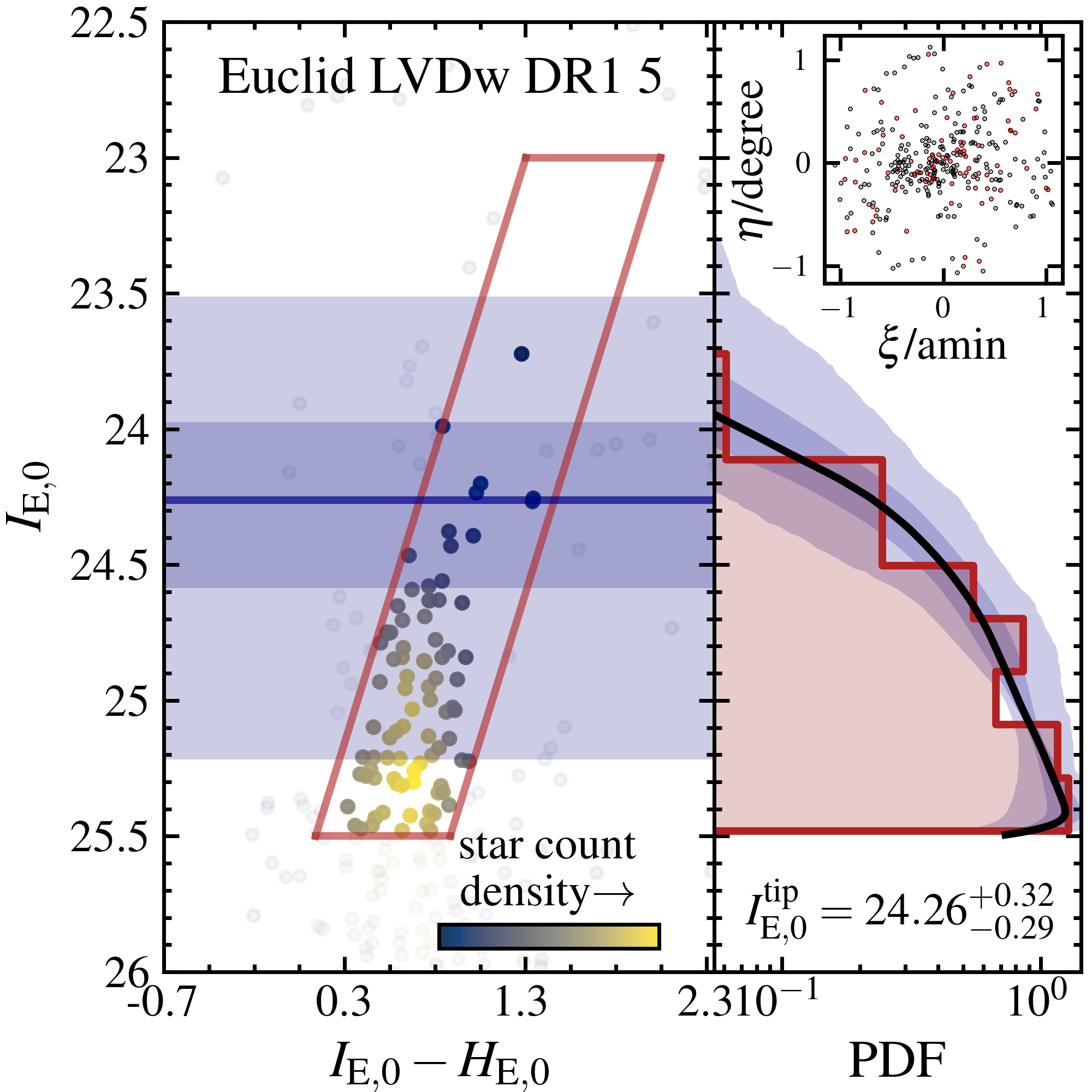}\\
    \includegraphics[width=.4\linewidth]{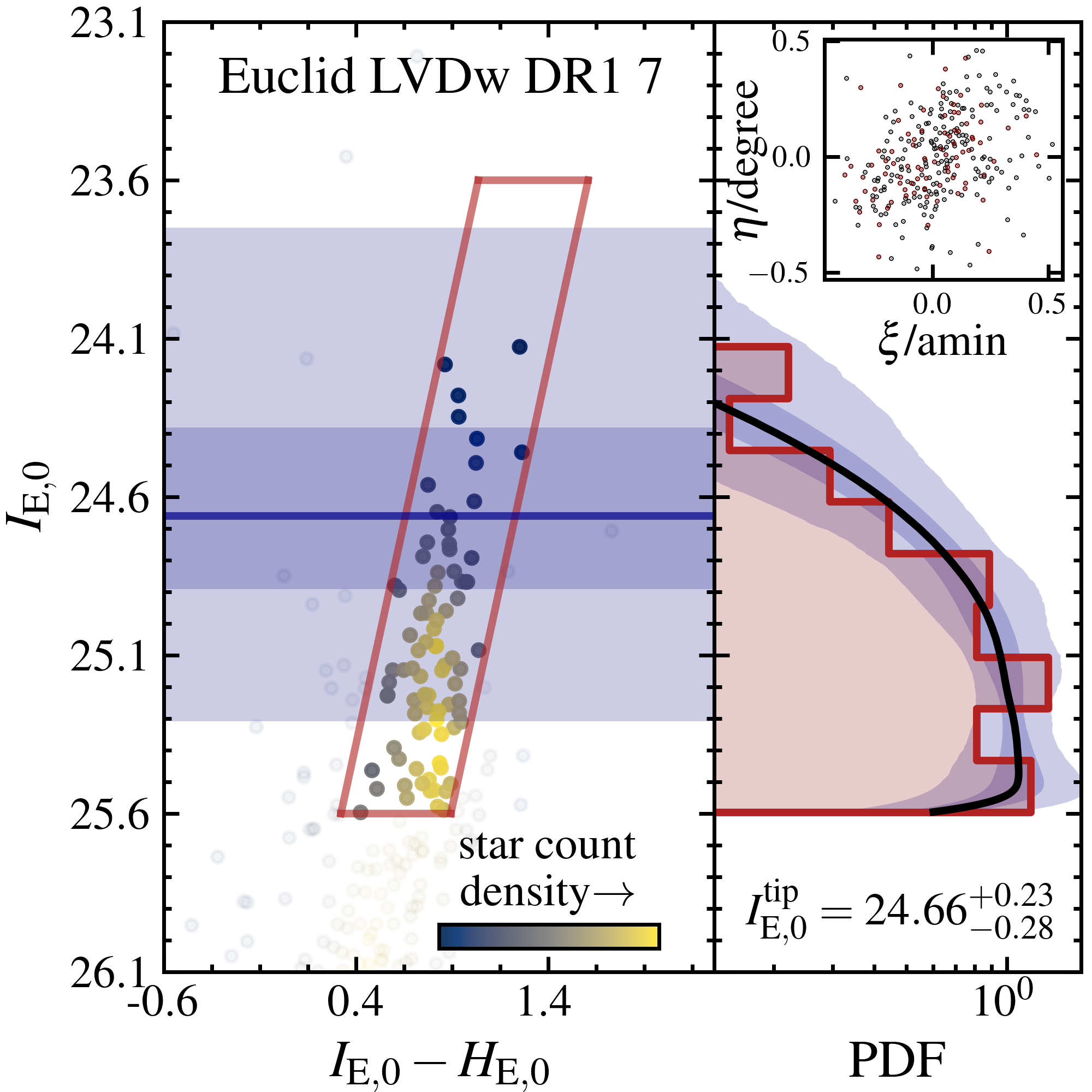}
    \includegraphics[width=.4\linewidth]{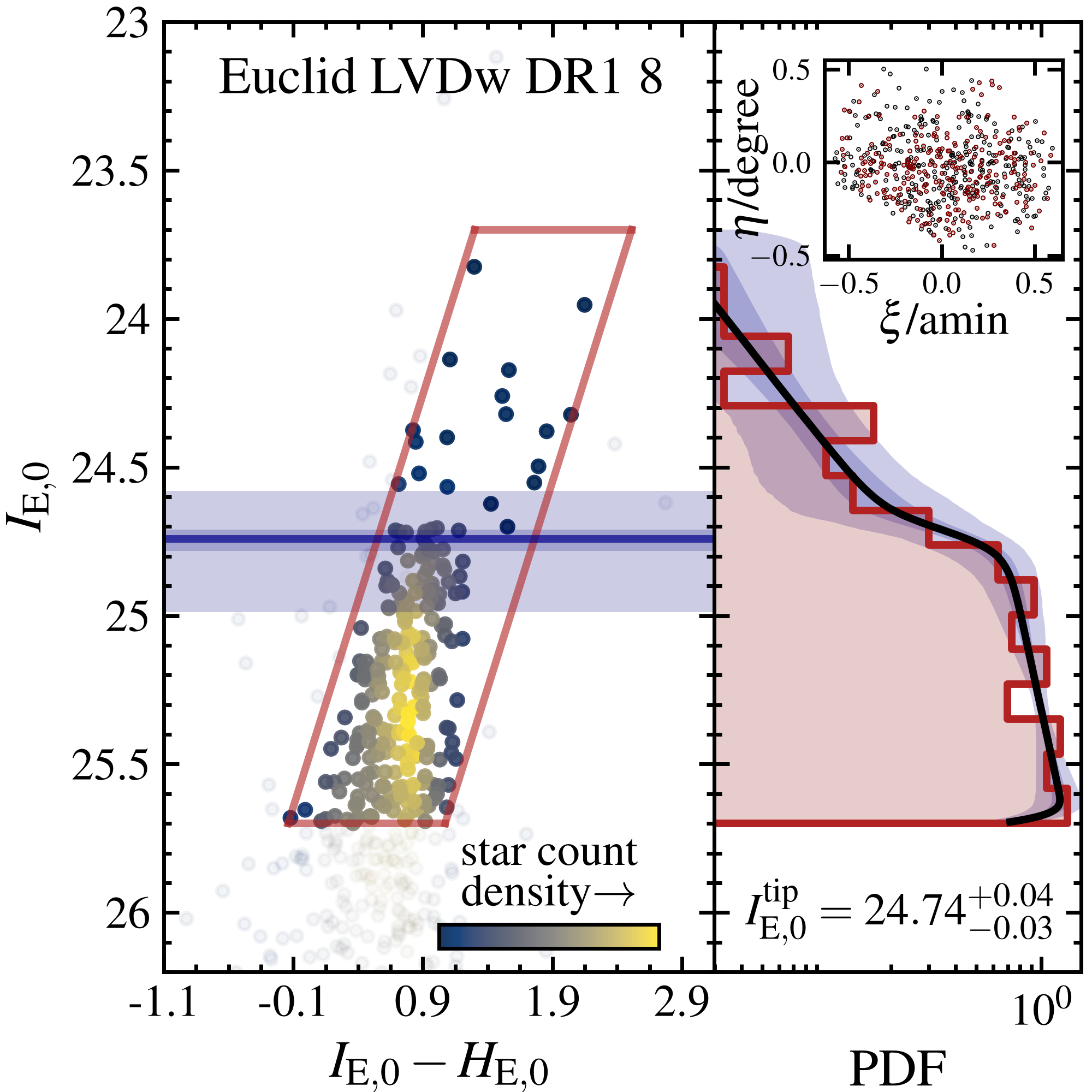}\\
    \includegraphics[width=.4\linewidth]{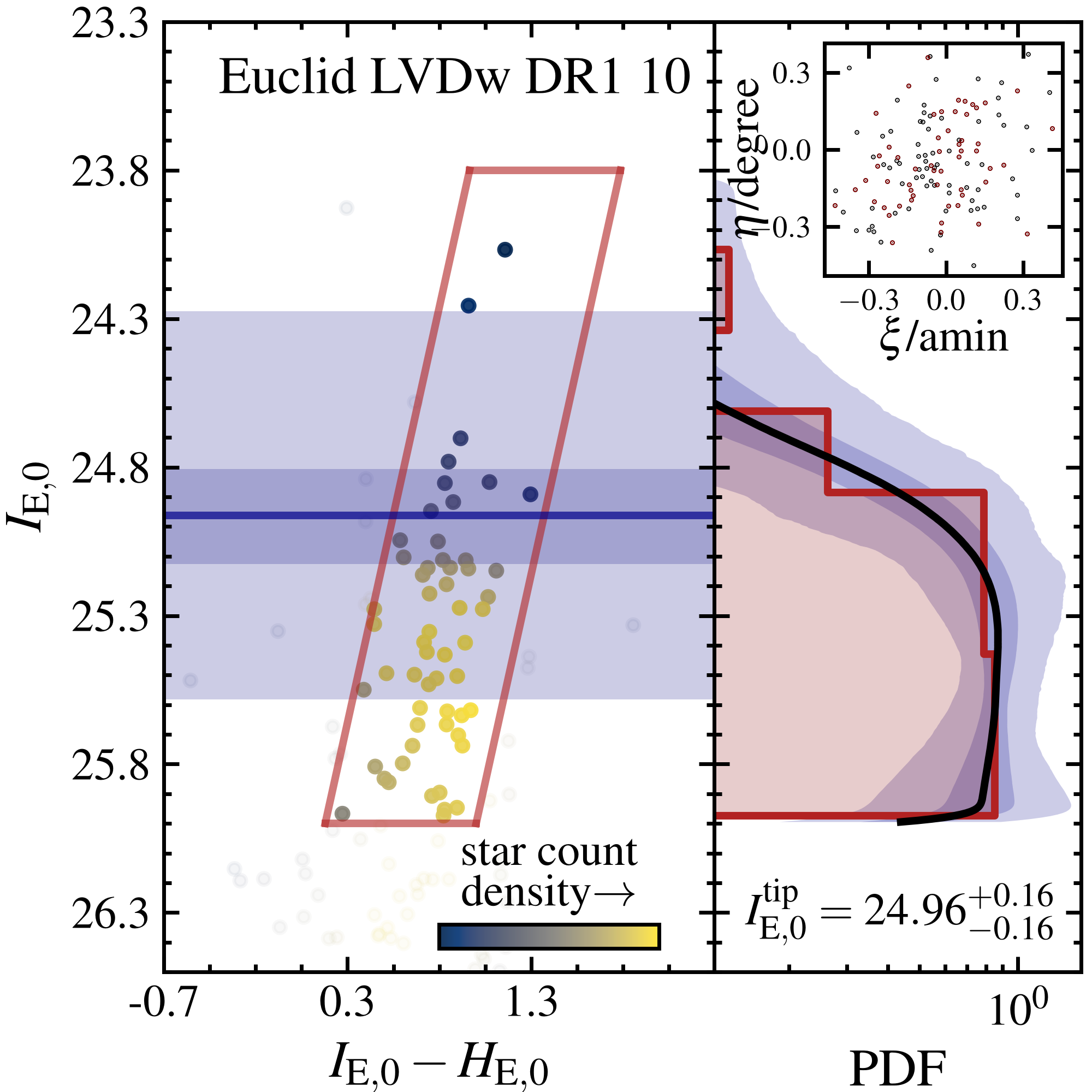}
    \includegraphics[width=.4\linewidth]{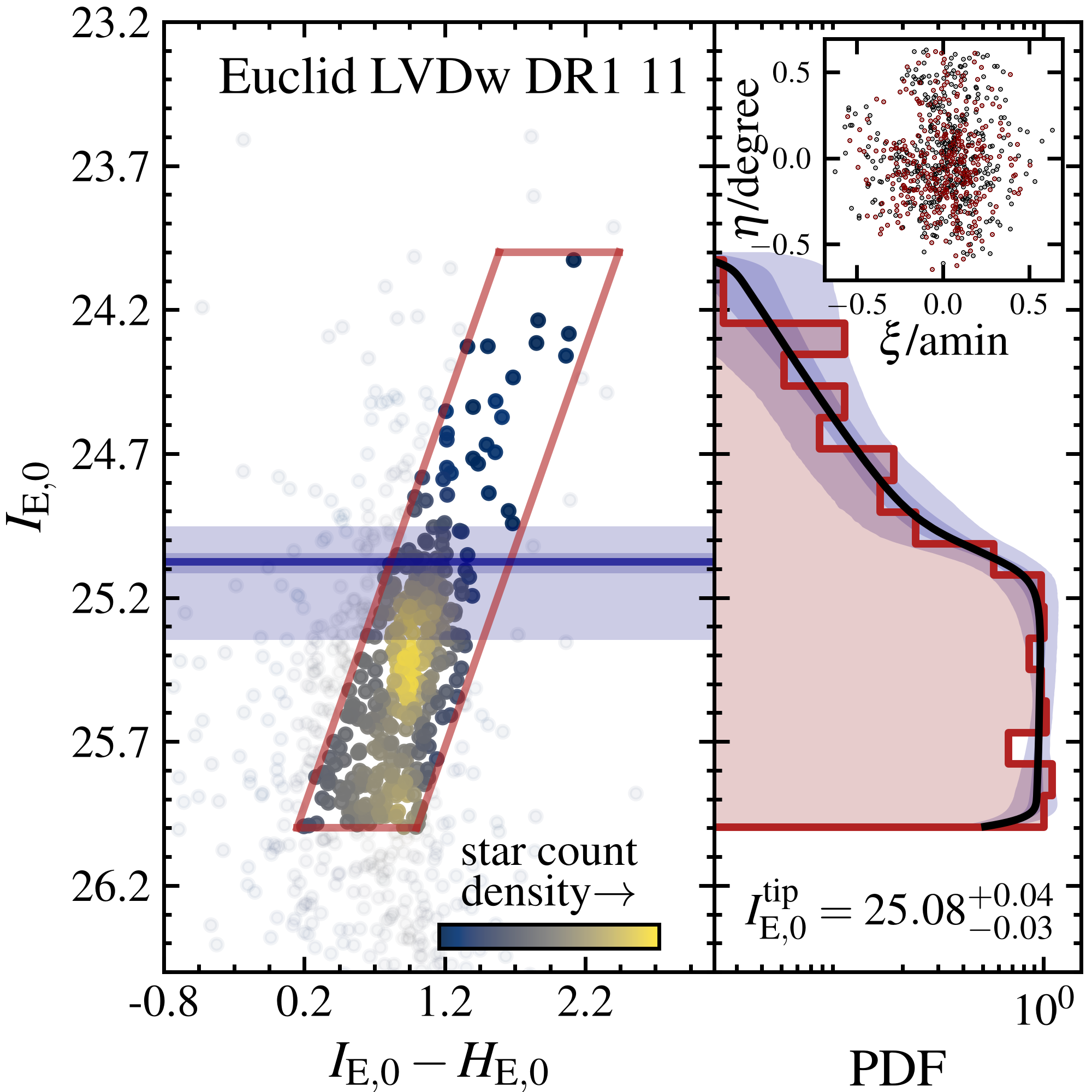}\\
    \caption{\massi{TRGB fit for \EuclidXII{} (top left panel), \EuclidVII{} (top right panel), \EuclidVIII{} (middle left panel), \EuclidX{} (middle right panel), \EuclidXI{} (bottom left panel), and \EuclidIX{} (bottom right panel). \emph{Left:} dereddened CMD of stars within the selection region (red parallelogram) used for the fit. The blue horizontal line marks the position of the RGB tip and the shaded bands show the 1$\sigma$ and 3$\sigma$ confidence intervals. \emph{Right:} Luminosity function of the selected stars as a function of $I_{\rm E,0}$ (red histogram), together with the best-fit Bayesian model (blue line) and its $1\sigma$ and $3\sigma$ confidence intervals. The inset shows the spatial distribution of the selected RGB+AGB candidates (red) along with the stars falling outside the selection region (black).}
    }
    \label{fig:trgb_figure}
\end{figure*}

\begin{figure}
    \centering
    \includegraphics[width=1.\linewidth]{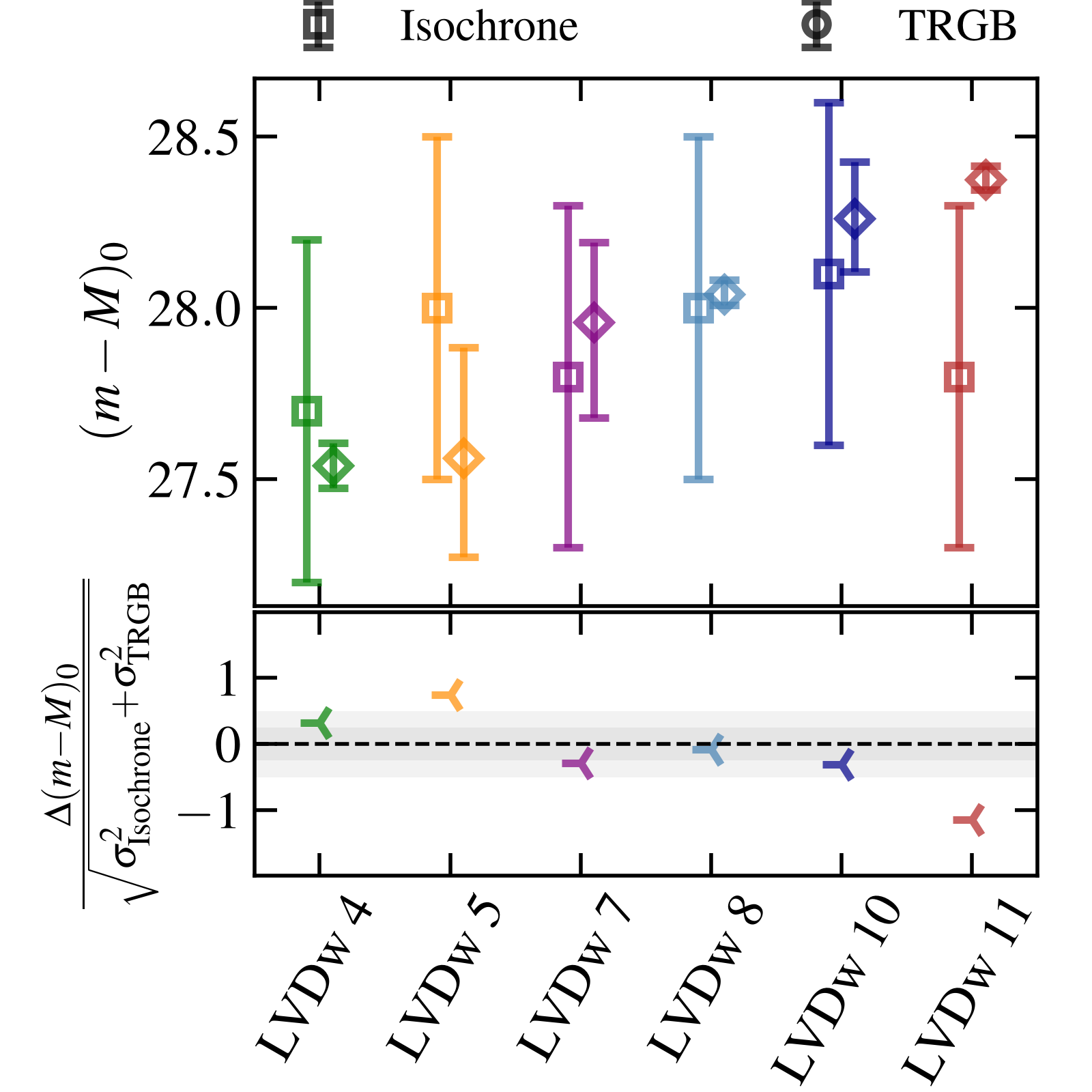}
   \caption{\massi{Distance moduli from the visual isochrone fit versus those from the TRGB for the six galaxies with both estimates.}}
    \label{fig:trgb_comparison}
\end{figure}

%\section{}
%
\massi{Figure~\ref{fig:trgb_figure} shows the TRGB measures for the six galaxies for which the stellar population was sufficiently well sampled for the method \citep{madore23}, namely \EuclidVII{} (top left), \EuclidVIII{} (top right), \EuclidIX{} (middle left), \EuclidX{} (middle right), \EuclidXI{} (bottom left), and \EuclidXII{} (bottom right). The measures have been performed strictly following \citet{Bellazzini&Pascale2024}.
Figure~\ref{fig:trgb_comparison} compares the distance moduli obtained from the visual isochrone fitting and from the TRGB method for the six galaxies for which both estimates are available. Overall, the two methods yield consistent results, agreeing within $1\sigma$ for five of the six galaxies. This indicates that, for distant systems where only the upper RGB is resolved, visual isochrone fitting already provides reliable distance estimates. The only exception is \EuclidIX{}, for which the two methods differ by $\simeq 1.1\sigma$. Notably, \EuclidIX{} is also the most populated galaxy in the sample (Fig.~\ref{fig:CMD_all}, fourth row, third column).
We attribute this discrepancy to the fact that, in a richer population such as that of \EuclidIX{}, a non-negligible presence of intermediate-age, bright AGB stars above the RGB is expected. In the visual fit, mistaking these AGB stars for RGB stars can pull the best-fit isochrone to brighter magnitudes (i.e.\ to a smaller distance modulus) in order to match them. This suggests that, for well-populated systems, purely visual isochrone fitting can be biased by a small number of evolved, non-RGB stars. This further motivates our choice to adopt the TRGB distance as the fiducial value whenever it is available.}

\end{appendix}

\label{LastPage}
\end{document}